\documentclass[11pt]{article}
\usepackage{titling}
\usepackage{graphicx}
\usepackage[T1]{fontenc}
\usepackage{parskip}
\usepackage{latexsym,amsmath,amssymb}
\usepackage{times}
\usepackage{zi4}
\usepackage[a4paper,left=2.5cm,right=2.5cm,top=3cm,bottom=3cm]{geometry}
\usepackage{algorithm,algorithmic} 
\usepackage{booktabs} 
\usepackage{colortbl} 
\usepackage{multirow} 
\usepackage{hyperref} 
\usepackage{xcolor} 
\usepackage{subcaption} 
\usepackage{longtable} 
\usepackage{pdfpages} 
\usepackage{changepage}

\usepackage{fancyhdr}
\usepackage[style=nature, backend=biber, giveninits=true, sortcites=true]{biblatex}
\definecolor{blue}{RGB}{75,104,184}
\definecolor{green}{RGB}{120,190,33}

\newcommand{\gr}[1]{{\color{black} #1}}
\newcommand{\as}[1]{{\color{black} #1}}

\begin{document}

\title{Network-based time series modeling for COVID-19 incidence in the Republic of Ireland}

\author{
Stephanie Armbruster \\
\small Department of Biostatistics \\
\small Harvard University, 655 Huntington Avenue, Boston, MA 02115, USA \and 
Gesine Reinert$^*$ \\ 
\small Department of Statistics \\
\small University of Oxford, 24-29 St Giles, Oxford OX1 3LB, UK \\ 
\\
\small $^*$Corresponding author: reinert@stats.ox.ac.uk
}

\date{}

\maketitle

\begin{abstract}
Network-based time series models have experienced a surge in popularity over the past years due to their ability to model temporal and spatial dependencies, arising from the spread of infectious disease.
The generalised network autoregressive (GNAR) model conceptualises time series on the vertices of a network; it has an autoregressive component for temporal dependence and a spatial autoregressive component for dependence between neighbouring vertices in the network. 
Consequently, the choice of underlying network is essential.
This paper assesses the performance of GNAR models on different networks in predicting COVID-19 cases for the 26 counties in the Republic of Ireland, over two distinct pandemic phases (restricted and unrestricted), characterised by inter-county movement restrictions.
Ten static networks are constructed, in which vertices represent counties, and edges are built upon neighbourhood relations, such as railway lines.  

We find that a GNAR model based on the fairly sparse Economic hub network explains the data best for the restricted pandemic phase while the fairly dense $21$-nearest neighbour network performs best for the unrestricted phase.
Across phases, GNAR models have higher predictive accuracy than standard ARIMA models which ignore the network structure. 
For county-specific predictions, in pandemic phases with more lenient or no COVID-19 regulation, the network effect is not quite as pronounced. 
The results indicate some robustness to the precise network architecture as long as the densities of the networks are similar.
An analysis of the residuals justifies the model assumptions for the restricted phase but raises questions regarding their validity for the unrestricted phase.
While 
\gr{generally performing better than} ARIMA models which ignore network effects, \gr{there is scope for further development of} the GNAR model 
to better model complex infectious diseases, including COVID-19. 

\noindent
Network-based time series, COVID-19, spatial models, networks
\end{abstract}

\setlength{\tabcolsep}{6pt}

\setlength\parindent{24pt}

\section{Introduction}
\label{chapter: introduction}
In recent years, statistical models which incorporate networks and thereby acknowledge spatial dependencies when predicting temporal data have experienced a surge in popularity (e.g.,\,\cite{knight2019generalised, knight2016modelling, urrutia2022sars}). 
Knight et al.\,\cite{knight2016modelling} developed a generalised network autoregressive (GNAR) time series model which incorporates a secondary dependence in addition to standard temporal dependence. 
The secondary dependence is captured in a network.  
In \cite{knight2016modelling}, the proposed network-based time series model is leveraged to predict mumps incidence across English counties during the British mumps outbreak in 2005. \gr{As graph to be  associated with the Mumps network time series, \cite{knight2016modelling} chose a ``county town'' for each county and connected all towns which were less than a radius of a fixed number of kilometers away from each other.} \\
Similar to mumps, COVID-19 is a highly infectious disease spread by direct contact between people \cite{nouvellet2021reduction}. 
Human movement networks have been extensively relied upon to explain COVID-19 patterns (e.g.\,\cite{jia2020population, kraemer2020effect, li2021assessing, mo2021modeling, nouvellet2021reduction, sun2021degree, wu2020nowcasting}). 
Therefore, it is a natural conjecture that such movement networks may help predict the spread of COVID-19. 
To investigate, this paper \begin{itemize}
    \item fits  GNAR models 
to predict the weekly COVID-19 incidence for all 26 counties in the Republic of Ireland, exploring different network constructions;
\item assesses the prevalence of a network effect in COVID-19 incidence in Ireland and the suitability of GNAR models  to predict epidemic outbreaks as complex as COVID-19;
\item investigates the influence of changes in inter-county mobility, due to COVID-19 restrictions, on the performance of GNAR models as well as on the model parameters and hyperparameters.
\end{itemize} 

\gr{The GNAR model is chosen because multivariate time series are often modelled by vector autoregressive (VAR) models. 
General VAR models are very flexible but require a large number of parameters to be estimated. 
The GNAR model which we employ here is a special case of a VAR model. It reduces the number of parameters to be estimated by restricting attention to edges in a network; in the case of a complete  graph, the VAR model and the GNAR model coincide. 
\as{\gr{Our} overview of network-based time series models, given in Supplementary Material \ref{app:lit_rev_network_models},  concludes that 
\gr{many} network-based time series models 
can be conceptualized as a special case of the GNAR model,  or are more restrictive \gr{with respect to the} 
temporal-spatial dependencies they can model.}
\gr{Moreover, as a VAR-type model,} the GNAR model inherits the well-understood VAR model framework, including parameter estimation via least squares, and model selection based on the BIC; for a survey see for example \cite{lutkepohl2005new}. 
These methods yield confidence sets for parameter estimation, which can inform analysis as well as policy development in a quantitative fashion. In contrast, deep learning approaches such as developed in \cite{park2024enhancing} for predicting rental and return patters at bicycle stations do not come with such theoretical guarantees.}

\gr{In addition to a distance-based network as chosen in \cite{knight2016modelling}, in this paper we explore a collection of network models based on potential movements of individuals.}  The \gr{movement} networks are constructed according to general approaches based on statistical definitions of neighbourhoods as well as approaches specific to the infectious spread of the COVID-19 virus. \gr{In abuse of notation, we call these networks {\it COVID-19 networks}, although they are only meant to reflect possible transmission routes of the disease.}  
For each network, we select the best performing hyperparameter values, to predict COVID-19 incidence by a GNAR model, using the Bayesian Information Criterion \as{(BIC)}.
By splitting the available Irish data into two phases of the pandemic, restricted and unrestricted, we are able to investigate the potential change in the temporal and spatial dependencies in COVID-19 incidence between the two phases.

Overall, our findings are that while there is a clear network effect, the performance of the optimal GNAR model varies little across different network architectures of similar network density. 
GNAR models indicate higher predictive accuracy than ARIMA models on a country level, since they account for inter-county dependencies.
On an individual county level, the variability of predictive performance is high, resulting in similar performance of ARIMA and GNAR models for some counties, while for others the GNAR model consistently outperforms the ARIMA model. 
The GNAR model seem better suited to predictive COVID-19 incidence in restricted pandemic phases than in unrestricted pandemic phases; the latter may be related to some of the model assumptions possibly requiring an adaptation as well as an increase in noise during unrestricted pandemic phases and high fluctuation of COVID-19 case numbers. \gr{Moreover, the classical VAR model, which is the GNAR model on the complete graph,  does not perform as well as the GNAR model with an underlying network that has fewer edges, illustrating the value of using a GNAR model.}

This paper is organised as follows. 
Section \ref{chapter: data} introduces the data set. 
The methodology for network construction and for network-based time series modeling is described in Section \ref{chapter:methodology}. 
Section \ref{chapter: exploration} provides an exploratory data analysis, while the model fit is shown in Section \ref{results_covid}. 
The conclusions for the different pandemic phases are found in Section \ref{chapter: phases}. 
The results are discussed 
in Section \ref{chapter: conclusion}.
The Supplementary Material includes a literature review on alternative models for incorporating temporal and spatial dependencies, visualisations of the COVID-19 networks, as well as details on the performance of GNAR models for predicting COVID-19 incidence.  

The data and code are provided \gr{at} \url{https://github.com/stephanieArmbru/Case_study_GNAR_COVID_Ireland.git}.

\section{The Irish COVID-19 data set}
\label{chapter: data}
By March 2023, the Republic of Ireland (abbreviated in this paper as {\it Ireland}) had recorded a total of 1.7 million confirmed COVID-19 cases and 8,719 deaths since the beginning of the pandemic \cite{COVID_dashboard}. 
The Health Protection Surveillance Centre identified four main variants of concern for the COVID-19 virus in Ireland \cite{hpsc_covid_variants}, in addition to the original variant: Alpha from 27.12.2020, Delta from 06.06.2021, Omicron I from 13.12.2021, and Omicron II from 13.03.2022 (Figure 1a in \cite{hpsc_covid_variants}).
The open data platform \cite{COVID_data_hub, COVID_data} by the Irish Government provides weekly updated multivariate time series data on confirmed daily cumulative COVID-19 cases for all 26 Irish counties, starting from the beginning of the pandemic in February 2020. 
A COVID-19 case is attributed to the county in which the patient has their primary residence \footnote{This attribution is only partially reliable due to a lack of validation during infection surges \cite{COVID_dashboard}.}. 
\as{To 
our knowledge, \gr{spatially} more detailed COVID-19 data is not available for Ireland. 
The limited granularity makes it difficult to implement \gr{fine-scale} 
spatial models. 
The GNAR model was originally demonstrated using county-level data, indicating its potential for modeling infectious diseases at lower resolutions.}
The cumulative case count is given for 100,000 inhabitants. 
Age profiles vary across counties and COVID-19 infection rates are age dependent. 
Hence, the cumulative case count is adjusted for age distribution according to the 2016 census of Ireland, to ensure inter-county comparability \cite{ireland_census}. 
In our dataset, the first COVID-19 case was registered in Dublin on 02.03.2020 and the last reported date is 23.01.2023, 
spanning a total of 152 weeks \cite{COVID_data}. 
From 20.03.2020 onward, COVID-19 cases were recorded in every Irish county. 
The daily COVID-19 data is aggregated to a weekly level to avoid modelling artificial weekly effects \cite{kubiczek2021challenges, sartor2020covid}.
Due to delayed reporting during winter 2021/22, the weekly COVID-19 incidences from 12.12.2021 to 27.02.2022 are averaged over a window of 4 weeks \cite{HPSC_covid_report_2020/21, COVID_report_2022_w24, wei2006time}.

The main COVID-19 regulations restricting physical movement and social interaction between Irish counties \cite{IM_timeline_pandemic_2020, independent_timeline_pandemic_video, IE_timeline_pandemic, IT_inter_county_travel} are used to naturally split the data into five sequential subsets, where the COVID-19 incidence is small at the beginning of the pandemic and shows a clear increasing trend over time.   
$\Bar{\sigma}$ denotes the average standard deviation in COVID-19 incidence across the 26 Irish counties within the considered data subset.
\begin{enumerate}
    \item[(1)] 
Start of the pandemic,  with gradually stricter movement restrictions and lockdowns (\textit{restricted} phase), from 27.02.2020 for 25 weeks ($\Bar{\sigma} = 19.17$);  
\item[(2)]  County-specific movement restrictions (\textit{less restricted} phase), from 18.08.2020 for 18 weeks ($\Bar{\sigma} = 35.05$);
\item[(3)]  Level-5 lockdown (\textit{restricted} phase),  with inter-county travel restrictions, from 26.12.2020 for 20 weeks ($\Bar{\sigma} = 148.79$);
\item[(4)] Allowance of non-essential inter-county travel (\textit{less restricted} phase), from 10.05.2021 for 43 weeks ($\Bar{\sigma} = 173.17$); 
\item[(5)] 
End of all restrictions (\textit{unrestricted} phase), from 06.03.2022 for 46 weeks ($\Bar{\sigma} = 101.46$).
\end{enumerate}

The datasets are grouped to avoid small sample sizes. 
Time gaps in either dataset are inserted as missing data. 
\begin{itemize}
    \item {\it restricted} dataset: dataset 1 and 3; representing pandemic situations with strict COVID-19 restrictions, including an inter-county travel ban \cite{IM_timeline_pandemic_2020}; contains 45 weeks of observation ($\sigma_r = 99.27$)
    \item {\it unrestricted} data set: dataset 2, 4 and 5; representing periods with fewer or no regulations, in particular no inter-county travel limitations \cite{IT_inter_county_travel}; contains 107 weeks of observation ($\sigma_{ur} = 129.62$)
\end{itemize}


\section{Methodology}
\label{chapter:methodology}
Let $\mathcal{G} = \{\mathcal{V}, \mathcal{E}\}$ denote a fixed, simple, undirected, unweighted network with vertex set $\mathcal{V}$ containing $N$ vertices and edge set $\mathcal{E}$; an edge between vertices $i$ and $j$ is denoted by $i \sim j$. 
The neighbourhood of a subset of vertices $A \subset \mathcal{V}$ is defined as the set of neighbours outside of $A$ to the vertices in $A$, $N(A) = \bigcup_{i \in A} \{ j \in \mathcal{V} \backslash A: i \sim j\}.$
The set of \textit{$r^{\,th}$-stage neighbours}, or the \textit{$r^{\,th}$-stage neighbourhood}, for vertex $i \in A$ is defined recursively as $N^{(0)}(i) = \{ i\}$ and
\begin{align*}
    N^{(r)}(i) = N\left( N^{(r-1)}(i) \right) \backslash \bigcup_{q = 1}^{r-1} N^{(q)}(i) \; . 
\end{align*}

\subsection{COVID-19 networks: constructions and properties} 
\label{chapter:methodology1}
\as{
The key to the GNAR model is the network. 
The true network underlying the data generating process \gr{(in this case, who infected whom in the spread of the disease)} is usually unknown. 
Ideally, expert knowledge can be leveraged to build a network that captures the relationship between \gr{vertices}, representing the subjects of interest. 
Networks to model the spread of an infectious disease, such as COVID-19, are frequently modelled off of human mobility patterns, which are considered to have a shaping influence on disease spread (e.g. \cite{colizza2006role,jia2020population,li2021assessing,mo2021modeling,sun2021degree}).
To 
our knowledge, detailed information on \gr{weekly} population flow between Irish counties is unavailable. 
Hence, we construct implicit COVID-19 transmission networks (COVID-19 networks hereafter) based on geographical approaches, in line with \cite{knight2016modelling}. 
}

In the \textit{Railway-based network}, an edge is established between two counties if there exists a direct train link between the respective county towns (without change of trains) and the county towns are closest to each other on this train connection. 
The \textit{Queen's contiguity} network connects each county to the counties it shares a border with \cite{sawada_spatial_autocorrelation}.
The \textit{Economic hub network} adds an additional edge between each county and its nearest economic hub to the Queen's contiguity network: Dublin, Cork, Limerick, Galway or Waterford \cite{ireland_cities}. 
To measure the distance to the nearest economic hub we use the \textit{Great Circle distance} $d_C({i, j})$, the shortest distance between two points on the surface of a sphere \cite{weisstein2002great}.
For two points $i, j$ with latitude $\delta_i, \delta_j$ and longitude $\lambda_i, \lambda_j$ on a sphere of radius $r > 0$, 
\begin{align*}
    d_C(i, j) &= r \cdot  cos^{-1}\left( cos(\delta_i) \cdot cos(\delta_j) \cdot cos(\lambda_i - \lambda_j)  - sin(\delta_i) \cdot cos(\delta_j) \right) \; .
\end{align*}
The \textit{K-nearest neighbours network} (KNN) connects a vertex with its $K$ nearest neighbours with respect to $d_C$ \cite{Bivand_2022_spdep,eppstein1997nearest}. 
The \textit{distance-based 
neighbour network} (DNN) constructs an edge between counties if their Great Circle distance $d_C$ lies within a certain range $[l, r]$ \cite{weisstein2002great}\gr{; this construction is similar to the one used in \cite{knight2016modelling}}. 
For the COVID-19 network, we set $l = 0$ and consider $r$ a hyperparameter, chosen large enough to ensure that no vertex is isolated. 
The maximum value for $r$ is determined by the largest distance between any two vertices, for which it returns a fully connected network \cite{bivand2013applied}. 

In addition to these geographical networks, the \textit{Delaunay triangulation} constructs geometric triangles between vertices such that no vertex lies within the circumsphere of any constructed triangle \cite{chen2004optimal}, thus ensuring that there are no isolated vertices. 
The \textit{Gabriel}, \textit{Sphere of Influence network} and \textit{Relative neighbourhood} are obtained from the Delaunay triangulation network by omitting certain edges. 
In a \textit{Gabriel network}, vertices $x$ and $y$ in Euclidean space are connected if they are {\it Gabriel neighbours}; that is, 
\begin{align*}
    d(x,y) \leq  \min \left(\left. \sqrt{(d(x,z)^2 + d(y,z)^2)} \, \right \vert \, z \in \mathcal{V} \right) \; 
\end{align*}
where  $d(x, y) = \sqrt{\sum_{i = 1}^n(x_i - y_i)^2}$ denotes the Euclidean distance.
In a \textit{Sphere of Influence network} (SOI), long distance edges in the Delaunay triangulation network are eliminated and only edges between {\it SOI neighbours} are retained, as follows. 
For $x \in \mathcal{V}$ and $d_x$ the Euclidean distance between $x$ and its nearest neighbour in $\mathcal{V}$, let $C_x$ denote the circle centred around $x$ with radius $d_x$.
For $y \in  \mathcal{V}$, the quantities $d_y$ and $C_y$ are defined analogously. 
Vertices $x$ and $y$  are SOI neighbours if and only if $C_x$ and $C_y$ intersect at least twice, preserving the symmetry property of the Delaunay triangulation \cite{bivand2013applied}.
The \textit{Relative neighbourhood network} only retains edges between relative neighbours, 
\begin{align*}
    d(x,y) \leq \min\left(\left. \max(d(x,z), d(y,z)) \, \right \vert \, z \in \mathcal{V}\right) \; .
\end{align*}
The Relative neighbourhood network is contained in the Delaunay triangulation, SOI and Gabriel network,  and  is the sparsest of the four networks \cite{bivand2013applied}. 
Finally, the \textit{Complete network} represents the homogeneous mixing assumption, where all countries are connected \cite{bansal2007individual}. 

\begin{figure}[h!]
\centering
\subcaptionbox{\as{Economic hub} network}{\includegraphics[width=0.325\textwidth]{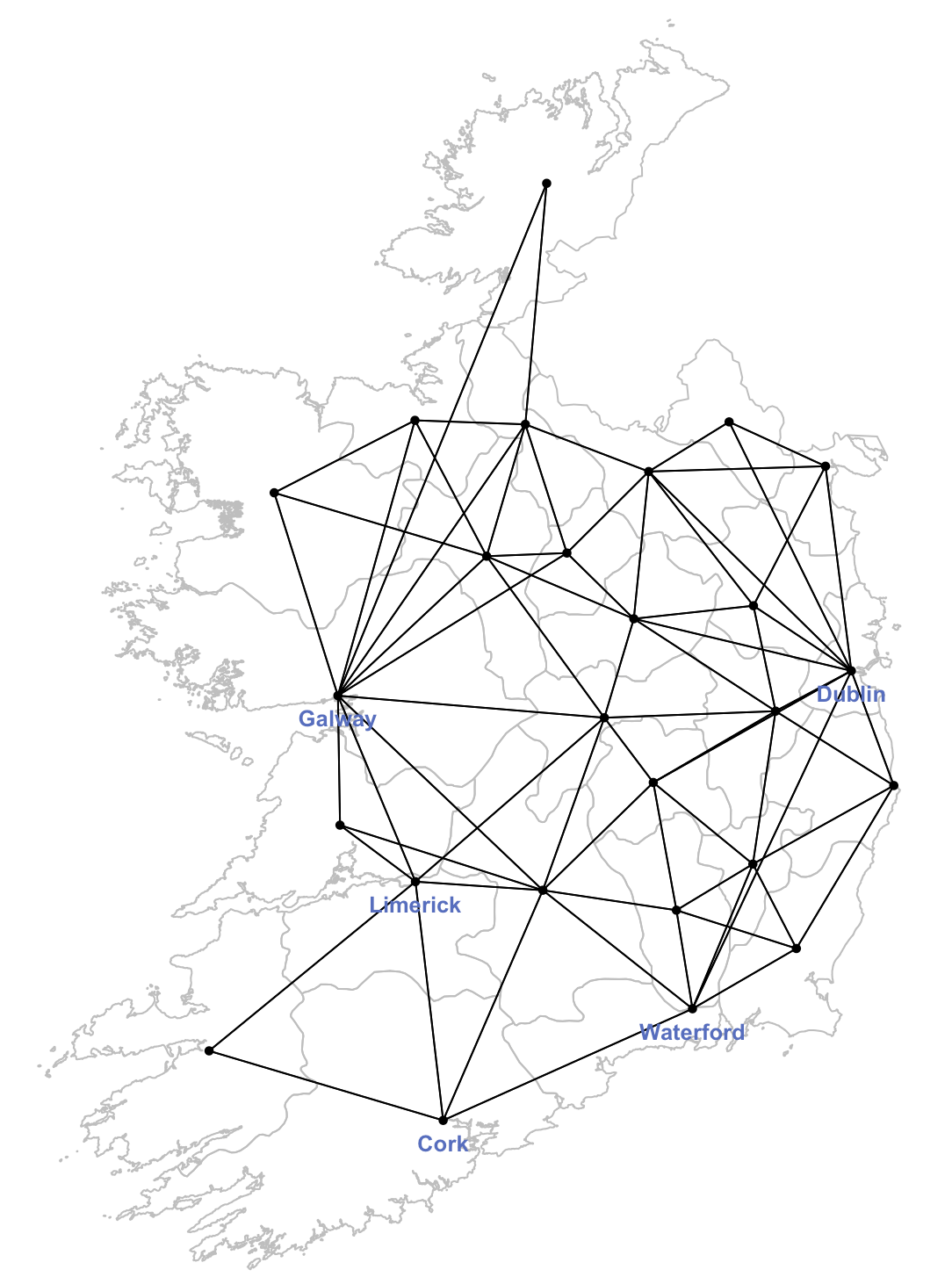}}%
\hspace{0.1cm} 
\subcaptionbox{KNN ($k = 11$) network}{\includegraphics[width=0.325\textwidth]{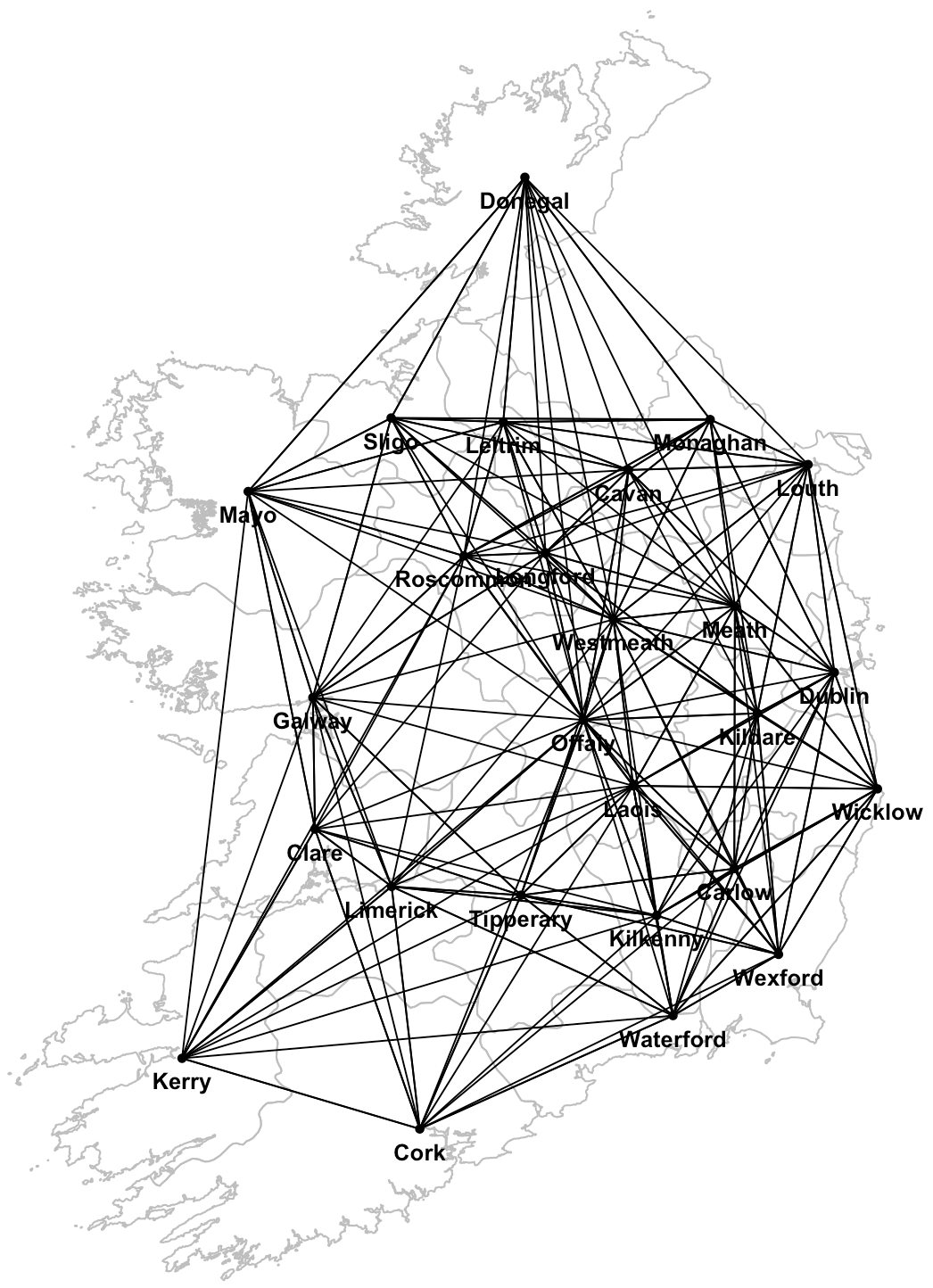}}%
\hspace{0.1cm} 
\subcaptionbox{KNN ($k = 21$) network}{\includegraphics[width=0.325\textwidth]{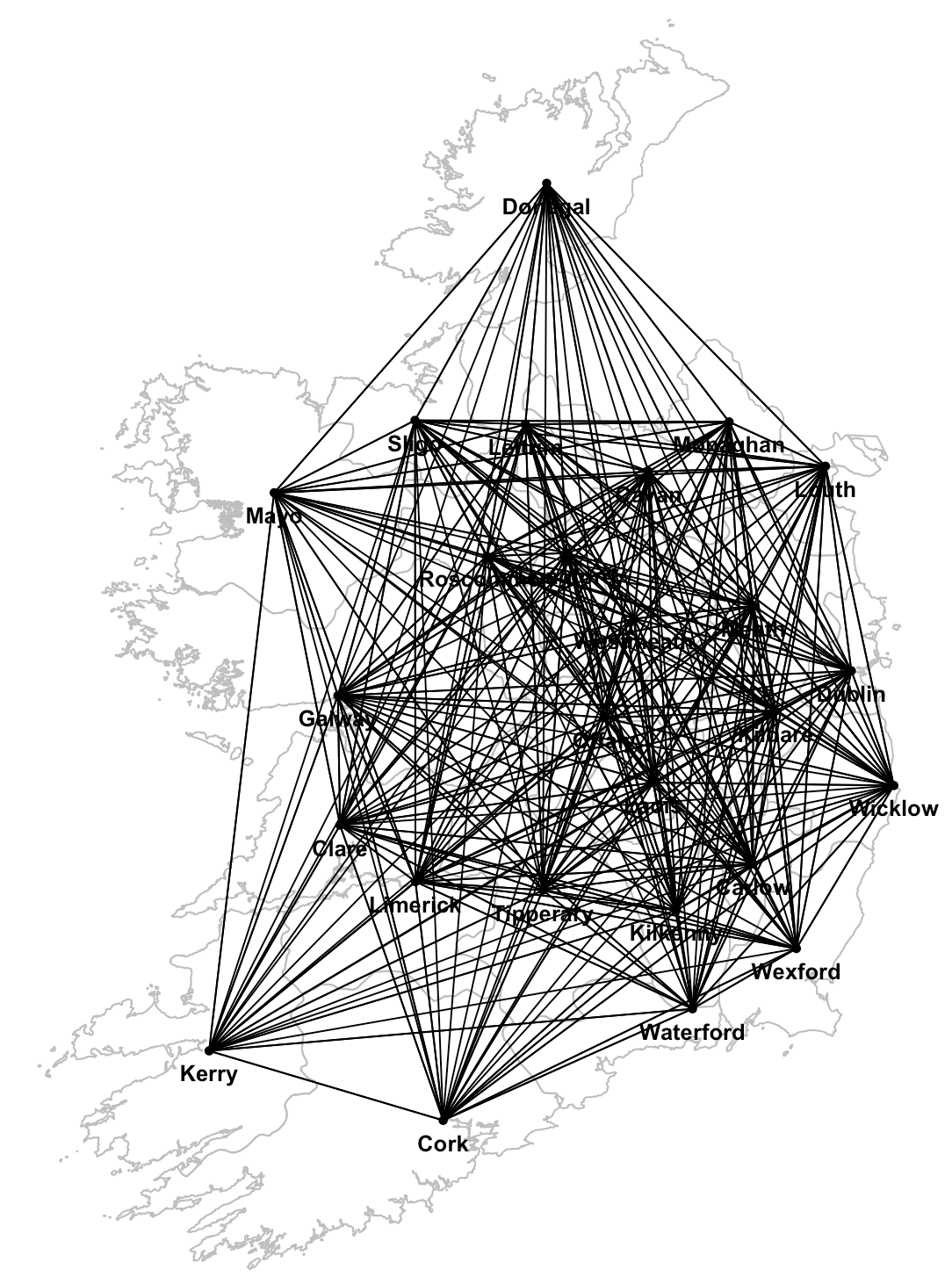}}%
\caption{Map of Ireland and COVID-19 networks; \as{economic hub towns marked in blue}}
\label{fig:network_maps}
\end{figure}

Figure \ref{fig:network_maps} shows the 
\gr{Economic hub} network and the KNN network \as{($k = 11$ and $k = 21$)} for Ireland. 
Figures of the other networks are found in the Supplementary Material \ref{app:networks}; network summaries are provided in Table \ref{tab:network_char}.

The networks are created based on 
the literature on spatial modelling; \cite{bivand2008applied} (pp. 239-251) suggests the Delaunay network and its variants,  Queen's contiguity network, distance-based networks, and $K$-nearest neighbour networks. In \cite{de2021airports} it was found that infrastructure has an effect on the spread of COVID-19 in Brazil; here we use the railway network as infrastructure network. The Economic hub network is motivated by the idea of including a proxy of commuter flows in the network construction, as commuting to work has been shown in \cite{mitze2022propagation} to be related to the spread of COVID-19 in Germany.

\subsection{Generalised network autoregressive models}
\label{chapter:methodology2}
Network-based time series models incorporate non-temporal dependencies in the form of networks in addition to temporal dependencies as established in time series models \cite{knight2019generalised,knight2016modelling,zhu2017network}. 
In contrast to standard time series methodology and spatial models \cite{box2015time,HamiltonJamesDouglas2020TSAe, wei2006time}, network-based time series models are not limited to geographic relationships but can incorporate any generic network. 
As COVID-19 is an infectious disease with spatial spreading behaviour, warranting constructing networks based on spatial information, we use terms relating to spatial dependence in our exposition. 
Other types of dependence could easily be incorporated in the model through networks which reflect the hypothesised dependence.

The \textit{global $\alpha$ generalised network autoregressive models} GNAR(p, s) models the observation $X_{i, t}$ for a vertex $i$ at time $t$ as the weighted linear combination of an autoregressive component of order $p$ and a network neighbourhood autoregressive component of a certain order, also called neighbourhood stage; for $i=1, \ldots, p$, the entry $s_i$ gives the largest neighbourhood stage considered for vertex $i$ when regressing on up to $p$ past values. 
In our analysis, $X_{i, t}$ denotes the 1-lag difference in weekly COVID-19 incidence over time $t$ for counties $i$. 
The effect of neighbouring vertices depends on some weight $\omega_{i, q}$.
A GNAR model GNAR(p-$s_1, \ldots, s_p$) has the following form, 
\begin{align}
\label{equ:GNAR}
X_{i, t} = \sum_{j = 1}^p \left( \alpha_{j} X_{i, t-j} + \sum_{r = 1}^{s_j} \sum_{q \in N^{(r)}(i)} \beta_{j, r} \, \omega_{i, q}^{(t)} \, X_{q, t-j} \right) + \varepsilon_{i, t} 
\end{align}
where $\varepsilon_{i,t} \sim N(0, \sigma^2_i)$ are uncorrelated\footnote{We define $\sum_{r = 1}^0 (.) := 0$.}. 
As weights $\omega_{i, q}$ we choose the normalised inverse shortest path length weight, where $d_{i, q}$ denotes the shortest path length (SPL) \cite{knight2016modelling}; in connected networks, $1 \leq d_{i, q} < \infty$ for $i \ne q$. 
For $i \in \mathcal{V}$ and $q \in N^{(r)}(i)$, we thus take, $\omega_{i, q} = {d_{i, q}^{-1}}/ ({\sum_{k \in N^{(r)}(i) } d_{i, k}^{-1}})$.

The general GNAR model relies on vertex specific coefficients $\alpha_{i, j}$, instead of vertex unspecific autoregressive coefficients, $\alpha_j$, indicating vertex specific temporal dependence.
This modification is comparable to including individual random effects in regression models. 

To fit a GNAR model, we must choose two hyperparameters, the lag $p$, or \textit{$\alpha$-order}, and the vector of neighbourhood stages, $s = (s_1, ..., s_p)$, also called \textit{$\beta$-order}. 
They can be determined either through expert knowledge, e.g.\,on the spread of infections, or through a criterion-based search \cite{knight2016modelling}. 
The model coefficients are computed via Estimated Generalised Least Squares (EGLS) estimation \cite{knight2016modelling, leeming2019new, Lutkepohl1991Itmt}\footnote{Additional information in Supplementary Material \ref{app:gls_estimation}}.

\subsection{GNAR model selection and predictive accuracy}
For our analysis of the Irish COVID-19 data, model selection, i.e.\,the choice of $\alpha$- and $\beta$-order, is performed by minimizing the \textit{Bayesian Information Criterion} (BIC) \cite{knight2016modelling}.
The BIC avoids overfitting by penalizing the observed likelihood $L$ by the dimensionality of the required parameters \cite{schwarz1978estimating}. 
For a sample $X$ of size $n$ and a parameter $\theta$ of dimension $k$, $ \text{BIC}(k, n) =  k \text{log}(n) - 2 \cdot \text{log}(L(X; \theta))$.

The GNAR package assumes Gaussian errors \cite{source_code_GNAR}; under this assumption, the BIC is consistent.
This assumption could be weakened; it can be shown that the BIC is consistent for the GNAR model \eqref{equ:GNAR} if the error term is i.i.d.\,with bounded fourth moments \cite{leeming2019new, lutkepohl2005new, lv2014model}. 

The predictive accuracy of a GNAR model is measured by the \textit{mean absolute scaled error} (MASE), due to its insensitivity towards outliers, its scale invariance and its robustness \cite{hyndman2006another}.
MASE is defined for each county $i$ as the ratio of absolute  forecasting error $\hat{\varepsilon}_{i, t} = 
|X_{i, t} - \hat{X}_{i, t}|$ divided by the mean absolute error between true and a naive 1-lag random walk forecast for the entire observed time period $[1, T]$ \cite{hyndman2006another,urrutia2022sars}; 
\begin{align*}
    |q_{i, t}| = \frac{|X_{i, t} - \hat{X}_{i, t} | 
    }{\frac{1}{T-1} \sum_{l = 2}^T |X_{i, l} - X_{i, l - 1}|} \;.
\end{align*}

\section{Data Exploration}
\label{chapter: exploration}

\subsection{The weekly incidence differences}
\label{weekly_ID}
\begin{figure}
    \centering
    \includegraphics[scale = 0.45]{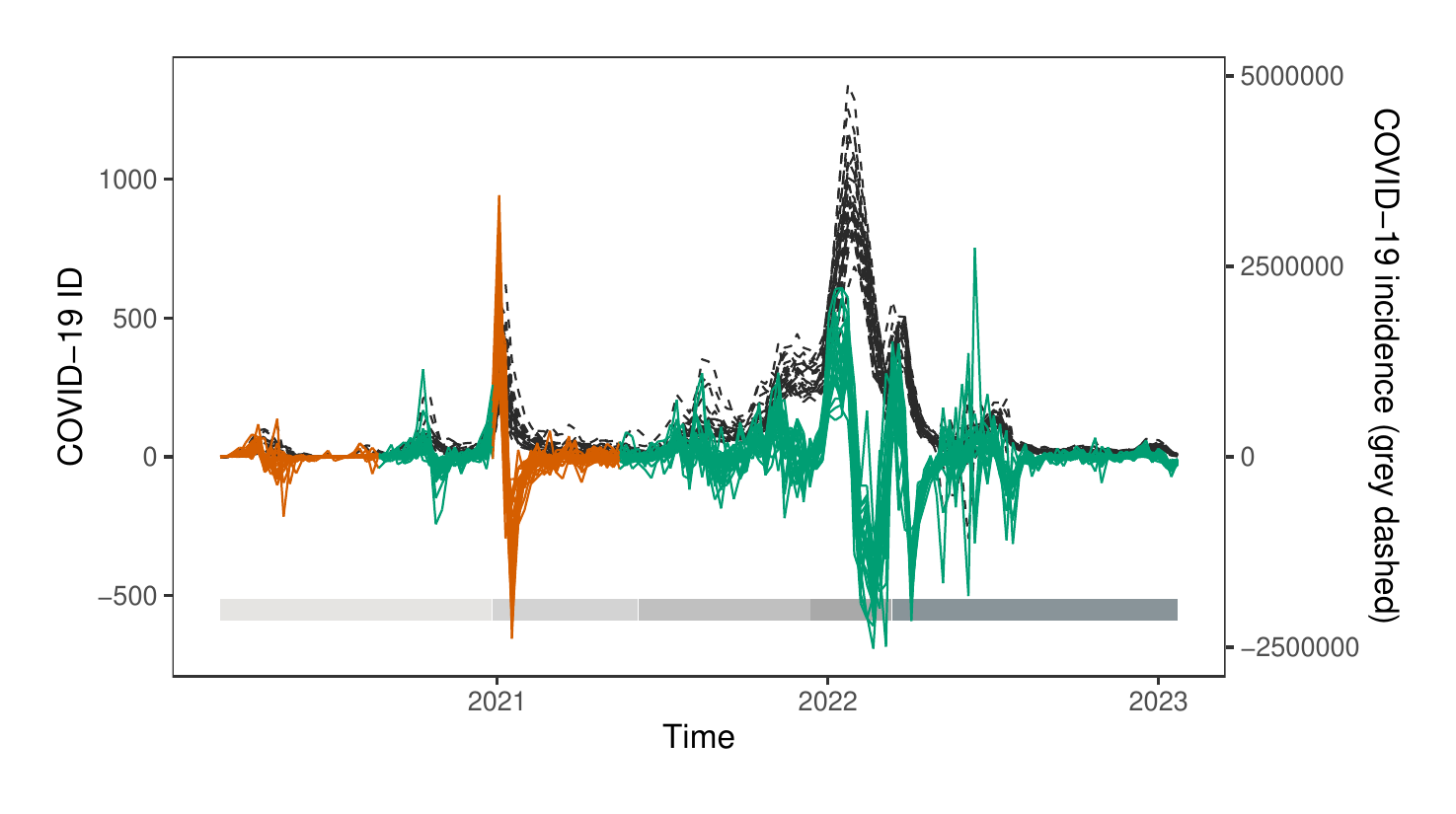}
    \caption[Weekly 1-lag COVID-19 ID with COVID-19 regulations and virus strains]{Weekly COVID-19 incidence difference (ID) and weekly COVID-19 incidence (dark grey dashed lines) for 100,000 inhabitants, from the start of the pandemic in March 2020 to mid June 2022, red for restricted phases, green for unrestricted; predominant COVID-19 virus variants shown by color scale at the bottom, in order: Original, Alpha, Delta, Omicron I and Omicron II}
    \label{fig:covid_weekly_cases}
\end{figure}

The GNAR model requires stationary data. 
Stationary data has no trend over time and is homogeneous, i.e. has time-independent variance \cite{knight2016modelling}. 
The weekly COVID-19 count is clearly not stationary, as it shows an increasing trend in Figure \ref{fig:covid_weekly_cases}. 
To remove any linear trend, we perform 1-lag differencing on the weekly COVID-19 incidence for the 26 Irish counties, resulting in the incidence difference, \textit{(1-lag) COVID-19 ID}, between two subsequent weeks \cite{montgomery2015introduction}. 
We assess the stationarity by applying a Box-Cox transformation to each data subset. 
As evident from Figure \ref{fig:stationarity} in the Supplementary Material \ref{app:stationarity}, the values for $\lambda$ achieving maximal likelihood fall close (enough) to 1, indicating that no further transformation is required.

\subsection{Constructed networks}
For the COVID-19 KNN network, neighbourhood sizes sequencing from $k = 1$ to the fully connected network, $k = 25$, by 2 steps are considered.
The minimal distance for the COVID-19 DNN network measures 90.3 km, between Kerry and Cork, and the maximal value 338.5 km, between County Cork and Donegal.
The KNN and DNN network parameters are chosen to minimise the BIC of the associated GNAR model. 
For the restricted pandemic phase, the KNN network has $k = 11$ and the DNN network $d = 325$. 
For the unrestricted pandemic phase, it is $k = 21$ and $d = 325$.

\begin{figure}[h!]
    \centering
    \includegraphics[scale = 0.4]{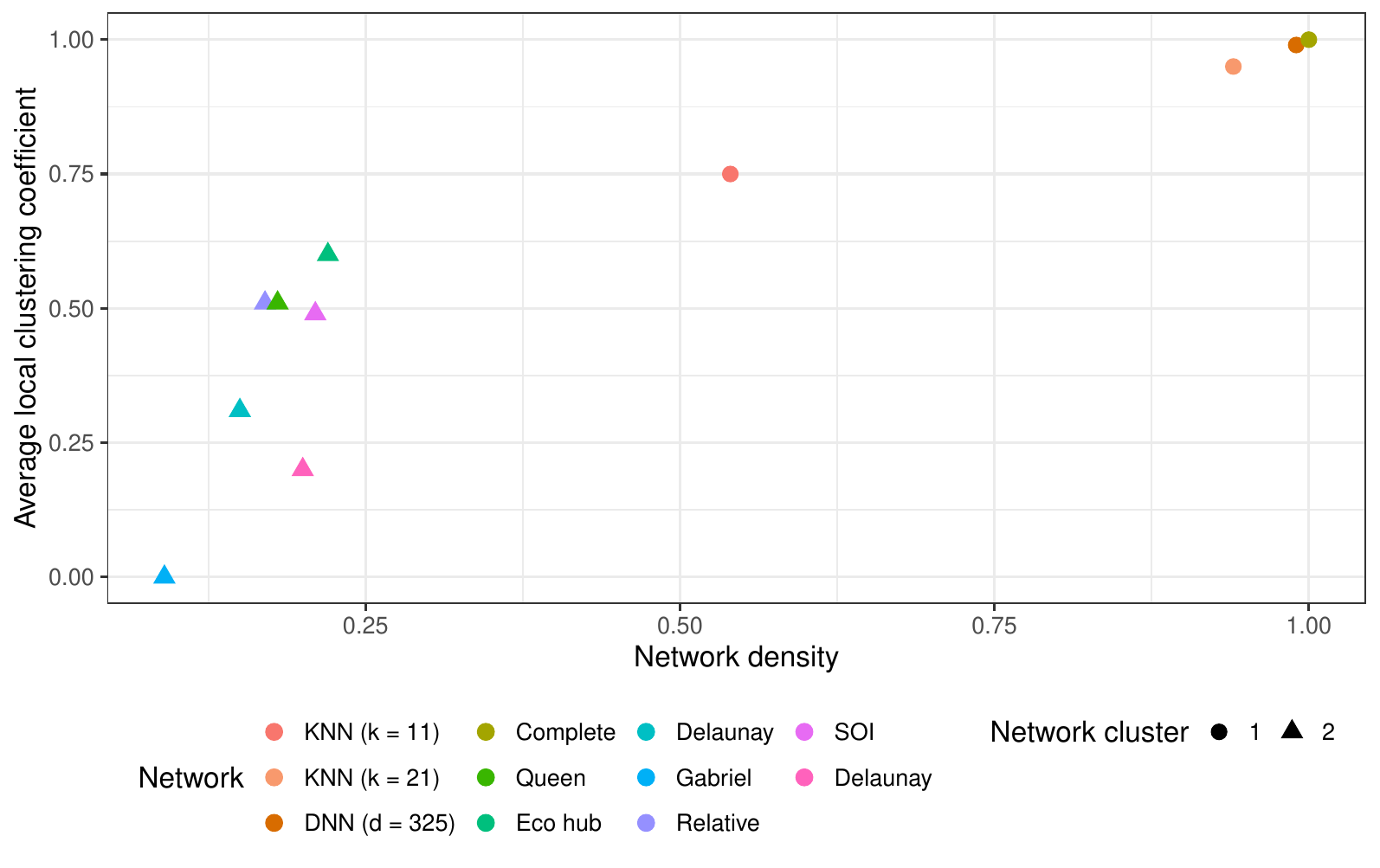}
    \caption{Network density against average local clustering coefficient for all 10 considered COVID-19 networks; 2 clusters detectable}
    \label{fig:density_clustering}
\end{figure}

There is considerable variability in network characteristics, Table \ref{tab:network_char}, in particular regarding the network density. 
The KNN and DNN networks \as{for the abovementioned hyperparameters} have much larger average degree than the other networks; the sparsest network is the Relative neighbourhood network. 
Consequently, the SPL is shortest in the denser DNN and KNN networks.  
The Railway-based network has the longest average SPL due to its vertex 
chains and the low number of shortcuts between counties.  
For the Queen's contiguity network, the introduction of shortcuts to the economic hubs leads to a decrease in average SPL, i.e. the disease spreads quicker.
The Gabriel network is sparser than the SOI network, with slightly longer average shortest path length. 
Deleting long edges in the Delaunay triangulation network to obtain the SOI network decreases the average degree and the average local clustering coefficient, but increases the average SPL.  
The Queen's network, the Economic hub network, the Delaunay network, the Gabriel network, and the SOI network show small world behaviour, i.e. high clustering with short SPL. 
To assess small world behaviour, the average SPL and average local clustering for a network is compared to a Bernoulli Random graph $G(n,m)$ with identical size $n$ and number of edges $m$ as the network under investigation. 
The Railway network has much larger average SPL than $G(n,m)$, while the dense KNN and DNN networks have almost the same average SPL and local clustering coefficient as the $G(n,m)$ network.
\begin{table}[ht]
\tiny
\centering
\begin{tabular}{l|rrrrrrrrrrr}
  \toprule 
  Metric & Railway & Queen & Eco. hub  & KNN (k = 11) & KNN (k = 21) & DNN (d = 325) & Delaunay & Gabriel & SOI & Rel. neigh.  \\
 \midrule 
 av. degree & 5 & 4.38 & 5.38 & 13.46 & 23.54 & 24.85 & 5.15 & 3.69 & 4.15 & 2.31 \\ 
  av. SPL & 4.15 & 2.74 & 2.27 & 1.46 & 1.06 & 1.01 & 2.41 & 3.02 & 2.86 & 3.99 \\ 
  av. local clust. & 0.2 & 0.51 & 0.6 & 0.75 & 0.95 & 0.99 & 0.49 & 0.31 & 0.51 & 0 \\ 
  \\ 
  BRG av. SPL & 2.08 & 2.24 & 2.05 & 1.46 & 1.06 & 1.01 & 2.1 & 2.24 & 2.28 & 3.38 \\ 
  BRG av. local clust. & 0.14 & 0.18 & 0.21 & 0.54 & 0.94 & 0.99 & 0.17 & 0.16 & 0.19 & 0.09 \\ 
   \bottomrule 
\end{tabular}
\caption{Overview of network characteristics for the Railway-based, Queen's contiguity, Economic (Eco.) hub, KNN ($k = 11$ and $k = 21$), DNN ($d = 325$), Delaunay triangulation, Gabriel, SOI, Relative neighbourhood (Rel. neigh.) network; including average (av.) degree, average (av.) shortest path length (SPL), 
average (av.) local clustering (clust.). The average shortest path length and average local clustering 
coefficient for a Bernoulli Random graph $G(n, m)$, with $n=26$, for each network is also reported.} 
\label{tab:network_char}
\end{table} 
\gr{Although there are differences in the detailed summary statistics, the networks can be clustered according to density and average local clustering coefficient; we use}
\as{
the kmeans algorithm, running 10 randomizations to ensure robustness over a range of $k = 1, \dots, 10$. The corresponding elbow plot implies two clusters of COVID-19 networks. As evident in Figure \ref{fig:density_clustering}, one cluster has high density and high average local clustering coefficient, while the second cluster has low density and low to medium average local clustering coefficient.}

\subsection{Spatial effects}
Intuitively, if spatial correlation is present in a network, the closer in SPL two vertices are, the more highly correlated their COVID-19 incidences.
Moran's I quantifies spatial correlation by estimating the average weighted correlation across space\cite{cliff1981spatial, moran1950notes, Zhou2008moran}. 
Let $t \in T$ and $x_i^{(t)}$ denote the COVID-19 ID for county $i$ at time $t$,
\begin{align*}
    I^{\,t} = \frac{\sum_{i = 1}^N \sum_{j = 1, i \neq j}^N w_{ij} \cdot (x^{(t)}_i - \Bar{x}^{(t)}) (x_j - \Bar{x}^{(t)})}{W_0 \cdot \frac{1}{N} \sum_i (x_i^{(t)} - \Bar{x}^{(t)})^2}
\end{align*}
where $W_0 = \sum_{i, j = 1}^N w_{ij}$ for normalisation based on SPL. 
{For non-neighbours, the weights are zero, i.e. 
$\forall r: \; j \not \in N^{(r)}(i): w_{ij} = 0$. \as{
\gr{Here we use as weights} $w_{ij} = e^{- d_{ij}}$ where $d_{ij}$ denotes the SPL between vertex $i$ and $j$ \cite{coscia2021pearson}.}}
The spatial dependency between counties varies strongly over time for every network, see Figure \ref{fig:morans_I_I} and Figure \ref{fig:morans_I_II} in Supplementary Material \ref{app:network_effect}.
Peaks in Moran's I coincide with peaks in the 1-lag COVID-19 ID at the beginning of the pandemic as well as during the winters 2020/21 and 2021/22. 
The introduction of restrictive regulations, e.g.\,lockdowns, shows a decreasing trend in Moran's I while the ease of restrictions from summer 2021 onward has lead to an increasing trend in Moran's I. 
This indicates a network effect in the data, which is associated with the inter-county mobility, and becomes particularly evident after the official end of pandemic restriction in March 2022. 
\as{To 
\gr{further assess the presence of} non-linear spatial correlation, we \gr{also} apply Moran's I to the ranks of the COVID-19 ID at each time point $t$ over the duration of data observation.}
\as{The rank-based Moran's I follows the same trajectory, with less extreme peaks, as evident in Figure \ref{fig:rank_morans_I}.}

\begin{figure}[h!]
\centering
\subcaptionbox{\as{Economic hub} network}{\includegraphics[width=0.4\textwidth]{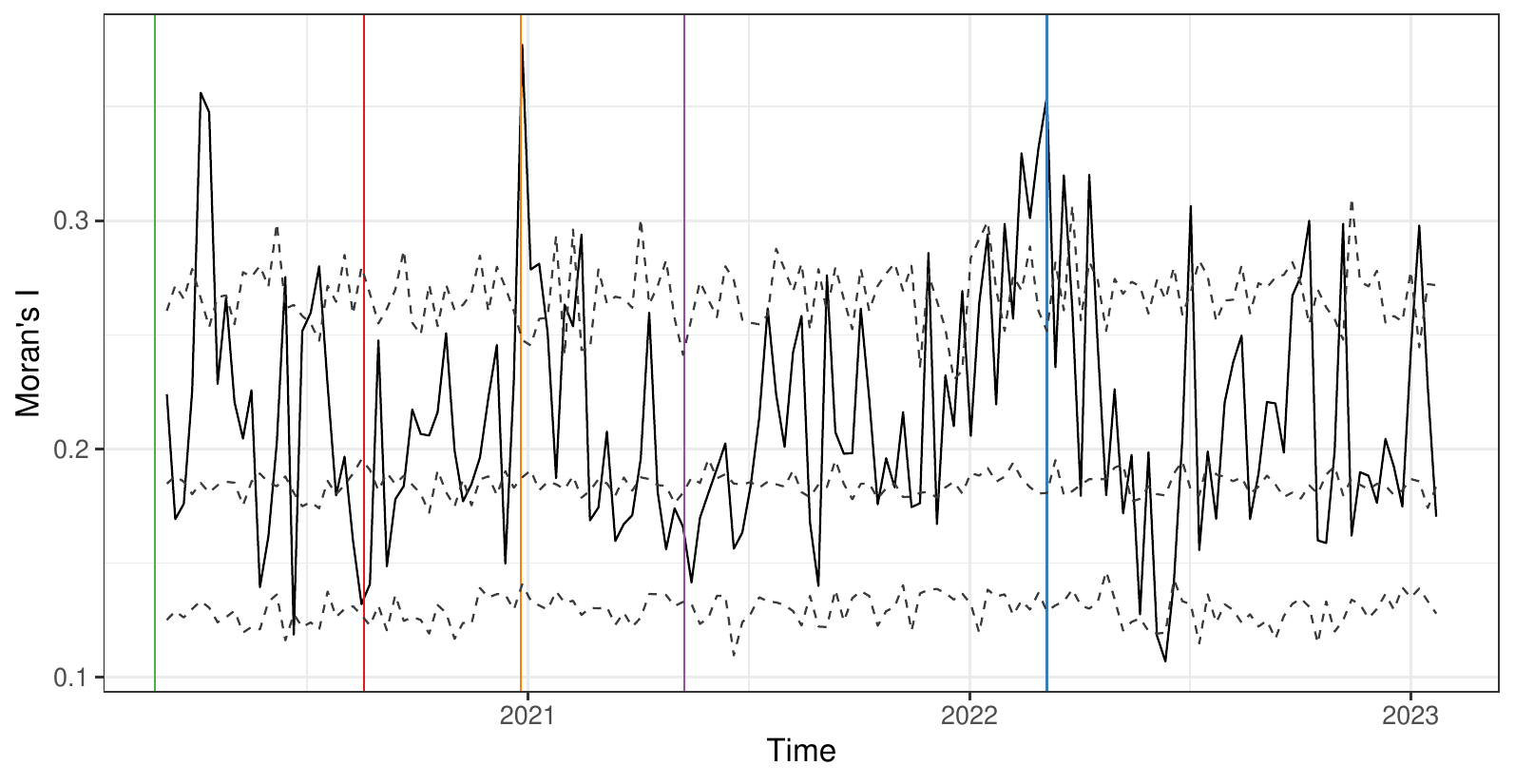}}%
\hspace{0.1cm}
\subcaptionbox{KNN network \as{($k = 21$)}}{\includegraphics[width=0.4\textwidth]{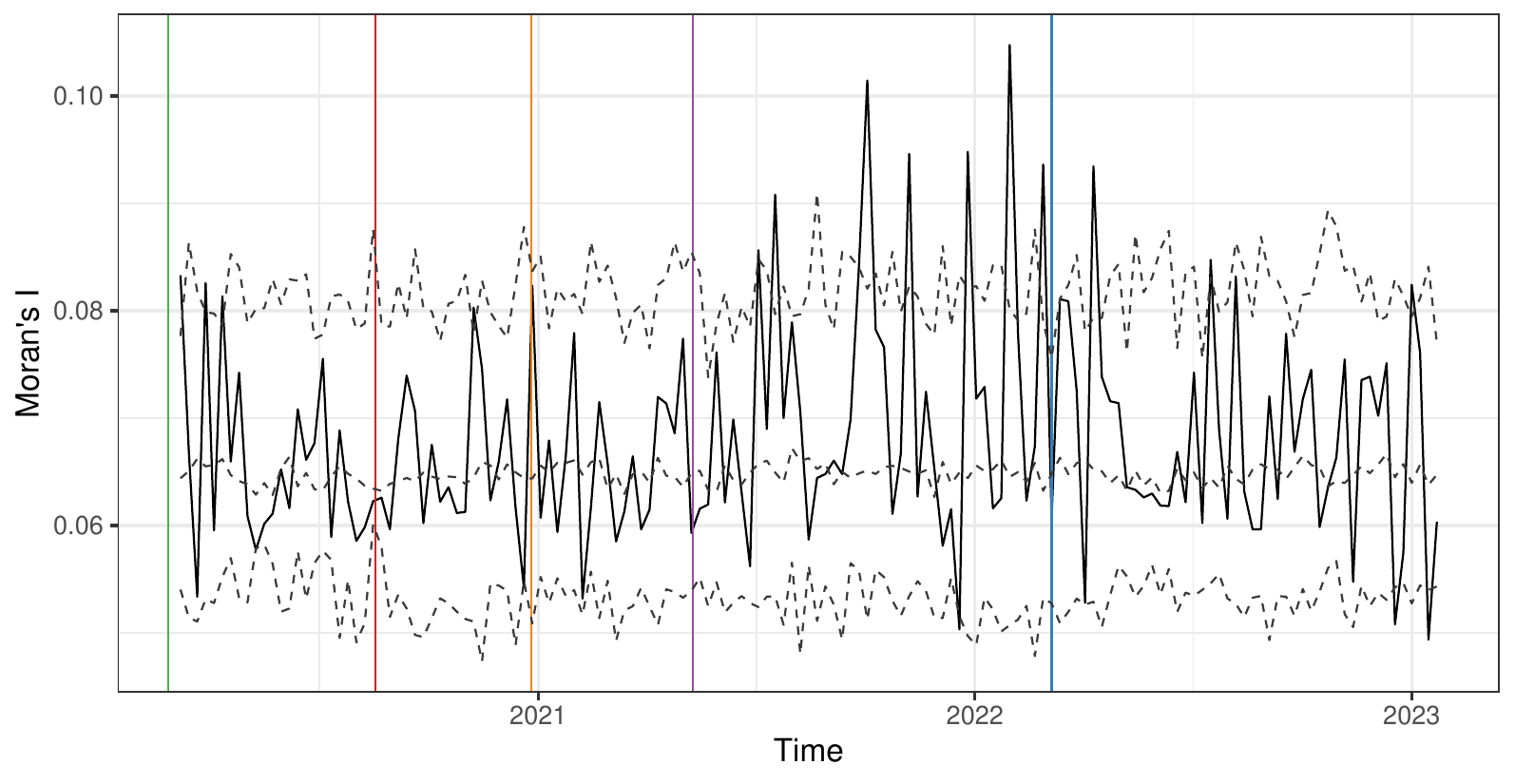}}%
\caption{Moran's I across time, \gr{with weights based on SPL;} main COVID-19 regulations by the Irish Government indicated by vertical lines; in order: initial lockdown, county-specific restrictions, Level-5 lockdown, allowance of inter-county travel, official end of all restrictions;  95\% credibility interval in grey dashed}
\label{fig:morans_I_I}
\end{figure}

\begin{figure}[h!]
\centering
\subcaptionbox{\as{Economic hub} network}{\includegraphics[width=0.4\textwidth]{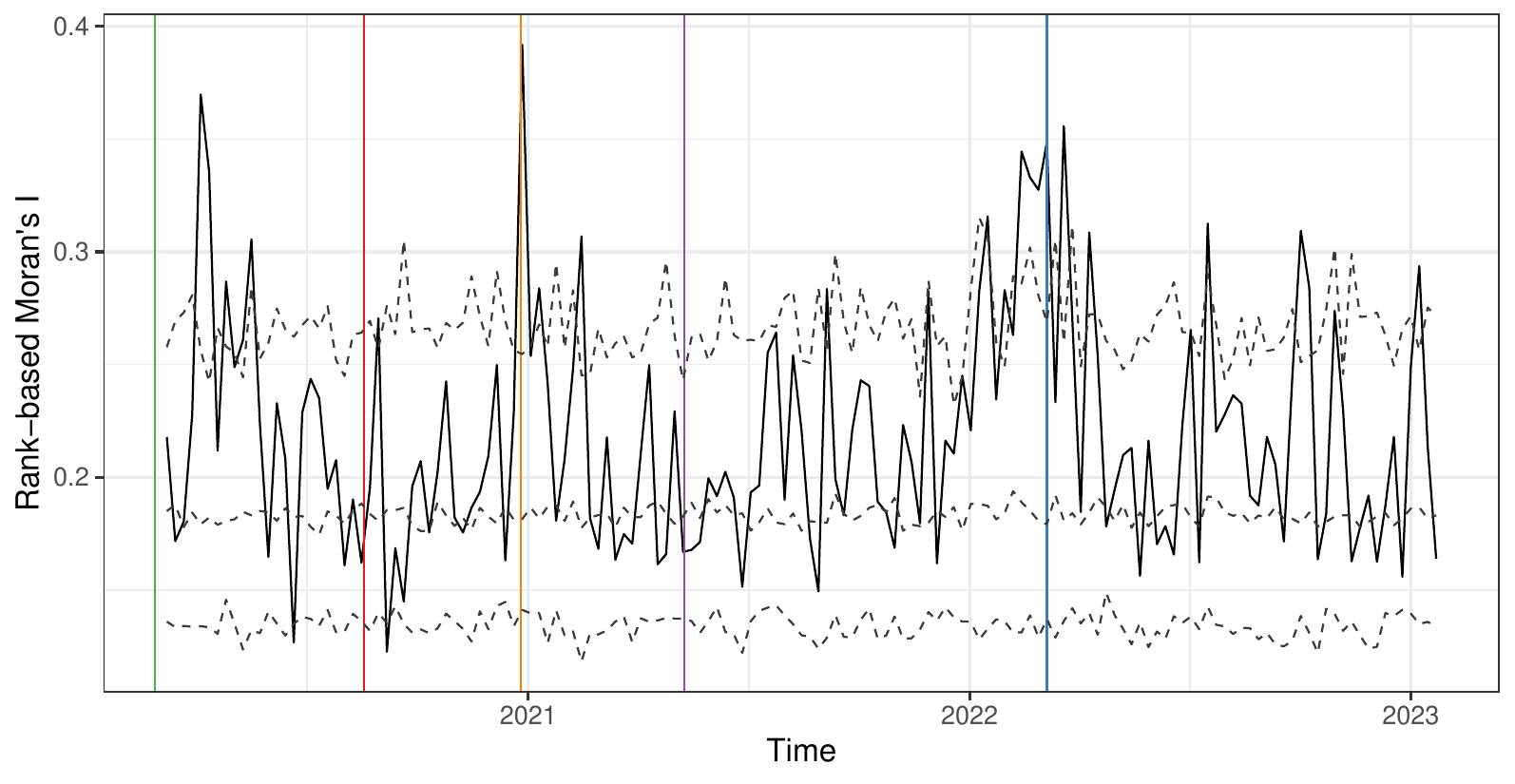}}%
\hspace{0.5cm}
\subcaptionbox{{KNN} network ($k = 21$)}{\includegraphics[width=0.4\textwidth]{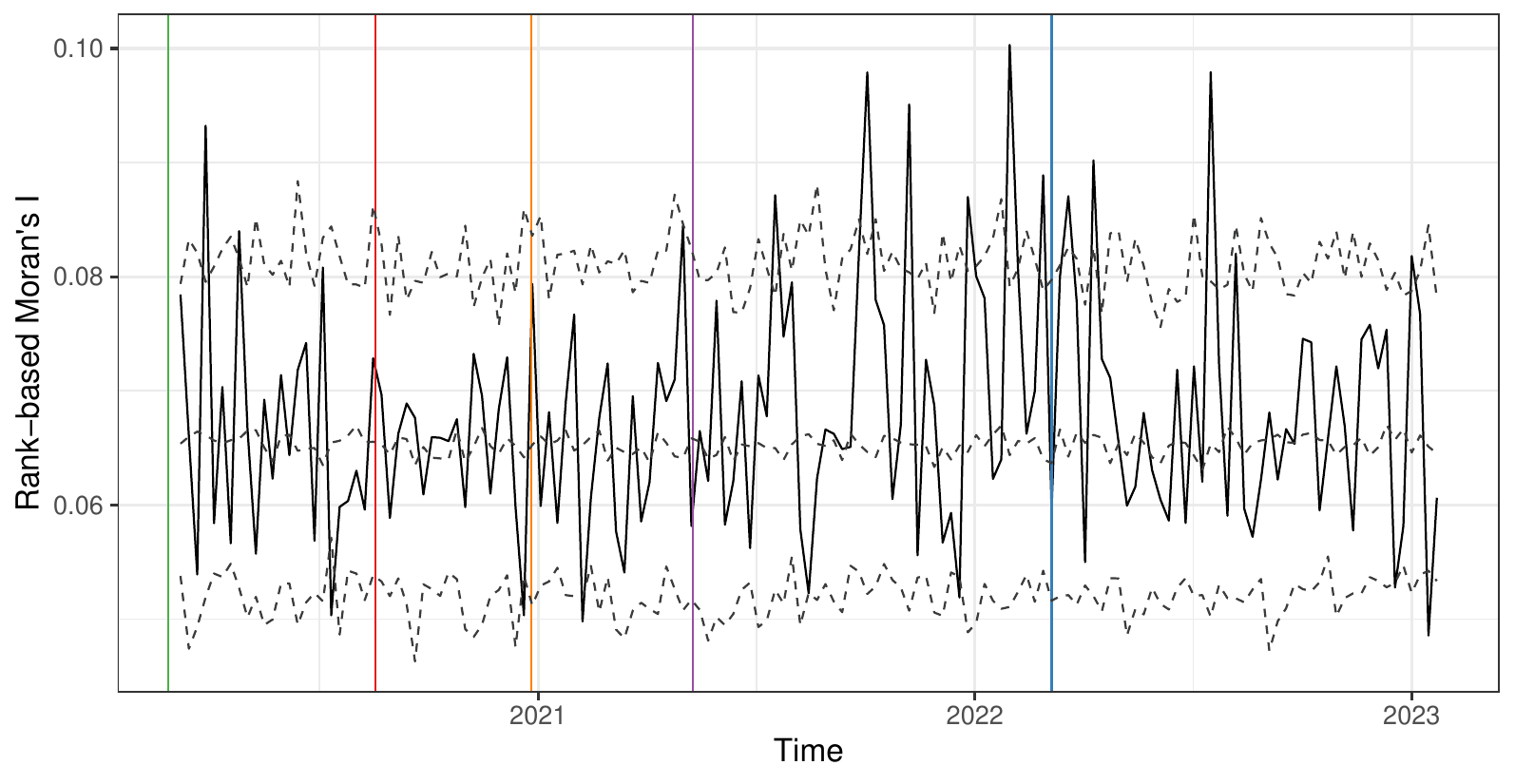}}%
\caption{\as{Rank-based Moran's I across time for the optimal network for the restricted and unrestricted pandemic phase; main COVID-19 regulations by the Irish Government indicated by vertical lines; in order: initial lockdown, county-specific restrictions, Level-5 lockdown, allowance of inter-county travel, official end of all restrictions; 95\% credibility interval in grey dashed}}
\label{fig:rank_morans_I}
\end{figure}

\gr{To statistically assess spatial dependence we}
perform a permutation test \as{for the standard as well as the rank based Moran's I} 
by permuting the COVID-19 cases between counties $R = 100$ times and computing Moran's I,
to account for exponential decay of influence over the network \cite{coscia2021pearson}. 
A date-specific 95\% credibility interval $\forall t = 1, \dots, T: \; [m_{t, l}, m_{t, u}]$ based on empirical quantiles ($q = 0.025, 0.5, 0.975$ quantiles) is constructed. 
Under the null hypothesis, assuming no correlation between network structure and COVID-19 incidence, 5\% ($0.05 \cdot T  \approx 8 $) of observed Moran's I values $m_{t}$ over time $t = 1, \dots, T$ are expected to lie outside the time dependent 95\% credibility interval, $[m_{t, l}, m_{t, u}]$. 
If the proportion $N_{m} = T^{-1} \sum_t \{  {\mathbb{I}(m_t  >  m_{t, u} ) + \mathbb{I}( m_t < m_{t, l})} \} $ of rejected tests over time is greater than expected under the null, we conclude that the network distance has an effect on the correlation between COVID-19 incidence\footnote{The constructed test is not a proper significance test in the statistical sense, given the dependence between tests over time. It rather provides a rough intuition regarding the spatial correlation in COVID-19 incidence assuming different underlying networks.}.
\gr{With exponential SPL weights, the} 
proportion of rejected tests for the restricted and unrestricted data set are: Railway-based network $N_m = (0.25, 0.142)$, Queen's contiguity network $N_m = (0.227, 0.217)$, Economic hub network $N_m = (0.25, 0.179)$, KNN network \as{($k = 11$ for restricted, $k = 21$ for unrestricted phase)} $N_m = (0.159, 0.151)$, DNN network \as{($d = 325$ for both phases)}  $N_m = (0.068, 0.094)$, Delaunay triangulation network $N_m = (0.205, 0.189)$, Gabriel network $N_m = (0.114, 0.132)$, SOI network $N_m = (0.182, 0.198)$, Relative neighbourhood network $N_m = (0.159, 0.189)$. 
The \as{proportions indicate a significant spatial correlation}. 
Depending on the network, the proportion for either the restricted or the unrestricted data set is larger.
\as{The rank-based Moran's I permutation test obtains similar proportions of rejected tests, indicating a significant nonlinear spatial correlation \gr{(data not shown)}.}

\section{Results}
\subsection{GNAR model fitting}
\label{results_covid}
To assess the benefit of accounting for network effects, the GNAR model is compared to a standard county-specific ARIMA model \gr{which is allowed to include seasonal effects} \footnote{Fitted for each county individually, the ARIMA models might differ in orders and parameters.}
\as{The GNAR model allows for more flexible spatial-temporal dependencies than other network-based time series models as detailed in Supplementary Material \ref{app:lit_rev_network_models}, including the ARIMA models, with a network specific selection of $\alpha$- and $\beta$-order. 
The models are selected by choosing the model with the lowest BIC.} 

On average, the ARIMA model achieves a \as{$BIC = 1846.88$} on the entire data set, \as{$BIC = 534.58$} for the restricted data set and \as{$BIC = 670.27$} for the unrestricted data set. 
Optimal\footnote{{\it Optimal} describes the best performing combination of $\alpha$- and $\beta$-order as well as global-$\alpha$ setting and weighting scheme which obtain the minimal BIC value. The a-priori range of $\alpha$-order spans \as{$\{1, ..., 7\}$}. 
\as{The maximum lag to consider follows from Schwert's rule \cite{ng1995unit}, applied to the minimum number of weeks across the individual five datasets ($w = 18$).}
The possible choices for the $\beta$-order \as{are listed in Supplementary Material \ref{app:beta_choices}.} \as{The maximum neighbourhood stage that can be included in the GNAR model is determined by the smallest maximum SPL across most network, which is 5. For the complete network, only 1st stage neighbourhood can be modelled, while for the Economic hub network the maximum neighbourhood stage is 4.}} GNAR models for each COVID-19 network achieve much lower BIC.  
For the restricted phase, the best phase-specific GNAR model yields $BIC= 58.91$ on the \as{Economic \gr{hub}} network, and for the unrestricted phase $BIC=190.07$ on the KNN ($k = 21$) network, see Table \ref{tab:best_model_pandemic_phases}. 
When fitted on the entire data set, the GNAR-5-11110 model with the KNN network ($k = 21$)\footnote{From hereon referred to as the KNN network.} achieves the lowest $BIC = 193.95$\footnote{For more detail on fitting a GNAR model to each COVID-19 network on the entire data set, see the Supplementary Material \ref{app:optimal_gnar}.}. 
All BICs are considerably smaller than those obtained from the ARIMA fit, with  the minimal BIC for the entire data set  much larger than the BIC for the restricted phase, and also larger than the BIC for the unrestricted phase, thus justifying the use of GNAR models, as well as the split of the data.

The nature of the virus suggests that the transmission of COVID-19 between Irish counties may depend strongly on the population flow between counties \cite{lotfi2020covid}. 
Protective COVID-19 restrictions taken by the Irish Government restricted and at times forbade inter-county travel in Level 3-5 lockdowns \cite{IG_5_level_plan,IT_inter_county_travel}. 
As supported by the positive and negative trends in Moran's I, the spatial dependence of COVID-19 incidence across counties is likely to have decreased during lockdowns and increased during periods in which inter-county travel was allowed \cite{wang2022prediction}. 
This motivates training a pandemic phase specific GNAR model.

\subsection{Pandemic phases} 
\label{chapter: phases} 
Table \ref{tab:best_model_pandemic_phases} summarises the optimal GNAR models and COVID-19 network for the restricted and unrestricted data set. 
For both phases, the best performing GNAR model select an autoregressive component of order \as{7}.
The average MASE are smaller for the restricted than the unrestricted pandemic phases, implying that  GNAR models are  more suited to predicting periods with strict regulations than periods with fewer or no restrictions.
\as{The variance for residuals and MASE is smaller for the GNAR model than the ARIMA model.}
The optimal network for the unrestricted pandemic phase is much denser than the optimal network for the restricted phase. 
As evident from Tables \ref{tab:best_models}, \ref{tab:best_model_subsets} in Supplementary Material \ref{app:optimal_gnar}, the BIC values for the optimal GNAR model lie within the \as{range $[58.91, 68.36]$} for the restricted data set and within the \as{range $[190.07, 192.56]$} for the unrestricted data set.
Figure \ref{fig:bic_density} in Supplementary Material \ref{app:optimal_gnar} illustrates that denser networks perform better for the unrestricted data set while sparse networks achieve lower BIC for the restricted data set, \as{on average}.  

\begin{table}[ht]
\small
\centering
\begin{tabular}{l|rrrrr}
  \toprule 
 Data subset & network & GNAR model & BIC  & $\Bar{\varepsilon}$ (sd) & av. MASE (sd) \\
 \midrule 
 \textit{Restricted} & Eco. hub & GNAR-7-3110000 & 58.91 & -1.60 (16.51) & 0.90 (0.73) \\ 
  \textit{Unrestricted} & KNN-21 & GNAR-7-1111000 & 190.07 & -13.46 (14.67) & 0.94 (0.74) \\ 
  \\ 
   \textit{Restricted} & ARIMA &  & 534.58 & -0.78 (19.08) & 1.22 (1.34) \\ 
  \textit{Unrestricted} & ARIMA &  & 670.27 & -16.65 (18.50) & 1.21 (0.96) \\ 
   \bottomrule  
\end{tabular}
\caption{Overview over the best performing model and network for 
restricted and unrestricted pandemic phases; average residual 
$\Bar{\varepsilon}$ and average (av.) MASE indicated for the predicted 5 
weeks at the end of the observed time period across all counties, 11.04.2021 - 09.05.2021 for the restricted dataset and 25.12.2022 - 23.01.2023 for the unrestricted dataset; standard deviation (sd) across counties in brackets.} 
\label{tab:best_model_pandemic_phases}
\end{table}

\begin{figure}[h!]
    \centering
    \includegraphics[width=0.8\textwidth]{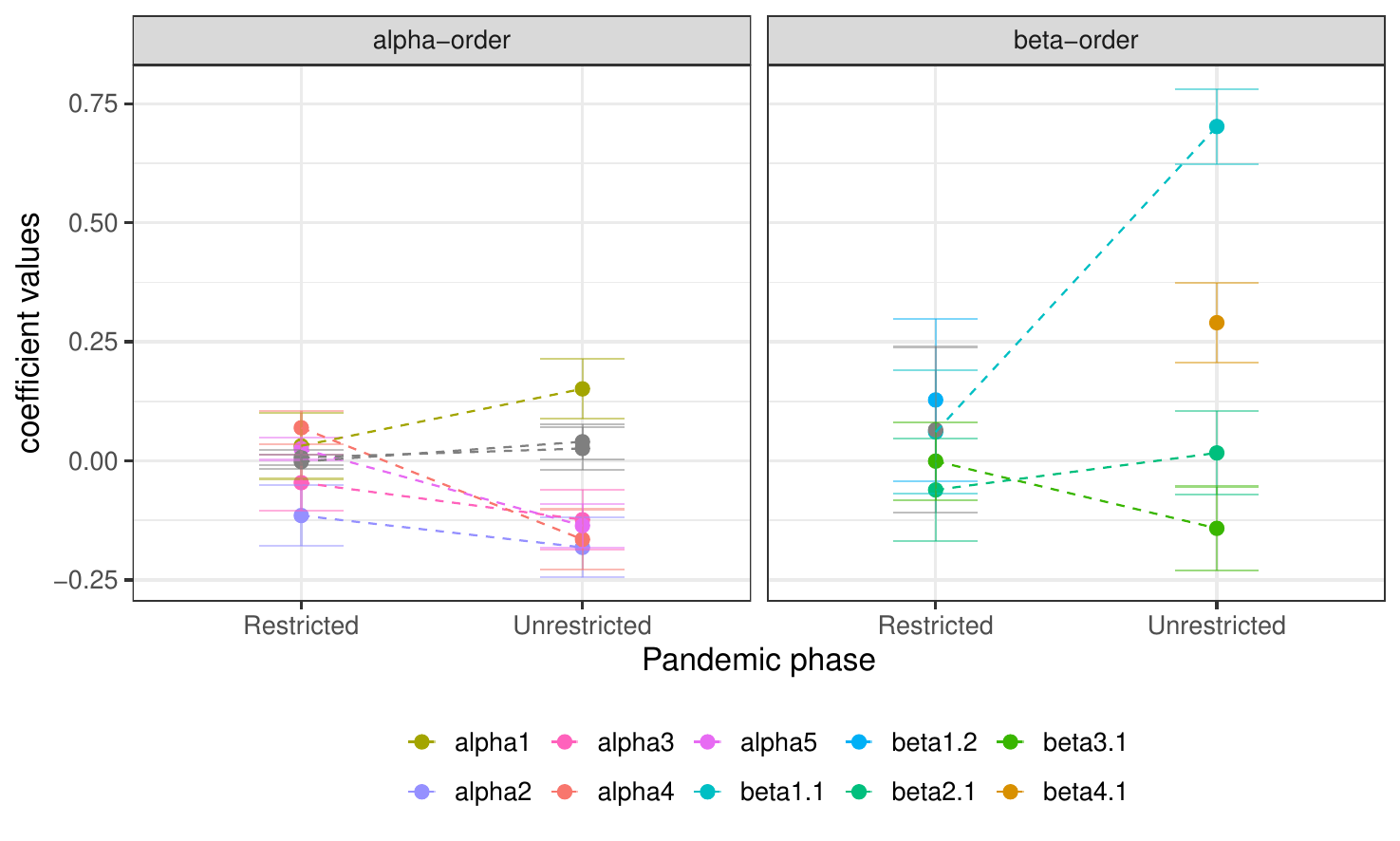}
\caption{Development of GNAR model coefficients for the restricted and unrestricted pandemic phase; restricted phase with Economic hub network, unrestricted phase with KNN ($k = 21$) network}
\label{fig:coef_change}
\end{figure}

A decrease in inter-county dependence due to COVID-19 restrictions should result in decreasing values for the $\beta$-coefficients in the GNAR model.
This hypothesis can only be partially verified, see Figure \ref{fig:coef_change}. 
The absolute value of $\beta$-coefficients increases from the restricted to the unrestricted phase, implying increased spatial dependence after COVID-19 restrictions have been eased or lifted. 
We note that the GNAR model picks up a decrease in temporal dependence in COVID-19 ID. 
As a disease spreads more freely due to lenient or no restrictions, it has been observed in other data studies that case numbers can grow more erratic and become less dependent on historic data \cite{firth2020combining,kraemer2020effect}.  
This effect, in addition to peaks and high volatility in COVID-19 ID observed during pandemic phases with less strict regulations, might contributed to the negative $\alpha$-coefficient values for the unrestricted data set. 
In general, an increase in noise during unrestricted pandemic phases might contribute to the decrease in predictive performance of the GNAR model.

Identical observations can be made when considering how the coefficients develop between the restricted and unrestricted phase for the  GNAR model that is  optimal for the entire data set, namely, GNAR(p = 5, s = (1, 1, 1, 1, 0)) with the KNN ($k = 21$) network; the $\beta$-coefficients increase in absolute value for the unrestricted phase compared to the restricted phase, see Figure \ref{fig:coef_change_optimal_model} in Supplementary Material \ref{app:gnar_coef_optimal}. 

\begin{figure}[h!]
\centering
\subcaptionbox{{Delaunay triangulation}, {Gabriel}, {Relative neighbourhood}, {SOI} and {Railway-based} network}{\includegraphics[width=\textwidth]{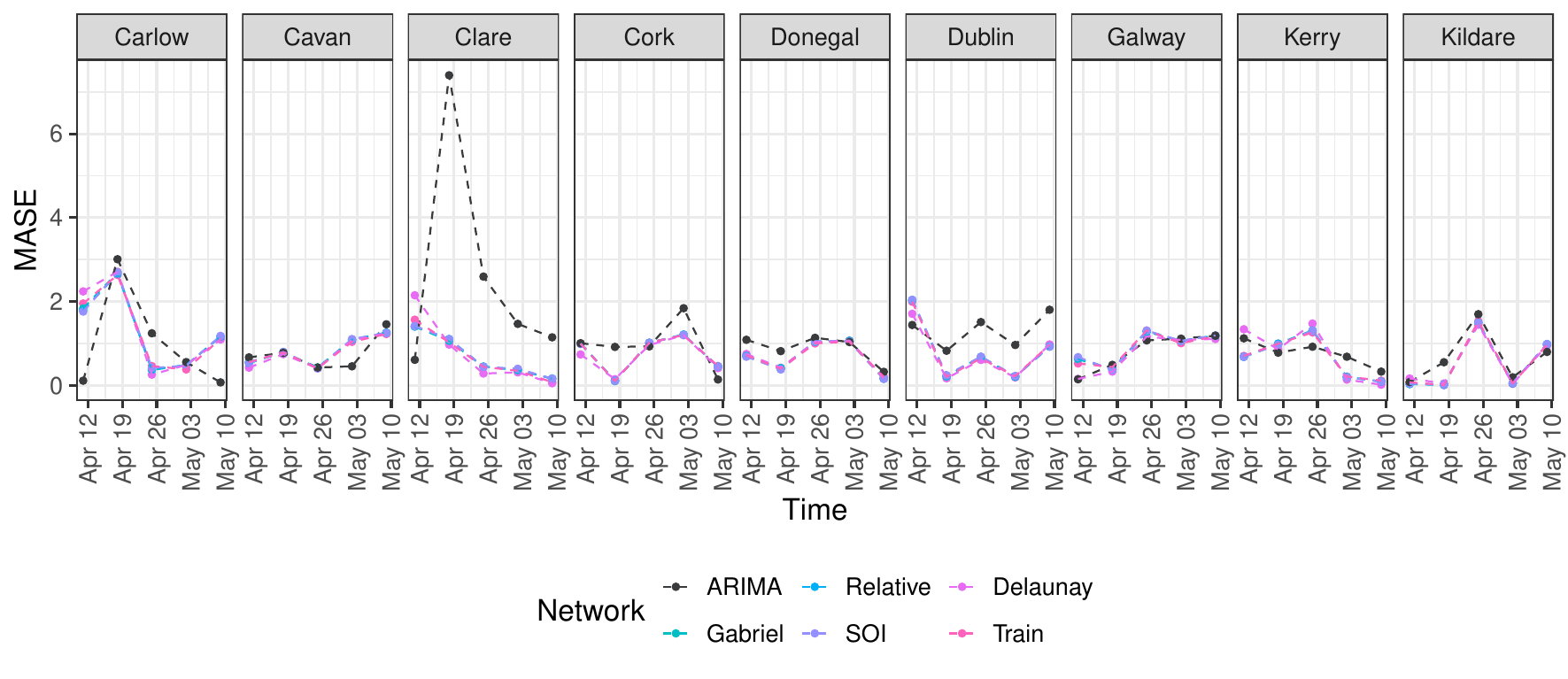}}%
\\
\subcaptionbox{{KNN $(k = 11)$}, {DNN $(d = 325)$}, {Complete}, {Queen's contiguity} and {Economic hub} network}{\includegraphics[width=\textwidth]{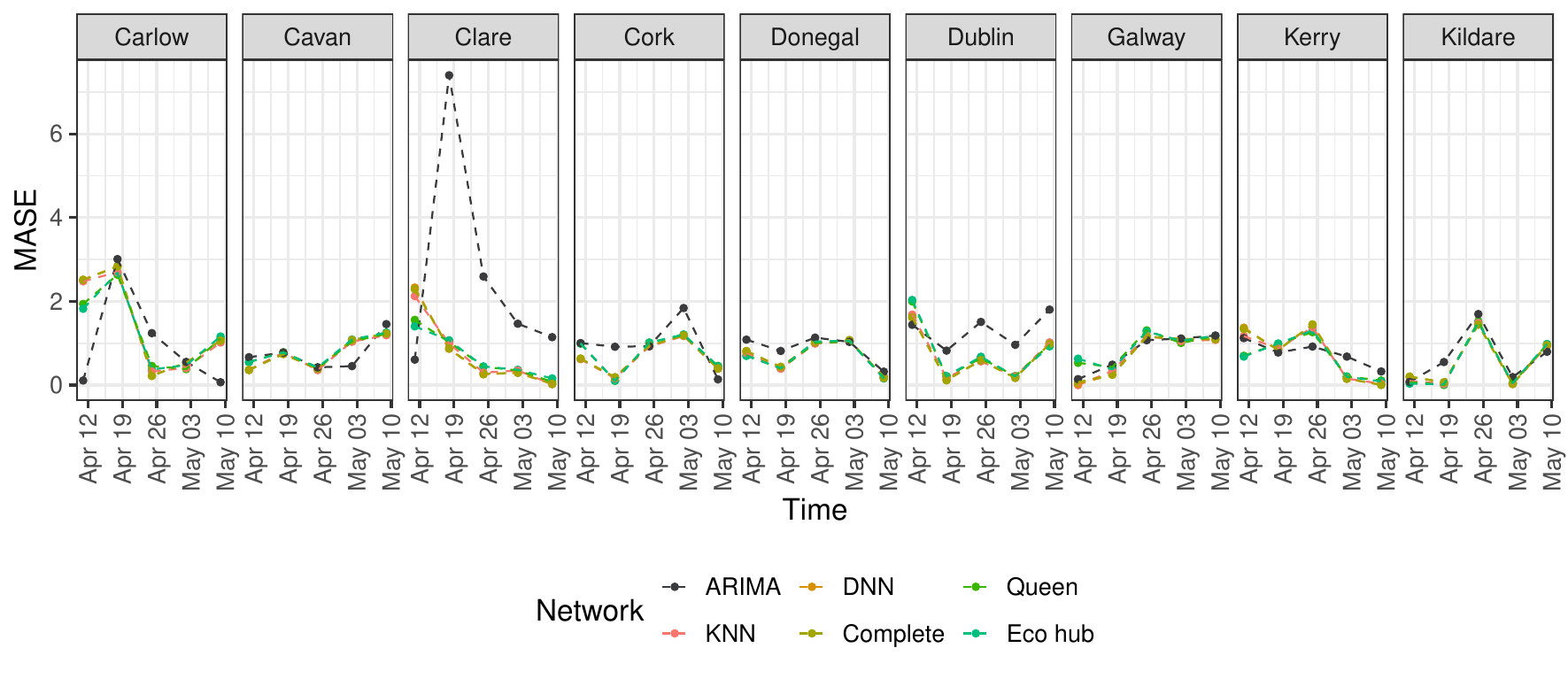}}%
\caption{MASE values for the restricted pandemic phase, for 9 selected counties}
\label{fig:mase_restricted_I}
\end{figure}

\begin{figure}[h!]
\centering
\subcaptionbox{
{Delaunay triangulation}, {Gabriel}, {Relative neighbourhood}, {SOI} and {Railway-based} network}{\includegraphics[width=\textwidth]{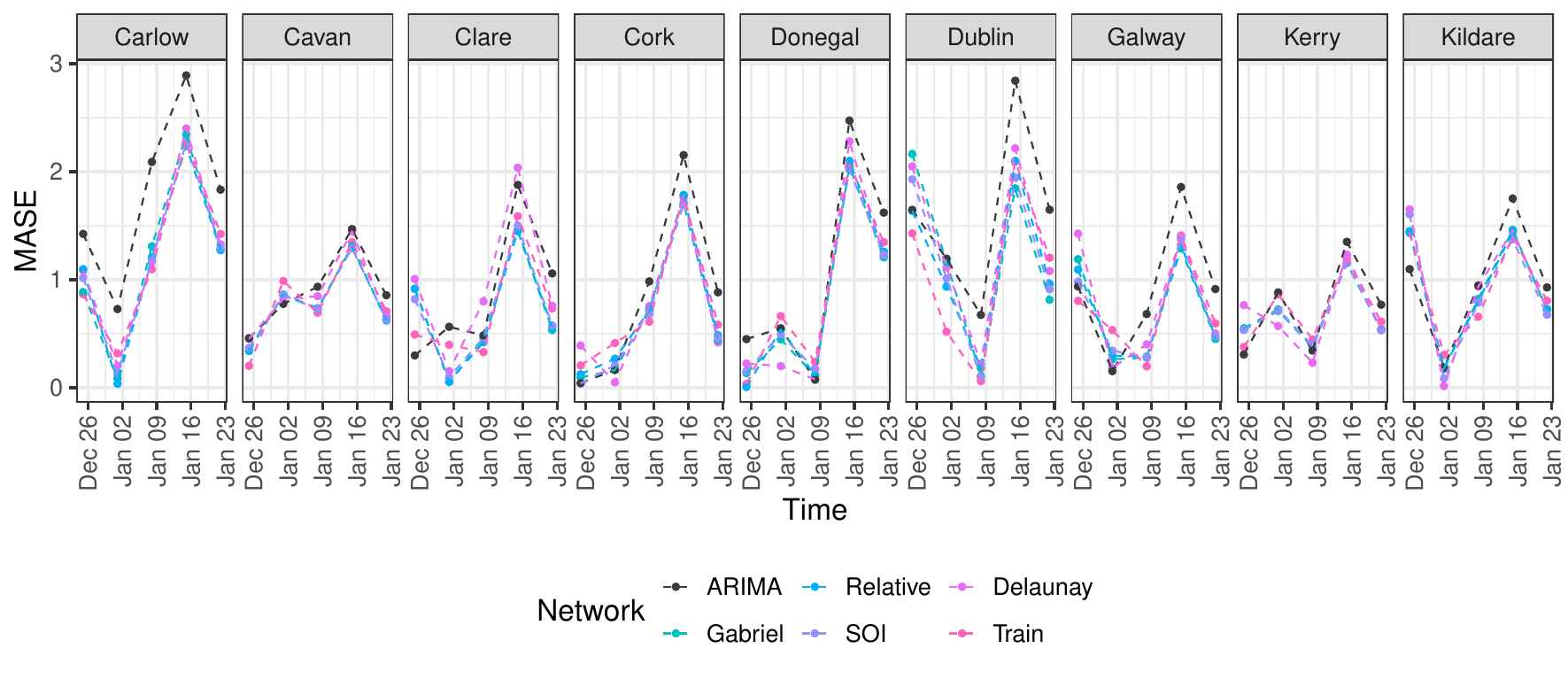}}%
\\
\subcaptionbox{{KNN $(k = 21)$}, {DNN $(d = 325)$}, {Complete}, {Queen's contiguity} and {Economic hub} network}{\includegraphics[width=\textwidth]{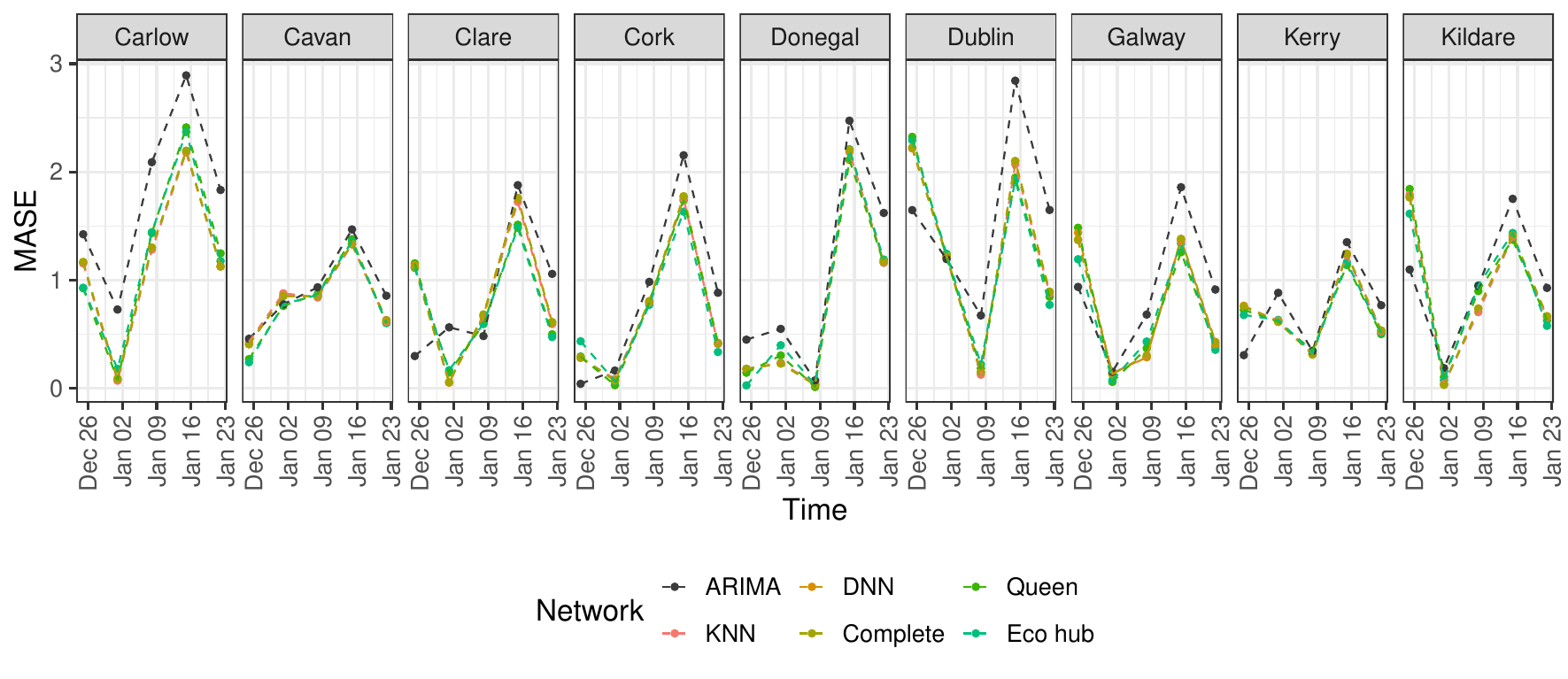}}%
\caption{MASE values for the unrestricted pandemic phase, for 9 selected counties}
\label{fig:mase_free_I}
\end{figure}

The predictive accuracy for both datasets is comparable and varies from county to county, see Figures \ref{fig:mase_restricted_I} and \ref{fig:mase_free_I} for 9 example counties; MASE values for the remaining counties follow similar patterns.

For the restricted phase, GNAR models achieve lower MASE than the ARIMA models except for counties \as{Cavan, Galway, Leitrim, Louth, Sligo, Tipperary, Roscommon}, for which the ARIMA model performs equally well. 
\as{For the unrestricted phase, the ARIMA model predicts particularly 
\gr{poorly} for counties Carlow, Kilkenney, Louth and Waterford, \gr{and} for the restricted phase for counties Cavan, Clare, Limerick and Wexford.} \gr{We note that Cavan, Leitrim and Sligo have a border with Northern Ireland, which could introduce some confounding factors.}

\clearpage

For the restricted phase, the predictions differ more strongly between the GNAR model and the ARIMA model, see Figure \ref{fig:predicted_restricted_I} in Supplementary Material \ref{app:predictions}. 
For the unrestricted phase, the GNAR and ARIMA models follow roughly the same trajectory while \as{the former achieves} smaller residuals for most counties. 



\subsection{Assessing the model assumptions}
The above model fits assume that the observations follow a Gaussian i.i.d.\,error structure. 
To assess this assumption, we test whether the residuals $\hat{\varepsilon}_{i,t}$ follow a normal distribution with a county-specific Kolmogorov-Smirnov test, aggregated over time.
In contrast, we obtain a majority of significant p-values across counties for the unrestricted phase (\gr{the counts of $p$-values are} $\# p \leq 0.025 = 21$, $\# p > 0.025 = 5$)\footnote{Insignificant p-values were established in counties Donegal, Dublin, Kilkenny, Laois, Offaly and Sligo}, raising doubts about the Gaussianity assumption in the unrestricted phase. 
We obtain primarily insignificant p-values across counties for the restricted phase ($\# p \leq 0.025 = 8$, $\# p > 0.025 = 18$)\footnote{The Kolmogorov-Smirnov test was significant for counties \as{Donegal, Galway, Kilkenny, Laois, Limerick, Offaly, Sligo and Wicklow.}}, so that the Gaussian assumption is not rejected.

Table \ref{tab:mase_res_p_counties} in Supplementary Material \ref{app:qq_plots} details the average MASE, average residual and p-value for each county, resulting for the two optimal GNAR models for the restricted and unrestricted data set. 
The Gaussian nature of residuals indicate suitability of the GNAR model to model restricted pandemic phases and ensure consistency in coefficient estimates.
For the unrestricted phase, the Gaussianity in the model assumptions could not be statistically verified.
These conclusions are supported by the county-specific QQ-plots in Supplementary Material \ref{app:qq_plots}. 
\as{The hyperparameters $\alpha$- and $\beta$-order were set data-adaptively to minimize the BIC criterion, which assumes Gaussian error. 
\gr{Under the assumption of Gaussian error, the BIC would be minimal for the correct higher order network dependence.} \gr{As we examined a large range of} $\beta$-order \gr{choices, this deviance from Gaussianity leads us to }propose investigating alternative error structures as future work.}

\begin{figure}[h!]
\centering
\subcaptionbox{Restricted}{\includegraphics[width=0.2\textwidth]{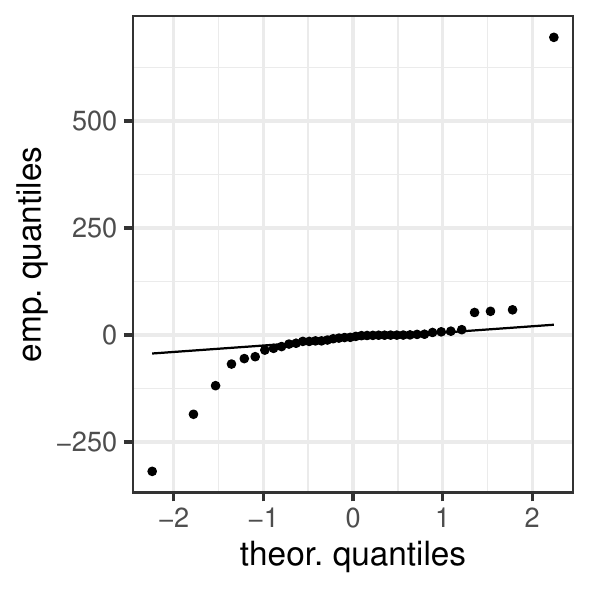}}%
\hspace{0.5cm}
\subcaptionbox{Unrestricted}{\includegraphics[width=0.2\textwidth]{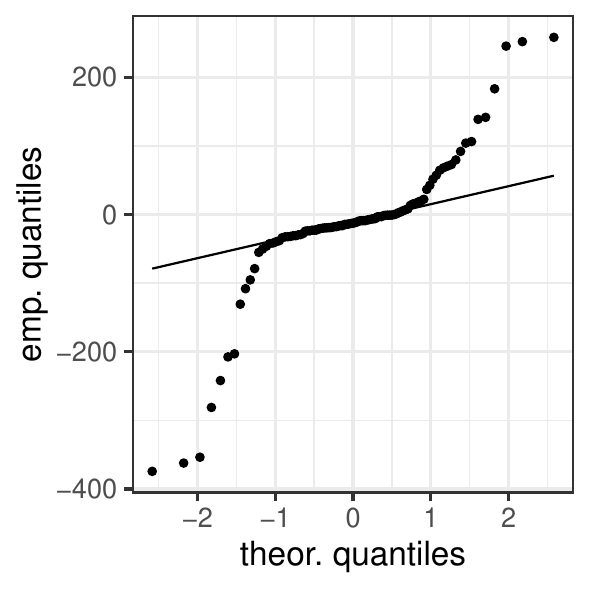}}%
\hspace{0.5cm}
\subcaptionbox{Restricted}{\includegraphics[width=0.2\textwidth]{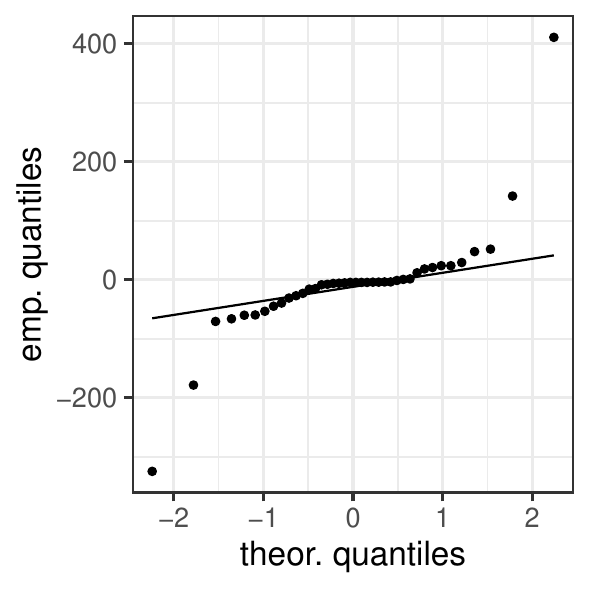}}%
\hspace{0.5cm}
\subcaptionbox{Unrestricted}{\includegraphics[width=0.2\textwidth]{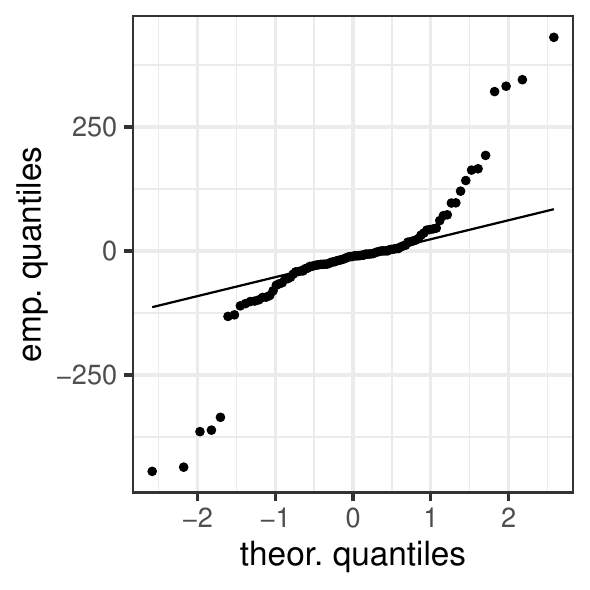}}
\caption{QQ-plot for the residuals from the best performing GNAR model and network for restricted and unrestricted pandemic phase for county Dublin (left) and Donegal (right)}
\label{fig:qq_dublin}
\end{figure}

The GNAR model further assumes that the errors are uncorrelated. 
To assess this assumption, the residuals are investigated according to their temporal as well as spatial autocorrelation using the Ljung-Box test and Moran's I based permutation test \cite{ljung1978measure,moran1950notes}.  
The former concludes significant temporal correlation for short-term lags in the GNAR residuals for each county.
Thus, there is evidence that the GNAR model insufficiently accounts for temporal dependence in COVID-19 incidence in subsequent weeks. 
The residuals show remaining spatial autocorrelation. 
The Moran's I based permutation test counts $N_m = 8$ Moran's I values outside the corresponding 95\% credibility interval (expected $0.05 \cdot 45 \approx 2$) for the restricted phase and $N_m = 16$ for the unrestricted phase (expected $0.05 \cdots 107 \approx 5$).
The reduction in spatial correlation for the restricted phases and the \as{Economic hub network} is greater ($N_m = 10$ on COVID-19 cases to $N_m = 8$ for residuals) than for the unrestricted phases and the KNN network ($N_m = 16$ for COVID-19 cases and residuals).
We conclude that there is evidence that the GNAR model may not sufficiently incorporate the spatial relationship in COVID-19 case numbers across counties. 
These possible violations of the model assumptions have to be taken into account when interpreting the model fit.

\section{Discussion}
\label{chapter: conclusion}
In general, a network model could can be a powerful tool to inform the spread of infectious diseases, see for example \cite{britton2019stochastic, overton2020using}. 
In this paper, we modelled the COVID-19 incidence across the 26 counties in the Republic of Ireland by fitting GNAR models, leveraging different networks to represent spatial dependence between the counties. 
While we do not assume that the disease only spreads along the network, we consider the edges to represent the main trajectory of the infection.
The analysis shows that there is a clear network effect, but networks of similar density perform similarly in predictive accuracy.
GNAR models perform better on data collected during pandemic phases with inter-county movement restrictions than data gathered during less restricted phases.
Sparse networks perform better for the restricted data set, while denser networks achieve lower BIC for the unrestricted data set. 

There are some caveats relating to the model. 
First, the time series for the restricted phase and for the unrestricted phase are actually concatenated time series; Figure \ref{fig:covid_weekly_cases} shows the original time series. The concatenation was carried out because of data availability, but it is possible that it obfuscates some potentially interesting phase-specific signals. \gr{As seen in Figure \ref{fig:covid_weekly_cases}, even after differencing the time series do not display clear stationarity.}
Further, COVID-19 is 
subject to ``seasonal'' effects, e.g. systematic reporting delays due to weekends and winter waves \cite{kubiczek2021challenges, nichols2021coronavirus, sartor2020covid}.
The GNAR model does not have a seasonal analogue which can incorporate seasonality in data, like SARIMA for ARIMA models \cite{shumway2000time}. 
Future work could introduce a seasonal component to the GNAR model, improving its applicability to infectious disease modelling.  \gr{There may be other spatio-temporal patterns such as non-linear effects which the GNAR model currently does not include.}
Moreover, the COVID-19 pandemic  had a strong influence on mobility patterns \cite{google_mobility_report, manzira2022assessing}, in particular due to restrictions of movement and an increased apprehension towards larger crowds. 
Considering only static networks may introduce a bias to the model \cite{bansal2007individual, mo2021modeling, perra2012random}. 
Future work could therefore explore how GNAR models can include dynamic networks to incorporate a temporal component of spatial dependency.  Alternative weighting schemes for GNAR models could be investigated to account for differences in edge relevance across time and network. 

Regarding the theory of GNAR models, alternative error distributions, such as a Poisson distributed error term \gr{as in \cite{armillotta2021poisson}}, could be explored given the indication of non-Gaussian residuals for the unrestricted pandemic phase. 
The stability of parameter estimation in GNAR models also warrants further investigation. 
The network constructions themselves could also be refined.
Simulations have shown that GNAR models are sensitive to network misspecifications. 
Omitting edges may result in bias in the GNAR coefficients. 
While this paper has carried some robustness analysis regarding network choice, future analysis could focus on more content-based approaches to constructing networks, e.g.\,building a network based on the intensity of inter-county trade, computed according to the gravity equation theory \cite{chaney2018gravity}. 
Many researchers have successfully modelled the initial spread of COVID-19 from Wuhan across China based on detailed mobility patterns, e.g.\,\cite{jia2020population, kraemer2020effect}.
Finally, our statistical analysis did not include information about the dominant strain.
With COVID-19 being an evolving disease, different strains may display different transmission patterns. 
If more detailed data become available then this question would also be of interest for further investigation. 
However, this study has illustrated that it may be of benefit to use a GNAR model for the spread of an infectious disease, in particular during movement restrictions, so that the spread is mainly local. 
It has also detailed a range of possible network choices, and it has provided a set of tests to assess the performance of the model and its fit.

\section*{Declarations} 

\paragraph{Funding}
G.R. acknowledges support from EPSRC grants EP/T018445/1, EP/X002195/1,  EP/W037211/1, EP/V056883/1, and EP/R018472/1.

\paragraph{Conflicts of interest/Competing interests} None.

\paragraph{Data Availability Statement} Data is available on the GitHub repository for the analysis as well as on the Open Data Platform by the Government of the Republic of Ireland: https://data.gov.ie/dataset/covid-19-hpsc-county-statistics-historic-data1.

\paragraph{Authors' contributions} All authors wrote and reviewed the manuscript. A.S. wrote code for the data exploration and analysis.

\paragraph{Ethics approval} Not applicable. 

\paragraph{Consent to participate} Not applicable. 

\paragraph{Consent for publication} All authors give consent for publication.

\paragraph{Acknowledgments} The authors would like to thank Emma Davies-Smith for insightful feedback, and Marina Knight, Guy Nason and Matt Nunes for helpful discussions. \gr{Moreover, we thank the anonymous referees for their helpful suggestions.}


\clearpage
{
\raggedright
\printbibliography[heading=bibintoc]
}
\newpage 
\appendix
{\LARGE \noindent \textbf{Supplementary Material}}

\section{COVID-19 networks}
\label{app:networks}
\begin{figure}[htp!]
\centering
\subcaptionbox{Railway-based network}{\includegraphics[width=0.325\textwidth]{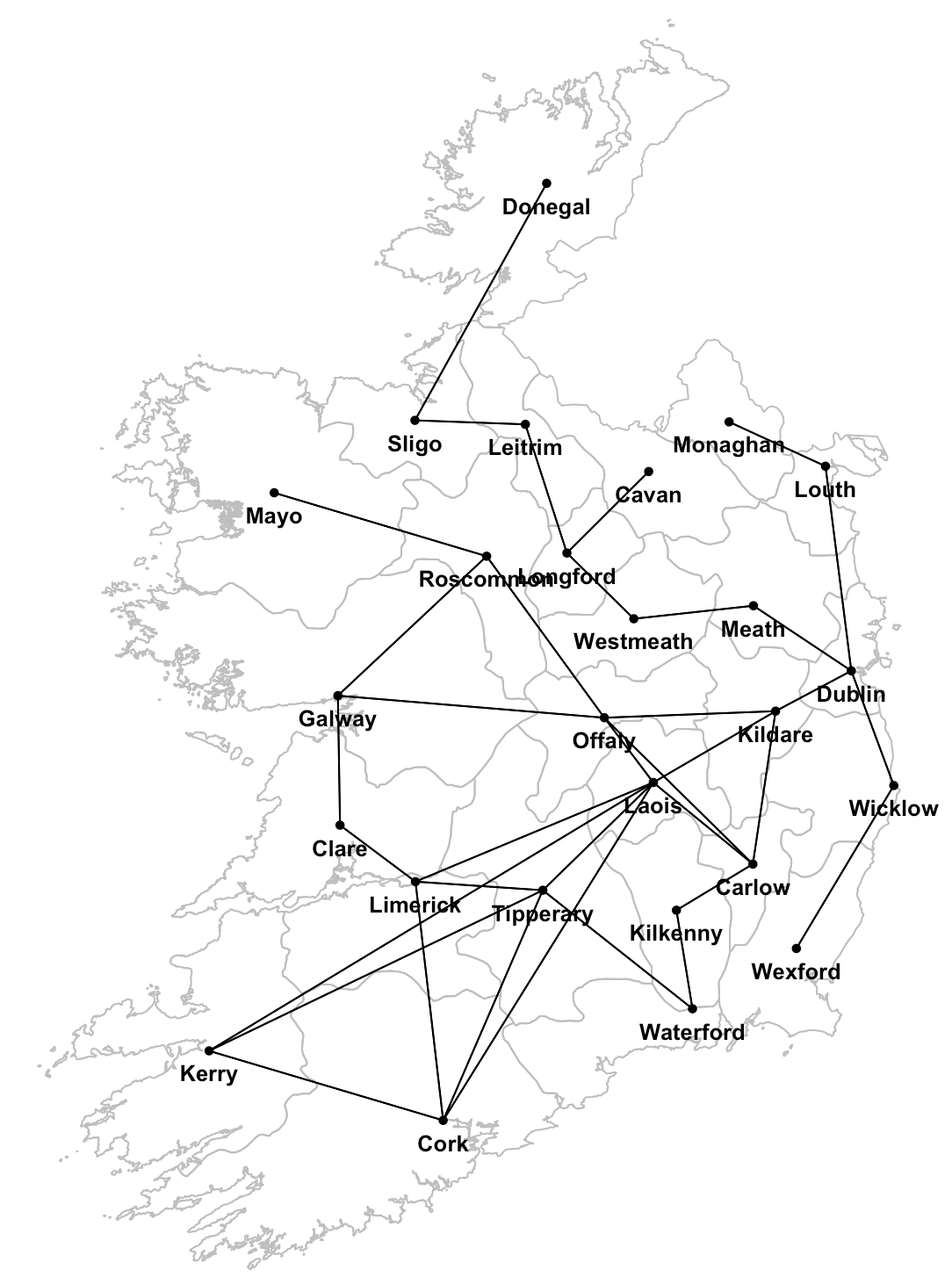}}%
\hspace{0.1cm}
\subcaptionbox{Queen's contiguity network}{\includegraphics[width=0.325\textwidth]{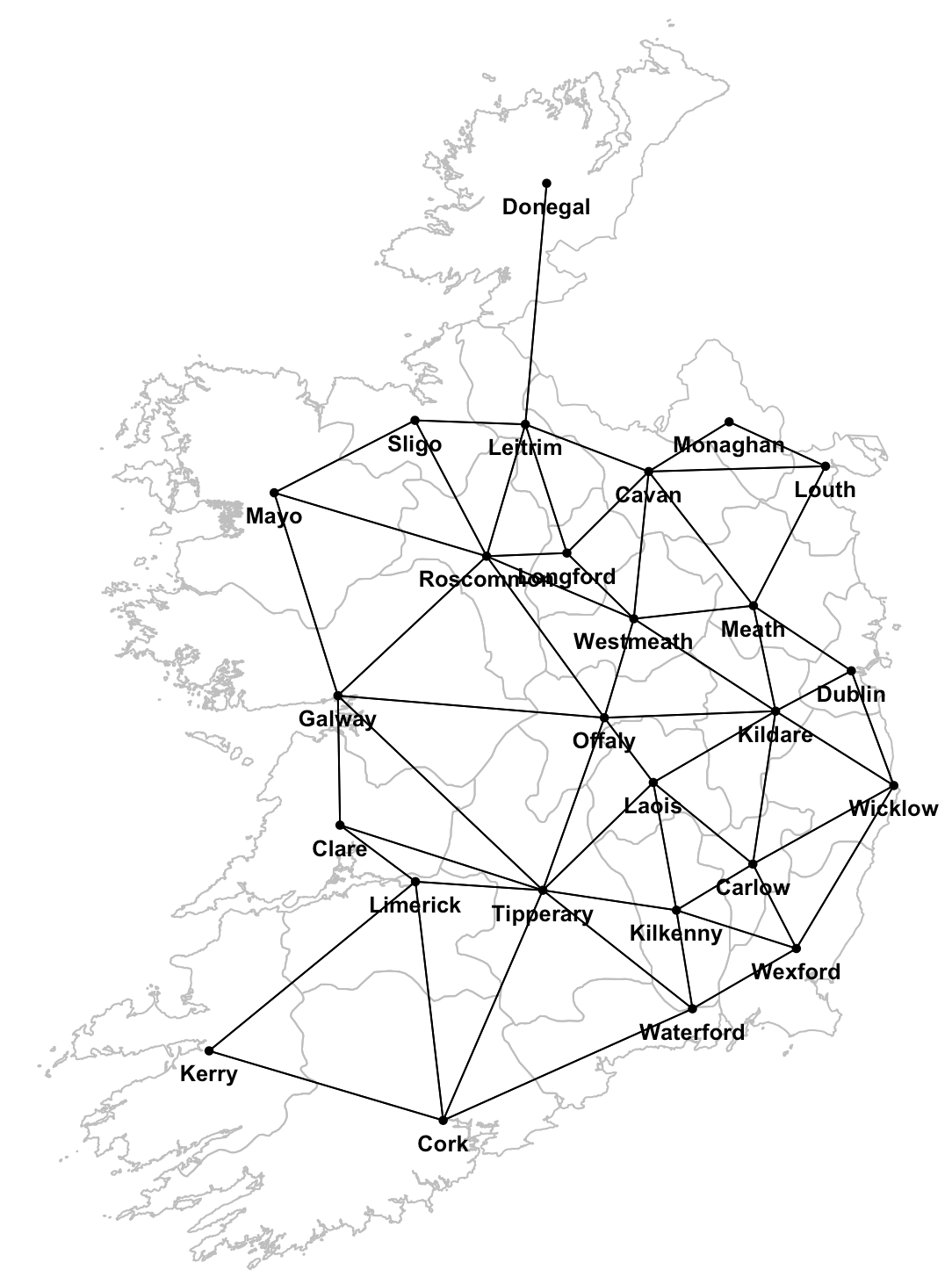}}%
\hspace{0.1cm}
\subcaptionbox{
{Delaunay triangulation} network}{\includegraphics[width=0.325\textwidth]{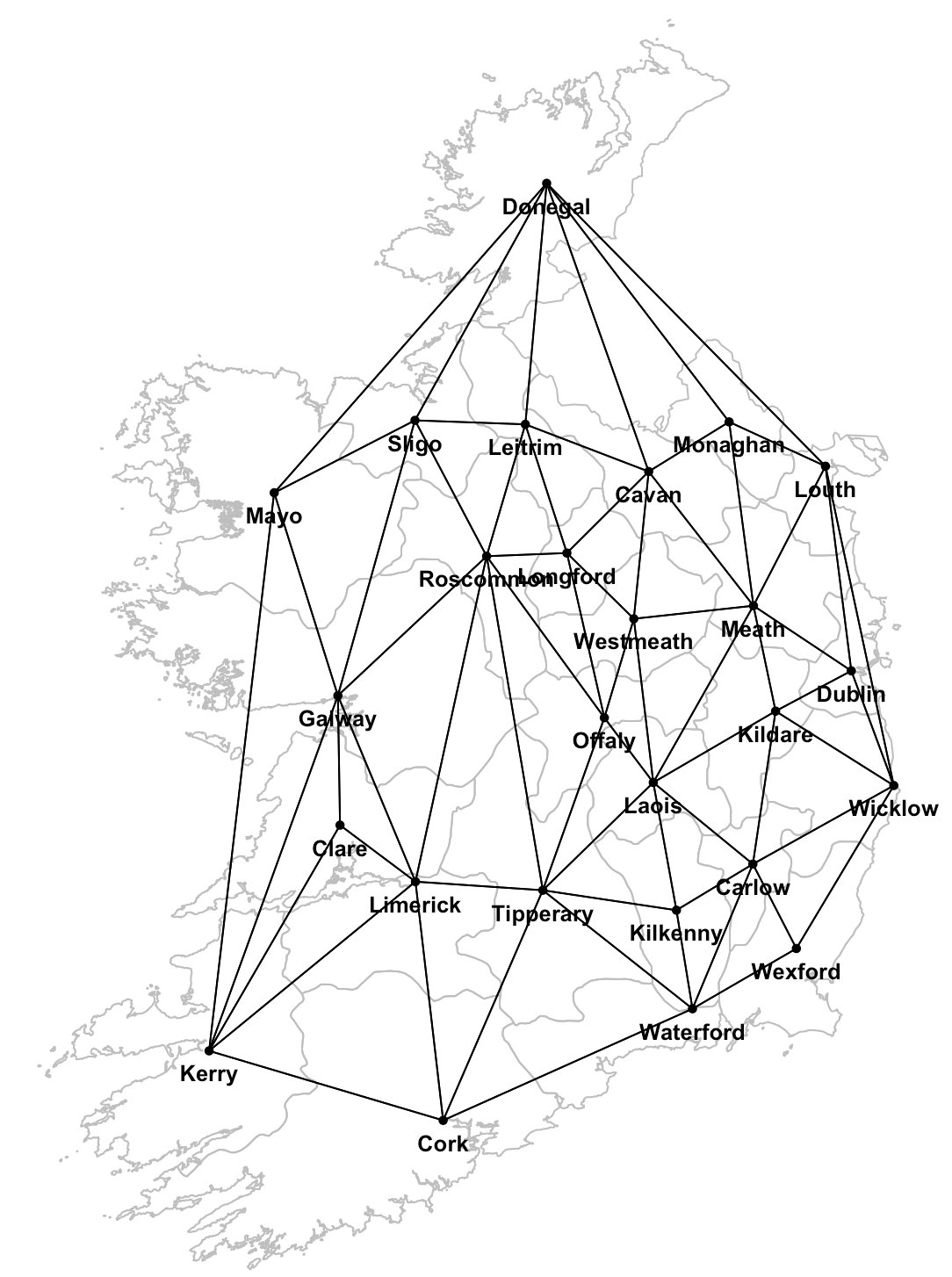}}%
\\
\subcaptionbox{
{Gabriel} network}{\includegraphics[width=0.325\textwidth]{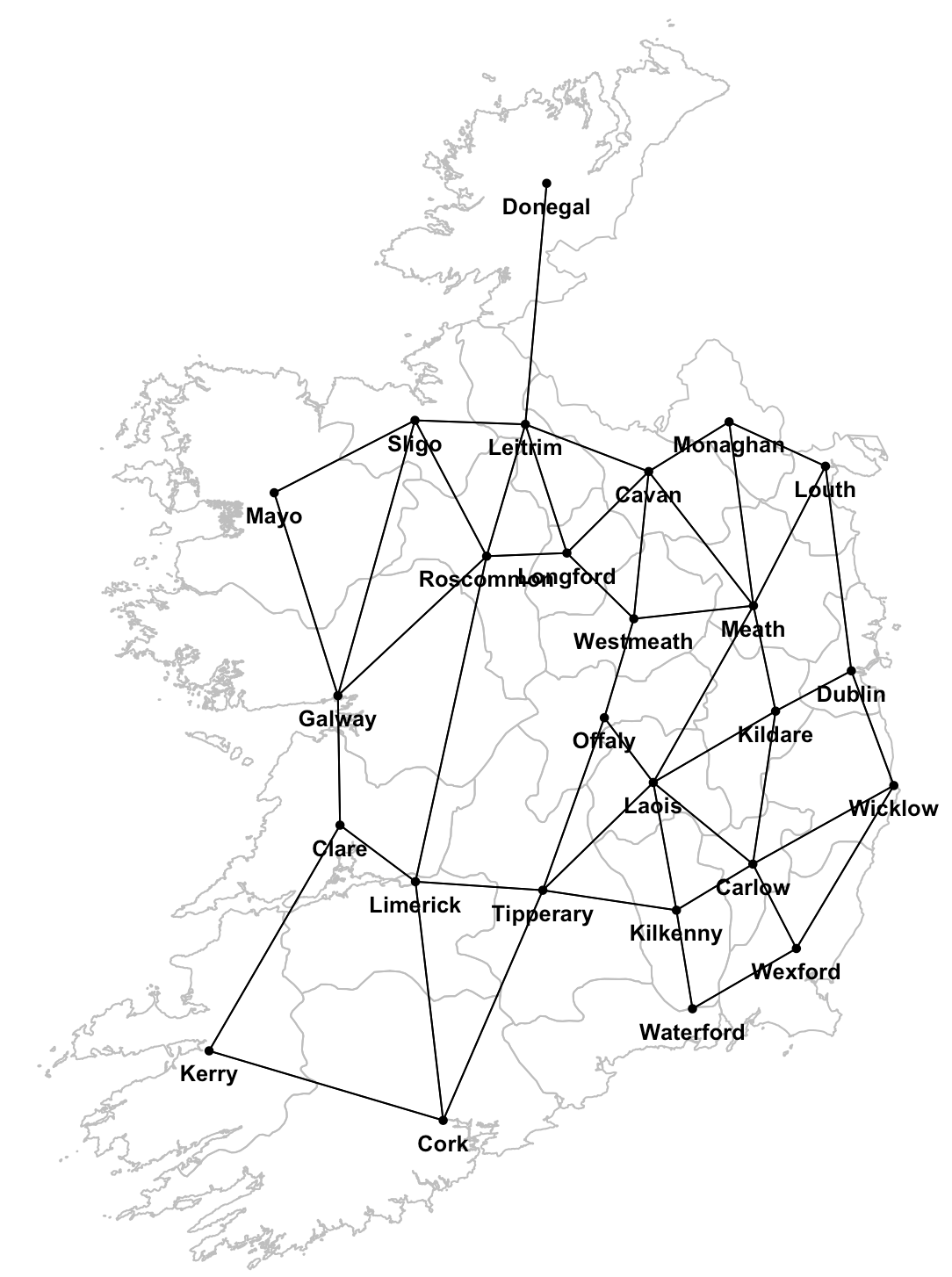}}%
\hspace{0.1cm} 
\subcaptionbox{
{SOI} network}{\includegraphics[width=0.325\textwidth]{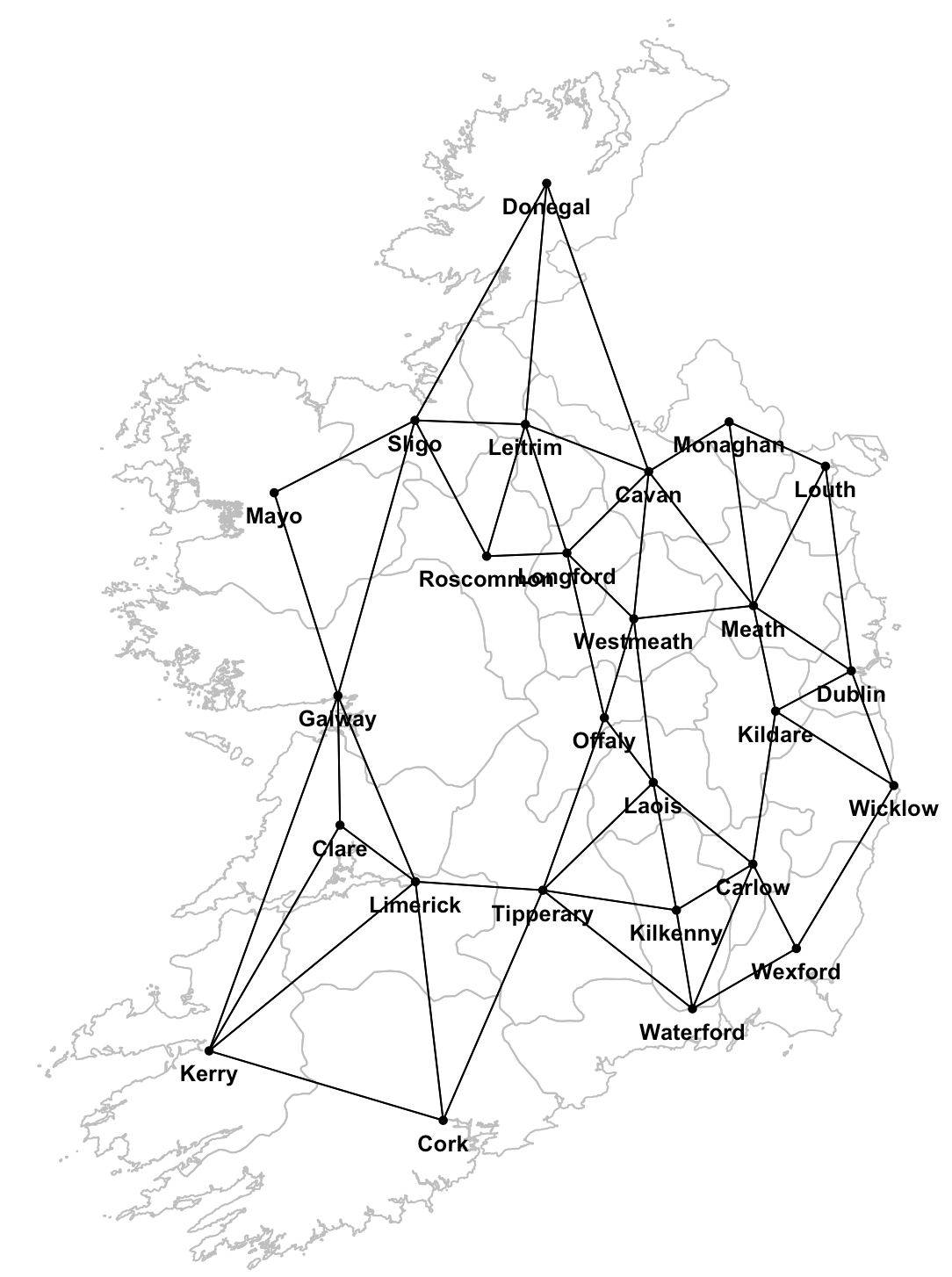}}%
\hspace{0.1cm}
\subcaptionbox{
{Relative neighbourhood} network}{\includegraphics[width=0.325\textwidth]{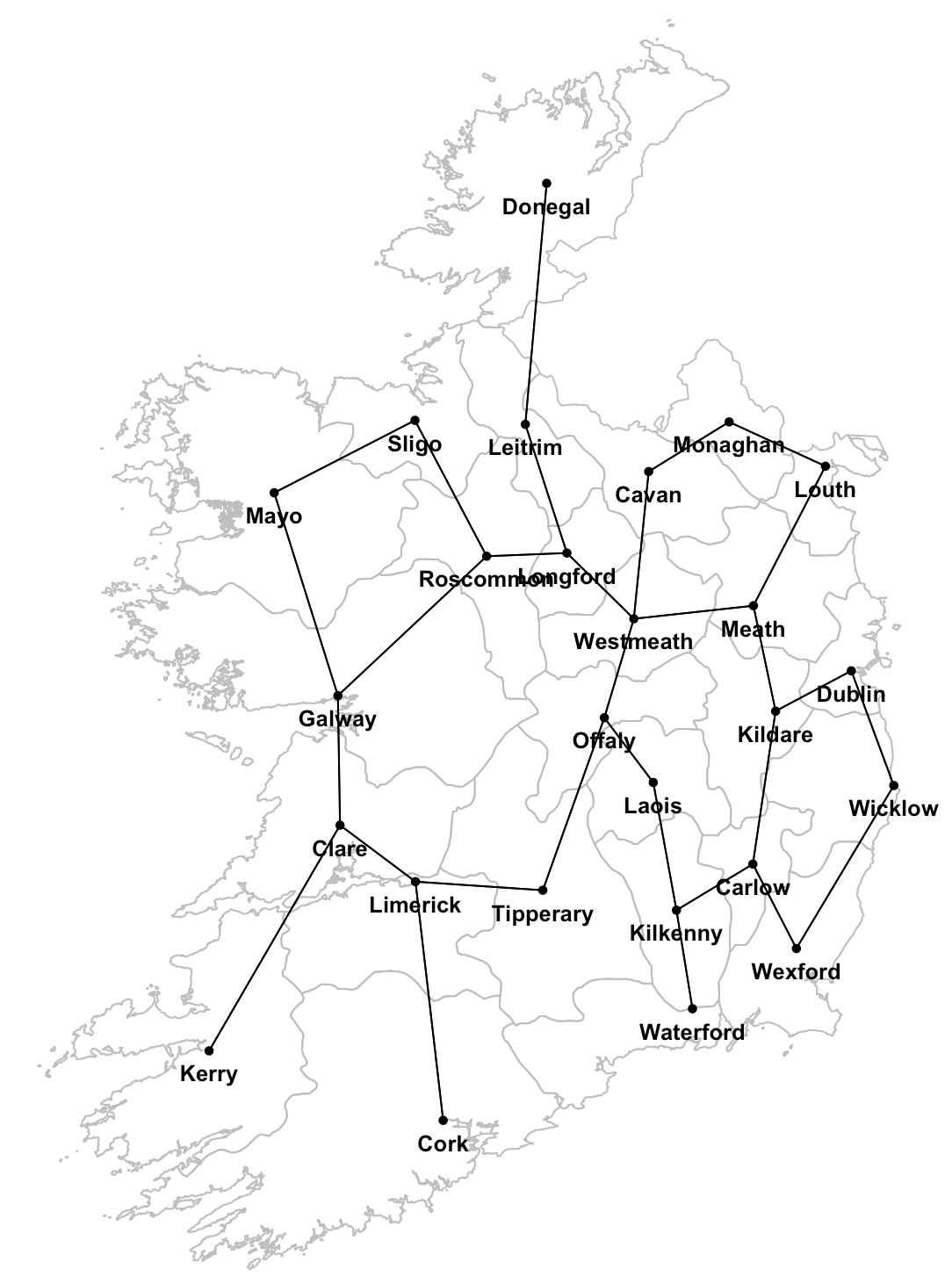}}%
\\ 
\subcaptionbox{
{DNN} network ($d = 325$)}{\includegraphics[width=0.325\textwidth]{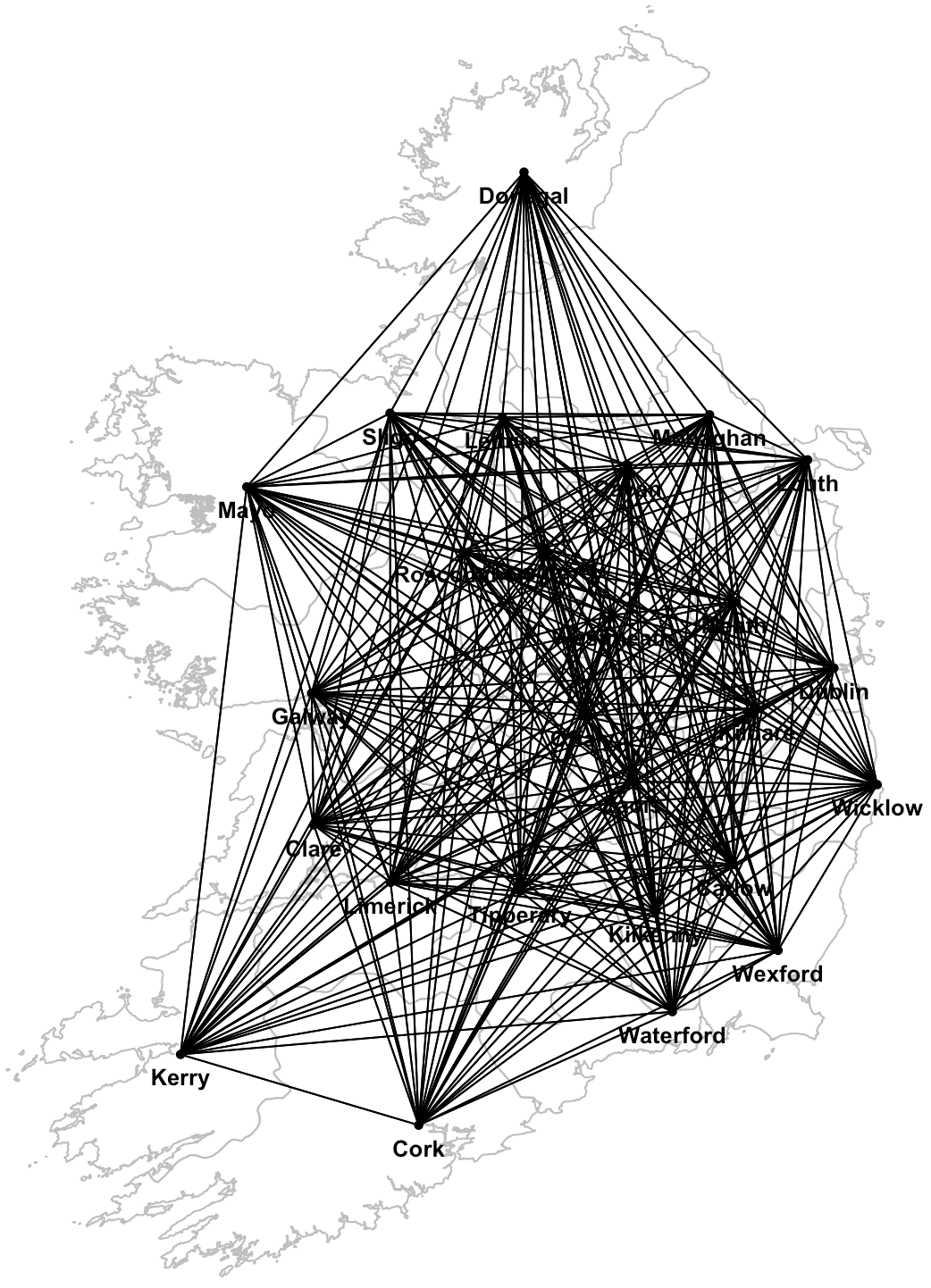}}%
\caption{Map of Ireland and COVID-19 networks - based on Great Circle distance}
\label{fig:network_maps_I}
\end{figure}

The COVID-19 networks are constructed according to either a geographical or a statistical definition of neighbourhood. 
The main text shows the \gr{Economic hub} 
network as well as the KNN network with \as{$k = 11$} and $k = 21$.
The remaining networks are illustrated in Figure \ref{fig:network_maps_I}. 

\subsection{Assessing the network effect via Moran's I}
\label{app:network_effect}
\begin{figure}[h!]
\centering
\subcaptionbox{{DNN} network ($d = 325$)}{\includegraphics[width=0.4\textwidth]{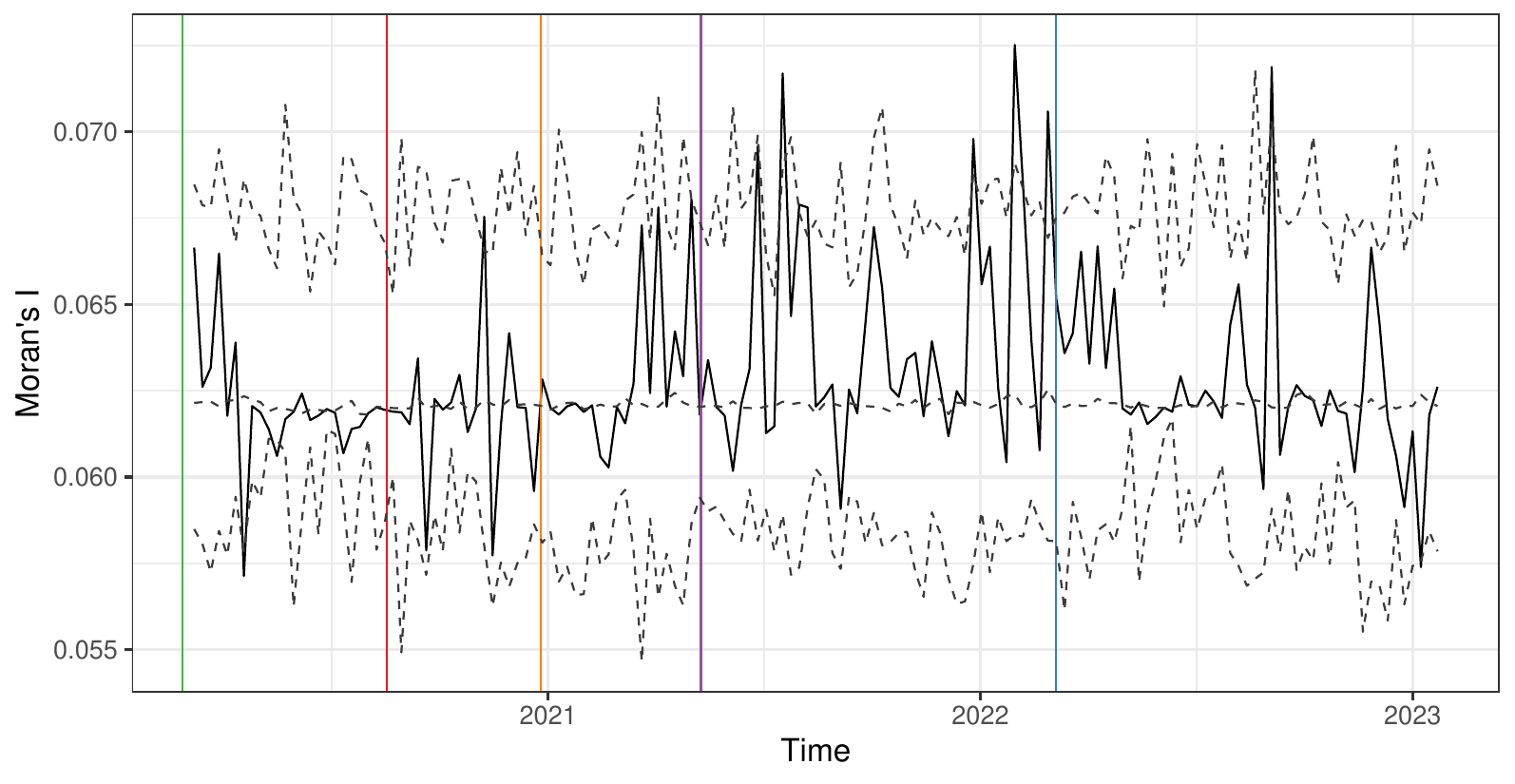}}%
\hspace{0.5cm}
\subcaptionbox{{Delaunay triangulation} network}{\includegraphics[width=0.4\textwidth]{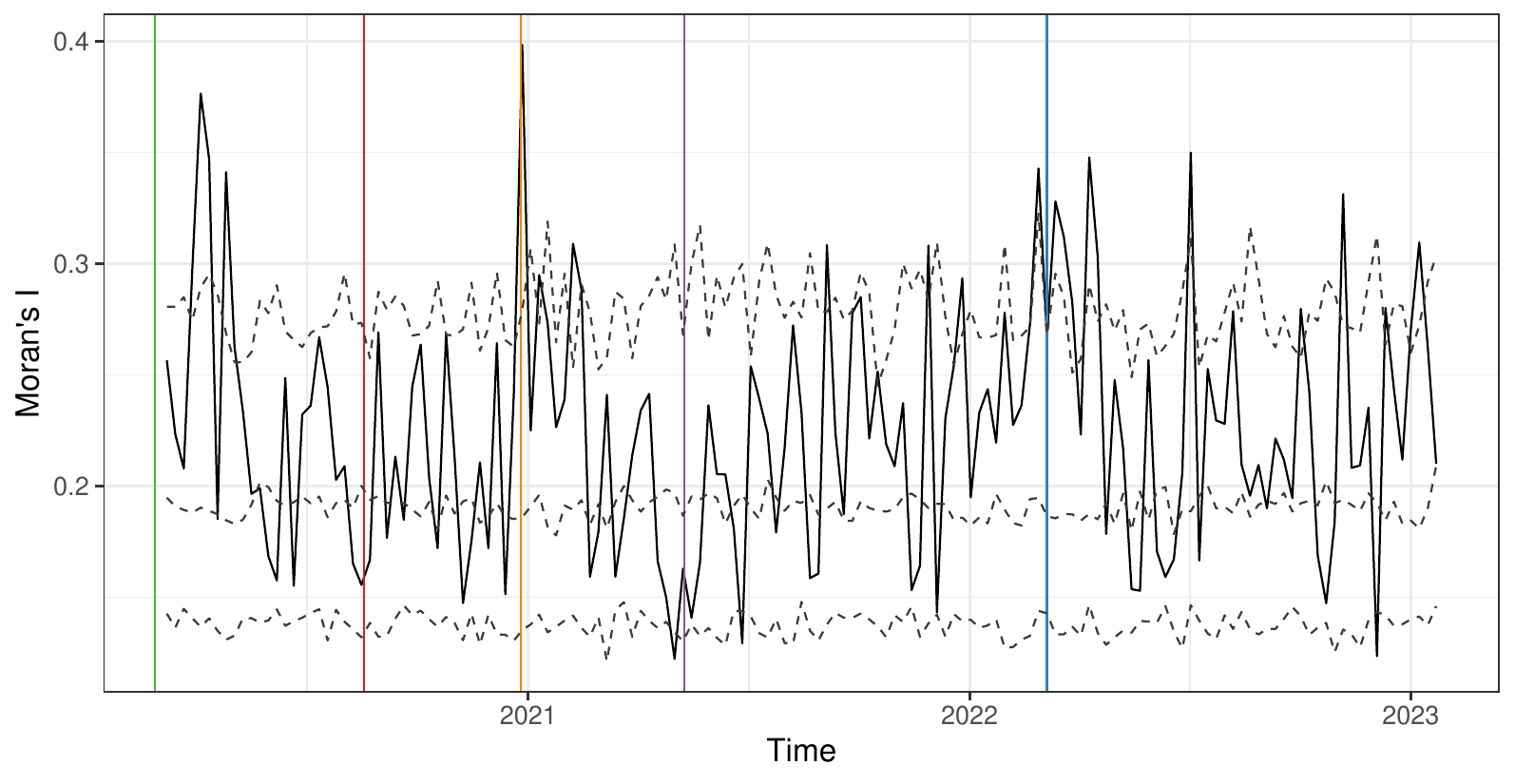}}%
\\
\subcaptionbox{{Gabriel} network}{\includegraphics[width=0.4\textwidth]{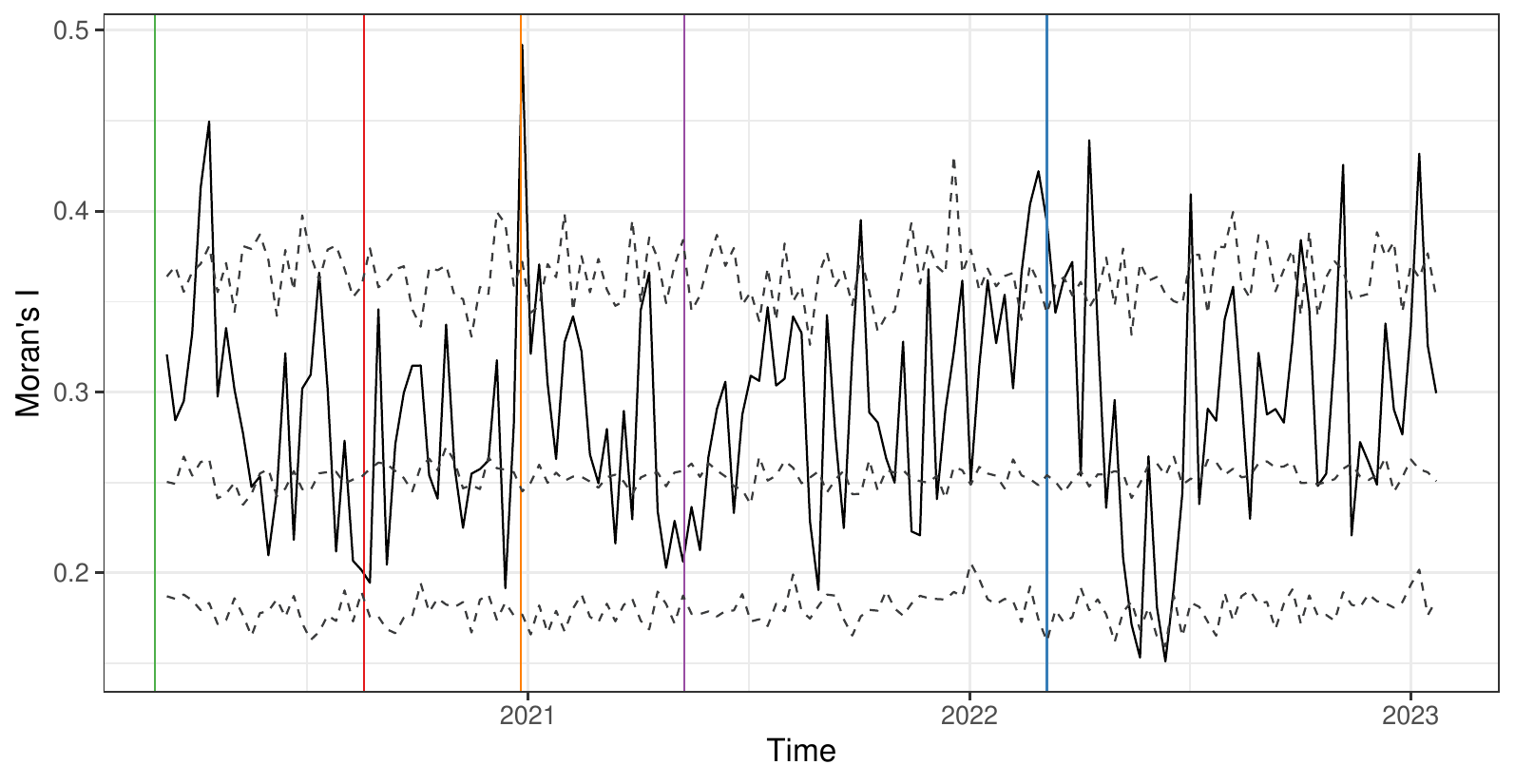}}%
\hspace{0.5cm}
\subcaptionbox{{SOI} network}{\includegraphics[width=0.4\textwidth]{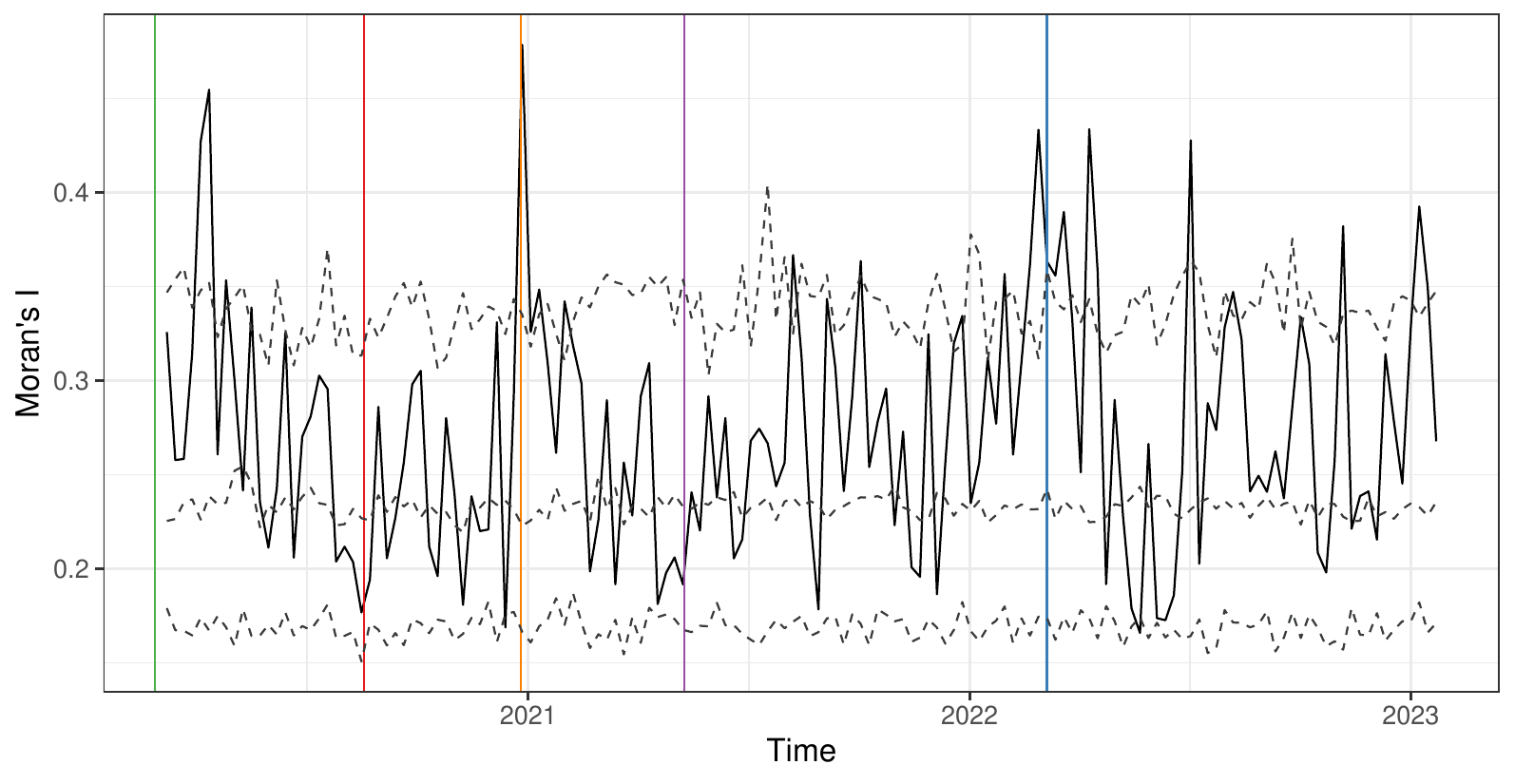}}%
\\
\subcaptionbox{{Relative neighbourhood} network}{\includegraphics[width=0.4\textwidth]{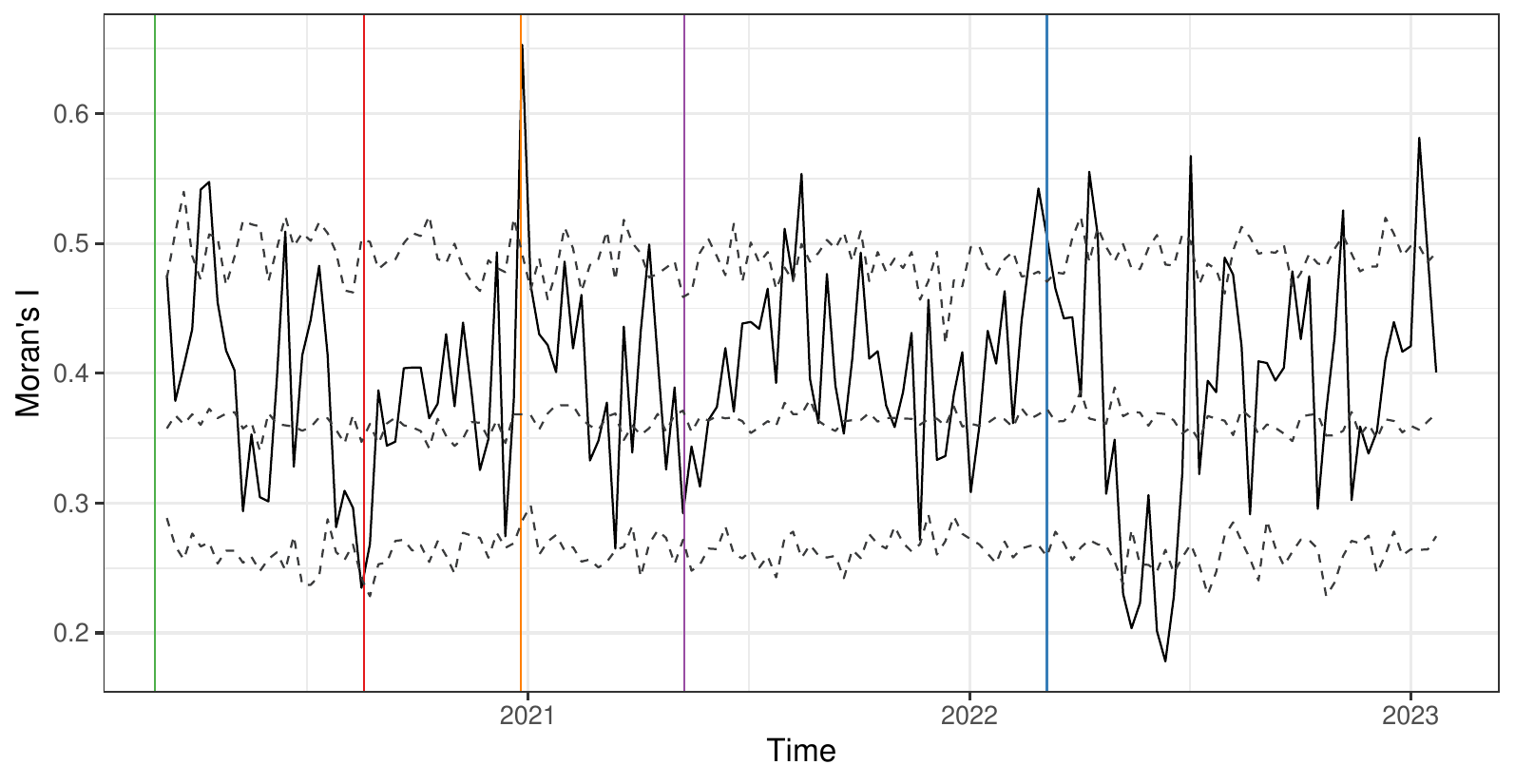}}
\hspace{0.5cm}
\subcaptionbox{{Queen's} network}{\includegraphics[width=0.4\textwidth]{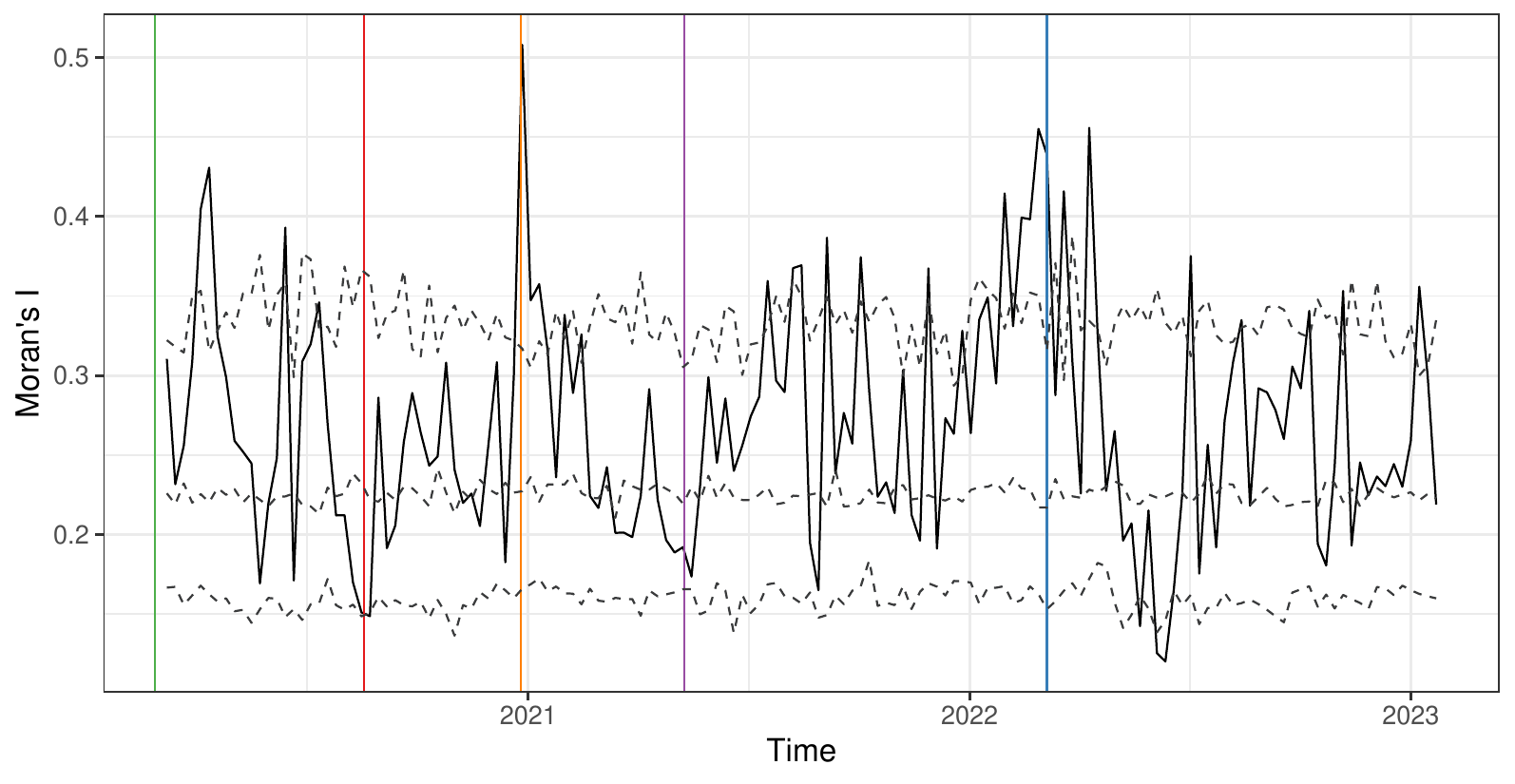}}
\caption{Moran's I across time, main COVID-19 regulations by the Irish Government indicated by vertical lines; in order: initial lockdown, county-specific restrictions, Level-5 lockdown, allowance of inter-county travel, official end of all restrictions; 95\% credibility interval in grey dashed}
\label{fig:morans_I_II}
\end{figure}

The main text shows Moran's I across time for \gr{Economic hub network and for} the KNN network.
Figure \ref{fig:morans_I_II} provides the results for the other networks. 
For the Delaunay triangulation, Gabriel, SOI and Relative neighbourhood network, Moran's I follows a similar pattern, due to the induced nature of the networks. 
The Economic hub network shows more extreme values in the trajectory of Moran's I compared to the Queen's network.

\clearpage
\section{GNAR models}

\subsection{Short introduction to Time Series Analysis}
\label{app:definition}
The temporal relationship within a time series is measured by the \textit{autocovariance} (or \textit{autocorrelation}), i.e. the covariance (or correlation) between $X_{t_1}$ and $X_{t_2}$ at some time points $t_1$ and $t_2$, observed for the same statistical unit (w.l.o.g. $t_1 < t_2$),
\begin{align*}
    AC(X_{t_1}, X_{t_2}) & = \mathbb{C}ov(X_{t_1}, X_{t_2}) \\
    & = \mathbb{E}\left( (X_{t_1} - \mu) (X_{t_2} - \mu) \right) \\
    ACor(X_{t_1}, X_{t_2}) & = \mathbb{C}orr(X_{t_1}, X_{t_2}) \\
    & = \frac{\mathbb{E}\left( (X_{t_1} - \mu) (X_{t_2} - \mu) \right)}{\sqrt{\mathbb{V}(X_{t_1})} \sqrt{\mathbb{V}(X_{t_2})}} \; .
\end{align*}

\textit{Stationarity} implies that, as time progresses, the distribution of the observations converges to a certain distribution, i.e. is independent of time.  \\
\textit{Weak stationarity} or \textit{Covariance-stationarity} is defined by a time-independent, constant mean and a time-independent covariance whose values only depends on the size of the lag $k$, not on time $t$ \cite{HamiltonJamesDouglas2020TSAe}, $\forall t \in \{1, ..., T\}, k \in \{0, ..., T-t \}$
\begin{align*}
    \mathbb{E}(X_t) & = \mu  \\
    \mathbb{C}ov(X_t, X_{t + k}) & = \gamma_k \; .
\end{align*}

A time series is \textit{strictly stationary} if the joint distribution for $(X_t, X_{t + k_1}, ..., X_{t + k_n})$ is independent of time $t$ and only depends on the time intervals $k_i$ between subsequent observations \cite{wei2006time}. \\

\subsection{Further background on network-based time series models}
\label{app:lit_rev_network_models}
Network time series models expand multivariate time series by incorporating non-temporal dependencies, represented by a network.
An edge indicates a certain relationship between the variables which are represented by the two vertices.  
The output variable at each vertex is modelled to depend on its own past values as well as the past values of its neighbouring vertices which are determined by the network underlying the model.
Many methods for multivariate time series analysis exist.
The following provides an incomplete overview. \\
\\
\noindent \textbf{Spatial Autoregressive and Moving Average models}. 
The development of network time series models has drawn inspiration from models for spatial observations.
\cite{ord1975estimation} developed the \textit{spatial autoregressive  model} (SAR) which models the value at a certain vertex as a weighted average across a vertex-specific re-defined location set and an additive random error.
Such model can be expanded to a \textit{mixed regressive-autoregressive model} by incorporating exogenous variables $Y$.  
\cite{ord1975estimation} also introduces an alternative model relying on autoregression in the error term.  
\cite{mur1999testing} extends these models to
\textit{spatial network autoregressive models} (SARMA) that include a weight matrix $W$ which encodes the spatial dependencies; see \cite{leenders2002modeling} for choices of $W$.
These models have been explored in numerous papers, e.g.\,\cite{doreian1980linear, doreian1989network, doreian1989models, dow1982network, loftin1983spatial, white1981sexual}.
However, all above mentioned models do not take any time dependence into account and assume a Gaussian error term with homoscedastic variance \cite{leeming2019new}.  
\\
\\
\textbf{Temporal extensions}. 
The \textit{m-STAR model} by \cite{hays2010spatial} models a spatial as well as a temporal dependence, restricted to 1-lag autoregression.
The sets of spatial weights represent network interdependence between vertices on different contextual levels.
The \textit{spatial temporal conditional autoregressive model} (STCAR) is a continuous Markov random field with a Gaussian conditional probability density function.
Its distributions rely on a space-time autoregressive matrix which accounts for both spatial and temporal dependencies. 
A general model choice is the \textit{vector autoregression model} (VAR). 
A VAR model regresses a vector at time on its values at previous time points, according to its order which is restricted by the size of the data set \cite{jiang2020autoregressive, leeming2019new}.
While general VAR models can model data which are represented by networks, they would typically include one parameter per edge in the network and are hence often too large to be of practical use. 
Hence, specifications of the VAR to network data have recently been developed, as follows. \\
\\
\noindent \textbf{Network autoregressive models}.
For network time series, \cite{zhu2017network} established a particular VAR model, the \textit{network autoregression model}.
It includes an intercept and exogenous variables while 
time dependence is restricted to lag 1. 
Network autoregression follows the tradition of SMA and SAR models but incorporates historic values of the vertex in question and its $1^{st}$-stage neighbourhood. 
To ensure less sensitivity to outliers in the data and to incorporate heteroscedasticity, the paper \cite{zhu2019network} expanded the model to \textit{network quantile autoregression} \cite{zhu2019network}. 
The paper \cite{zhu2020grouped} further develops the \textit{grouped network autoregressive model} (groupNAR), which breaks the homogeneity by attributing each vertex to a group and estimating group-specific coefficients.  
The classification of the vertices is learned simultaneously with the parameter estimation.
The number of classes has to be pre-specified. 
Similarly to the 1-stage network autoregressive model, the network autoregression model NAR(p,s) includes historic values of the observation itself as well as its neighbours while assuming stationarity and spatial network homogeneity, with a general choice of lags and of neighbourhood stages.  

The \textit{network autoregressive (integrated) moving average models} (NARIMA) resemble restricted vector autoregressive (VAR) models with restriction reflecting the underlying network \cite{knight2019generalised, knight2016modelling}.
NARIMA models facilitate dimensionality reduction and reduces the computational complexity for calculating model coefficients from $\mathcal{O}(n^2)$ for VAR models to $\mathcal{O}(n)$ for NARIMA models.
NARIMA models also allow for great flexibility in modelling spatial and temporal dependencies \cite{knight2019generalised, leeming2019new}. 
A NARIMA model consists of a NAR component, $X_t$, and a moving average component of order q (MA(q)), $\sum_{l = 1}^q \eta_l \varepsilon_{i, t-l}$ 
\begin{align}
\label{equ:NARMA}
X_{i, t} = \sum_{j = 1}^p \left( \alpha_j X_{i, t-j} + \sum_{r = 1}^{s_j} \sum_{q \in N^{(r)}(i)} \beta_{j, r} X_{q, t-j} \right) + \sum_{l = 1}^q \eta_l \varepsilon_{i, t-l} + \varepsilon_{i, t} \; .
\end{align}

The relevance of vertices is acknowledged by vertex-specific weights \cite{knight2016modelling}.
The choice of weights strongly depends on the scenario and content the model is applied to \cite{leenders2002modeling}.
\begin{align}
\label{equ:NARMA_weighted}
X_{i, t} = \sum_{j = 1}^p \left( \alpha_j X_{i, t-j} + \sum_{r = 1}^{s_j} \sum_{q \in N^{(r)}(i)} \beta_{j, r} \omega_{i, q} X_{q, t-j} \right) + \sum_{l = 1}^q \eta_l \varepsilon_{i, t-l} + \varepsilon_{i, t} \; .
\end{align}
The gNARIMA model generalises (\ref{equ:NARMA}) by incorporating time dependent weights $\omega_{i, q, t}$ which allow excluding some of the vertices some of the time \cite{knight2016modelling}. 
All above models do not incorporate any exogenous variables and can only integrate one network \cite{knight2016modelling, leeming2019new}. 
The GNAR model \eqref{equ:GNAR} used in this paper is an adaptation of the gNARIMA model \ref{equ:NARMA_weighted} with autoregressive weights $\alpha_{i,j}$ which are allowed to depend on the vertex itself, but without a moving average component.

\subsection{The GNAR model in matrix form, and generalised least squares estimation}
\label{app:GNAR_matrix_form}
\label{app:gls_estimation}
We can rewrite the GNAR model \eqref{equ:GNAR} in matrix form
\begin{align*}
    X = B Z + \varepsilon 
\end{align*}
with $X = [X_{p+1}, ..., X_T]$ and $Z = [Z_p, ..., Z_{T-1}]$ where $Z_t^T = [X_t, ..., X_{t-p+1}]$.
The matrix $B = [\phi_1, ..., \phi_p]$ summarises $\alpha$- and $\beta$-coefficients in the form $\phi_j  = \text{diag}(\alpha_{\alpha_{i, j}}) + 
\sum_{r = 1}^{s_j} \beta_{j, r} \, W^{(r)}$, where $W^{(r)}$ denotes the weight matrix with $[W^{(r, }]_{l, m} = \omega_{l, m} \cdot \mathbb{I}_{m \in N^{(r)}(l)}$.
The random error matrix is expressed by $\varepsilon = [\varepsilon_{p+1}, ... , \varepsilon_{T}]$, where the vector $\varepsilon_t = (\varepsilon_1, ..., \varepsilon_n) \sim (0, \Sigma_{\varepsilon})$ is iid.\,with variance $\Sigma_{\varepsilon} = \sigma^2 \cdot I_{N \times N}$ \cite{knight2019generalised}.

The GNAR model implies restrictions $R \in \mathbb{R}^{pN^2 \times M}$ on the parametrisation,
\begin{align}
\label{equ:restrictions}
    vec(B) = \phi = R \gamma
\end{align}
where $vec(B)$ describes the reformatting of matrix $B$ into a vector by stacking its columns, the vector $\phi$ denotes the unrestricted coefficient vector for a VAR model and the vector $\gamma$ consists of the $M$ unrestricted parameters.
For a vertex-specific GNAR model, $M = N p +  \sum_{j = 1}^p s_j$ and for a global-$\alpha$ model, $M = p +  \sum_{j = 1}^p s_j$ \cite{knight2019generalised}.
The Least Squares (LS) estimation finds $\hat{\phi}$ such that it minimises the sum of squares \cite{Lutkepohl1991Itmt}, 
\begin{equation}
\label{equ:multivariate_ls}
    \hat{\phi}  = \text{argmin}_{\phi} \, tr\left( (X - BZ)^T \Sigma_{\varepsilon}^{-1} (X - BZ) \right) \notag  = \left( (Z Z^T)^{-1} Z \otimes I_N \right) \cdot vec(X) , 
\end{equation}
where $\otimes$ denotes the Kronecker product. 
For the re-parametrisation imposing constraints in (\ref{equ:restrictions}), the Generalised Least Squares estimation (GLS) computes
\begin{align}
\label{equ:gls}
    \hat{\gamma} = \left( R^T (ZZ^T \otimes {\Sigma}_{\varepsilon}^{-1}) R \right)^{-1} R (Z \otimes {\Sigma}_{\varepsilon}^{-1} ) \cdot vec(X)
\end{align}
\cite{knight2019generalised, nason2021quantifying}.
The GLS estimator is consistent and asymptotically follows a normal distribution if $\{X_t\}_t$ is stationary and $\varepsilon_t$ a \textit{standard white noise process}. 
It is identical to the Maximum Likelihood estimator if we assume $\varepsilon_t$ to be Gaussian \cite{Lutkepohl1991Itmt}. 
However, the GLS estimator requires knowledge of the error covariance matrix $\Sigma_{\varepsilon}$ which is usually unknown. 
The Estimated GLS estimator (EGLS) substitutes the true covariance matrix in the estimation \eqref{equ:gls} with a consistent estimator $\hat{\Sigma}_{\varepsilon}$ which converges to $\Sigma_{\varepsilon}$ in probability as $n \to \infty$ \cite{Lutkepohl1991Itmt}, 
\begin{align}
\label{equ:egls}
    \Tilde{\gamma} = \left( R^T (ZZ^T \otimes \hat{\Sigma}_{\varepsilon}^{-1}) R \right)^{-1} R (Z \otimes \hat{\Sigma}_{\varepsilon}^{-1} ) vec(X) \; .
\end{align}
The matrix $R^T (ZZ^T \otimes \hat{\Sigma}_{\varepsilon}^{-1}) R$ must be non-singular which holds with probability 1 for continuous $X_t$ \cite{Lutkepohl1991Itmt}. 
The estimate $\Tilde{\gamma}$ is consistent and asymptotically normal if $\varepsilon$ is standard white noise.
For a stationary time series with standard white noise, the EGLS estimate \eqref{equ:gls} and GLS estimate \eqref{equ:gls} are asymptotically equivalent \cite{Lutkepohl1991Itmt}.  
Under stationarity and standard white noise error, a possible choice for a consistent estimator $\hat{\Sigma}_{\varepsilon}$ is 
\begin{align}
\label{equ:estimate_sigma}
    \hat{\Sigma}_{\varepsilon} = \frac{1}{T} (X - \hat{B}Z) (X - \hat{B}Z)^T
\end{align}
where $\hat{B}$ follows from the unconstrained LS estimate (\ref{equ:multivariate_ls}) and a corresponding transformation to obtain matrix $\hat{B}$ \cite{Lutkepohl1991Itmt}. 
We obtain $\Tilde{B}$ by inserting the EGLS estimate for $\gamma$ into (\ref{equ:restrictions}) \cite{knight2019generalised},  
\begin{align*}
    \widehat{vec(B)} = R \Tilde{\gamma} \; .
\end{align*}

\subsection{Choices for the order of the model}
\label{app:beta_choices}
We iterate through all possible combinations of $\alpha$-order \as{$p \in $ {1, ..., 7}} and $\beta$-order \as{
    1-7, (1, 1)-(5, 1), (2, 2), (1, 1, 1)-(5, 1, 1), (2, 2, 1), (2, 2, 2)-(5, 2, 2), 
    (1, 1, 1, 1)-(5, 1, 1, 1), (2, 2, 1, 1), (2, 2, 2, 1), (3, 2, 2, 1), (4, 2, 2, 1), (5, 2, 2, 1), (2, 2, 2, 2), 
    (1, 1, 1, 1, 1), (2, 1, 1, 1, 1)-(5, 1, 1, 1, 1),(2, 2, 1, 1, 1)-(5, 2, 2, 1, 1), 
    (2, 2, 2, 1, 1), (2, 2, 2, 2, 1), (2, 2, 2, 2, 2) as well as higher $\alpha$-order options by inserting zeros for higher lags.}

\medskip
In addition to the BIC, a second model selection criterion is the \textit{Akaike Information Criterion} (AIC); it  works similarly  to the BIC but penalises the number of parameters differently \cite{akaike1974new, bozdogan1987model};
\begin{align*}
    \text{AIC}(k, n) = 2 k - 2 \cdot \text{log}(L(X; \theta)) \;.
\end{align*} 
The AIC is not necessarily consistent (\cite{lutkepohl2005new}, Corollary 4.2.1) and therefore the BIC is preferred for model selection \cite{leeming2019new}.
We found the AIC values very similar to the BIC values in our data analysis and hence do not report them in the main text.
\section{Stationarity in COVID-19 Data Subsets}
\label{app:stationarity}

\begin{figure}[h!]
\centering
\subcaptionbox{Data subset 1}{\includegraphics[width=0.4\textwidth]{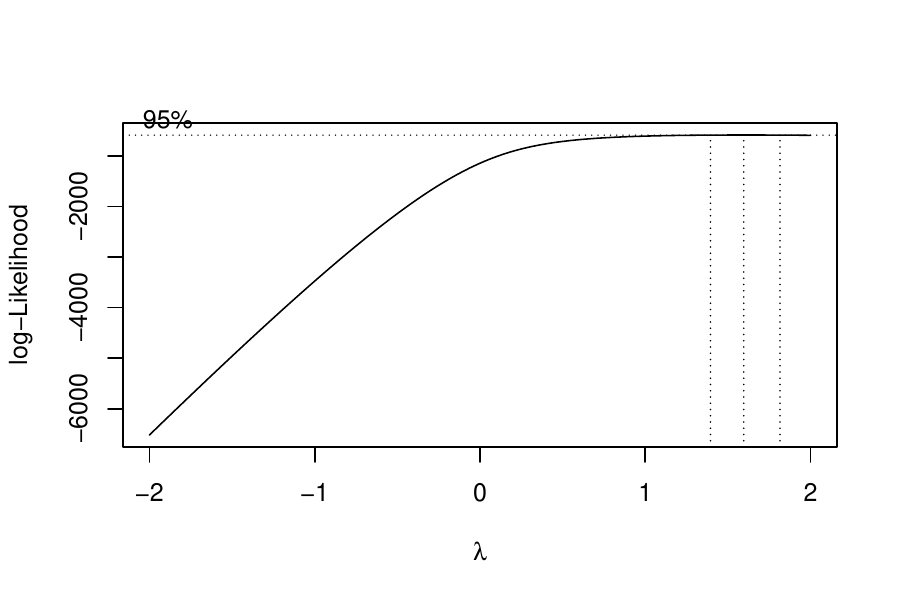}}%
\hspace{1em} 
\subcaptionbox{Data subset 2}{\includegraphics[width=0.4\textwidth]{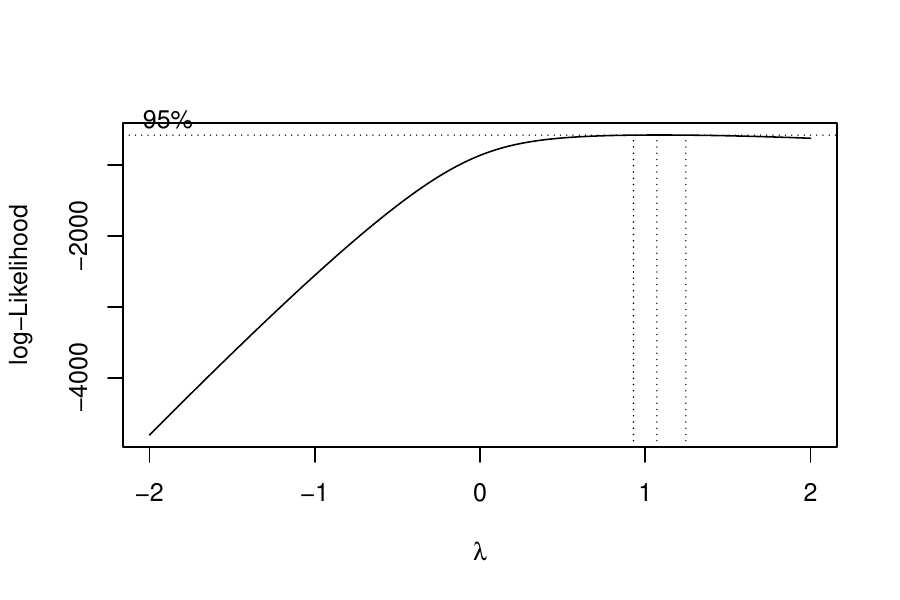}}%
\\
\subcaptionbox{Data subset 3}{\includegraphics[width=0.4\textwidth]{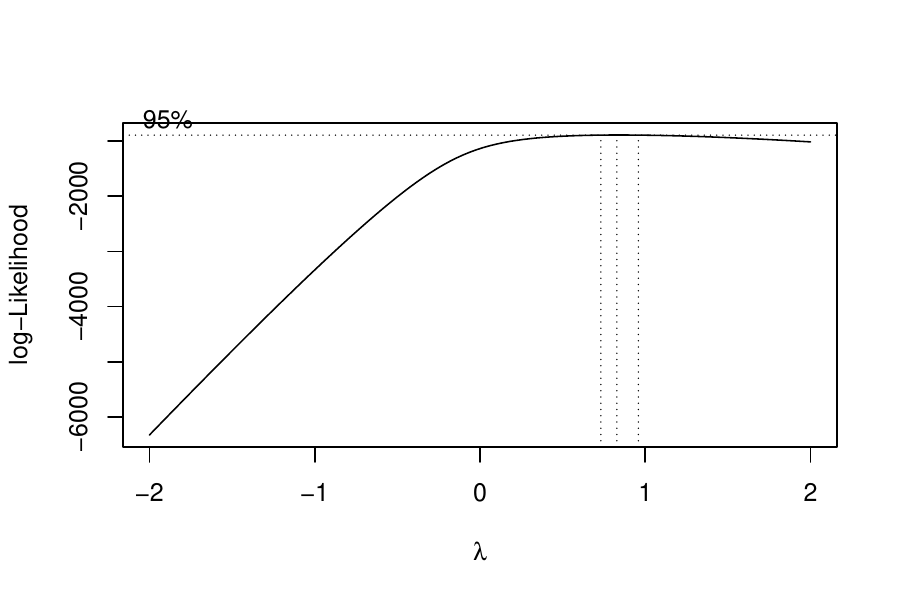}}%
\hspace{1em} 
\subcaptionbox{Data subset 4}{\includegraphics[width=0.4\textwidth]{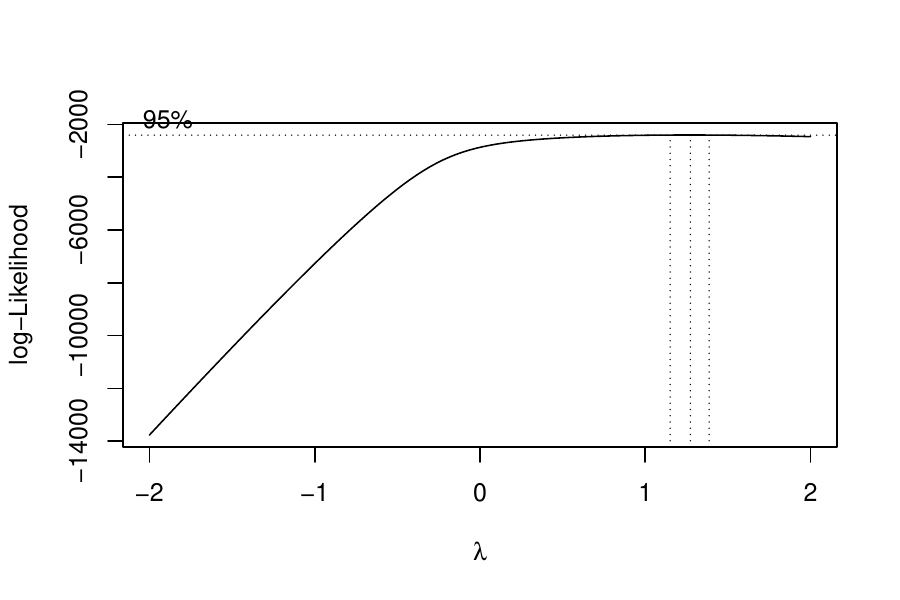}}%
\\
\subcaptionbox{Data subset 5}{\includegraphics[width=0.4\textwidth]{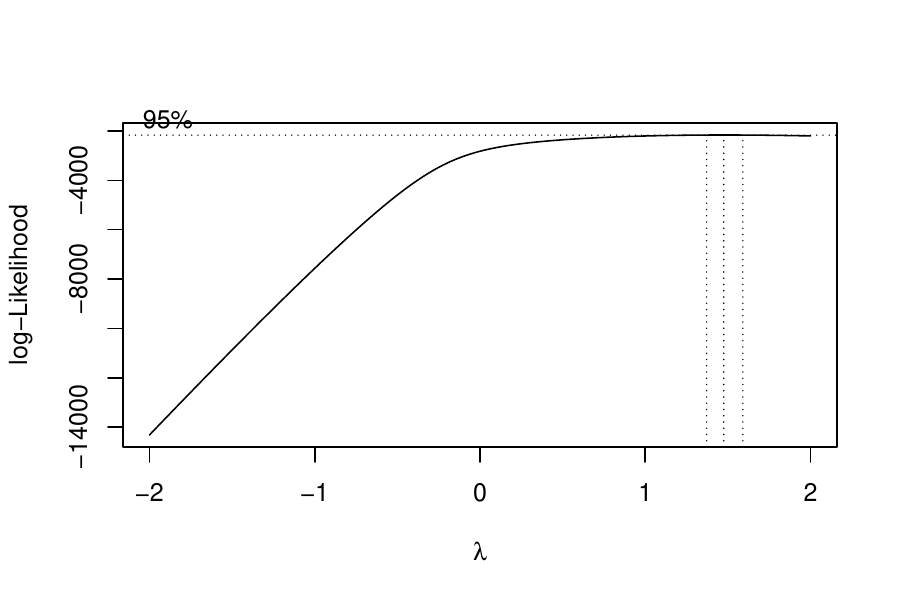}}%
\caption{Box-Cox transformations for each subset to assess stationarity}
\label{fig:stationarity}
\end{figure}

A Box-Cox transformation of the data can often improve stationarity. 
For datasets 2, 3, and 4, see Figure \ref{fig:stationarity} (a) and (c), the optimal $\lambda$ values are close to 1. 
For datasets 1 and 5, see Figure \ref{fig:stationarity} (b), (d) and (e), the values lie around $\lambda \approx 1.5$. 
These findings suggest that the 1-lag differenced COVID-19 incidence across subsets is already approximately stationary, indicating no need for further transformation.

\section{Training GNAR models}
\label{app:gnar_entire_data}
\subsection{Selecting optimal GNAR models across networks}
\label{app:optimal_gnar}
The models are fitted according to three weighting schemes: (1) shortest path length (SPL), (2) inverse distance weighting (IDW), (3) inverse distance and population density weighting (PB).
Table \ref{tab:best_models} shows the results for the entire data set.
Overall, the best performing models have comparatively large $\alpha$-order and include at least the $1^{st}$-stage neighbourhood.

The KNN network ($k = 21$) with SPL weighting achieves the absolute lowest BIC ($BIC = 193.9510$), followed by the DNN network ($d =325$; $BIC = 194.1066$) and the Complete network ($BIC = 194.1308$).
The best performing sparser network is the Delaunay triangulation network ($BIC = 194.4935$). 
As a general trend, it is noticeable that models with larger $\beta$-order are preferred for sparser network and models with smaller $\beta$-order for denser networks, indicating a certain trade-off between parameter "complexity" and network density. 
The SPL weighting outperforms both alternative weighting schemes for \as{most} network.
\begin{table}[h!]
\small
\centering
\begin{tabular}{l|rcrr}
Network & Weighting scheme & GNAR model &
                                          BIC \\
 \midrule 
 Delaunay & SPL & GNAR-4-4100 & 194.49 \\ 
  Gabriel & PB & GNAR-4-5000 & 195.13 \\ 
  SOI & SPL & GNAR-7-4220000 & 194.63 \\ 
  Relative & SPL & GNAR-4-5000 & 195.13 \\ 
  Complete & SPL & GNAR-5-11110 & 194.13 \\ 
  Queen & IDW & GNAR-5-50000 & 194.97 \\ 
  Eco. hub & IDW & GNAR-4-4000 & 195.47 \\ 
  Rail & SPL & GNAR-4-4000 & 195.48 \\ 
  \textbf{KNN (k = 21)} & \textbf{SPL} & \textbf{GNAR-5-11110} & \textbf{193.95} \\ 
  DNN (d = 325) & SPL & GNAR-5-11110 & 194.11 \\ 
   \bottomrule 
\end{tabular}
\caption{\as{Overview over best performing GNAR models for 
each COVID-19 network, with best performing weighting scheme;
best performing GNAR model in bold face.}} 
\label{tab:best_models}
\end{table}

{Table \ref{tab:best_model_subsets} indicates that the GNAR models have an overall better fit for the restricted data set than for the unrestricted data set. 
The large $\alpha$ values for each model indicate a strong temporal dependence in COVID-19 ID. 
} 

\begin{table}[h!]
\centering
\begin{tabular}{l|rrr}
  \toprule 
 Network & data subset & best model &
                                          BIC \\
 \midrule 
Train & restricted & GNAR-7-1000000 & 63.85 \\ 
  Queen & restricted & GNAR-7-1100000 & 65.85 \\ 
  \textbf{Eco. hub} & \textbf{restricted} & \textbf{GNAR-7-3110000} & \textbf{58.91} \\ 
  KNN-11 & restricted & GNAR-7-2222000 & 59.12 \\ 
  DNN-325 & restricted & GNAR-7-2100000 & 61.03 \\ 
  Delaunay & restricted & GNAR-7-2111000 & 64.22 \\ 
  Gabriel & restricted & GNAR-7-4100000 & 67.13 \\ 
  Relative & restricted & GNAR-7-4000000 & 68.36 \\ 
  SOI & restricted & GNAR-7-2222200 & 62.06 \\ 
  Complete & restricted & GNAR-7-1100000 & 65.61 \\ 
  \\
  Train & unrestricted & GNAR-7-5000000 & 191.76 \\ 
  Queen & unrestricted & GNAR-7-3000000 & 192.20 \\ 
  Eco. hub & unrestricted & GNAR-7-3000000 & 192.56 \\ 
  \textbf{KNN-21} & \textbf{unrestricted} & \textbf{GNAR-7-1111000} & \textbf{190.07} \\ 
  DNN-325 & unrestricted & GNAR-7-1111000 & 190.28 \\ 
  Delaunay & unrestricted & GNAR-7-4111000 & 190.92 \\ 
  Gabriel & unrestricted & GNAR-7-4000000 & 192.05 \\ 
  Relative & unrestricted & GNAR-7-5000000 & 191.53 \\ 
  SOI & unrestricted & GNAR-7-4000000 & 190.99 \\ 
  Complete & unrestricted & GNAR-7-1111000 & 190.31 \\ 
   \bottomrule 
\end{tabular}
\caption{Overview over the best performing GNAR model for each network on 
the restricted and unrestricted data set; best model \gr{in} bold \gr{face}} 
\label{tab:best_model_subsets}
\end{table}

\begin{figure}[h!]
    \centering
    \includegraphics[scale = 0.55]{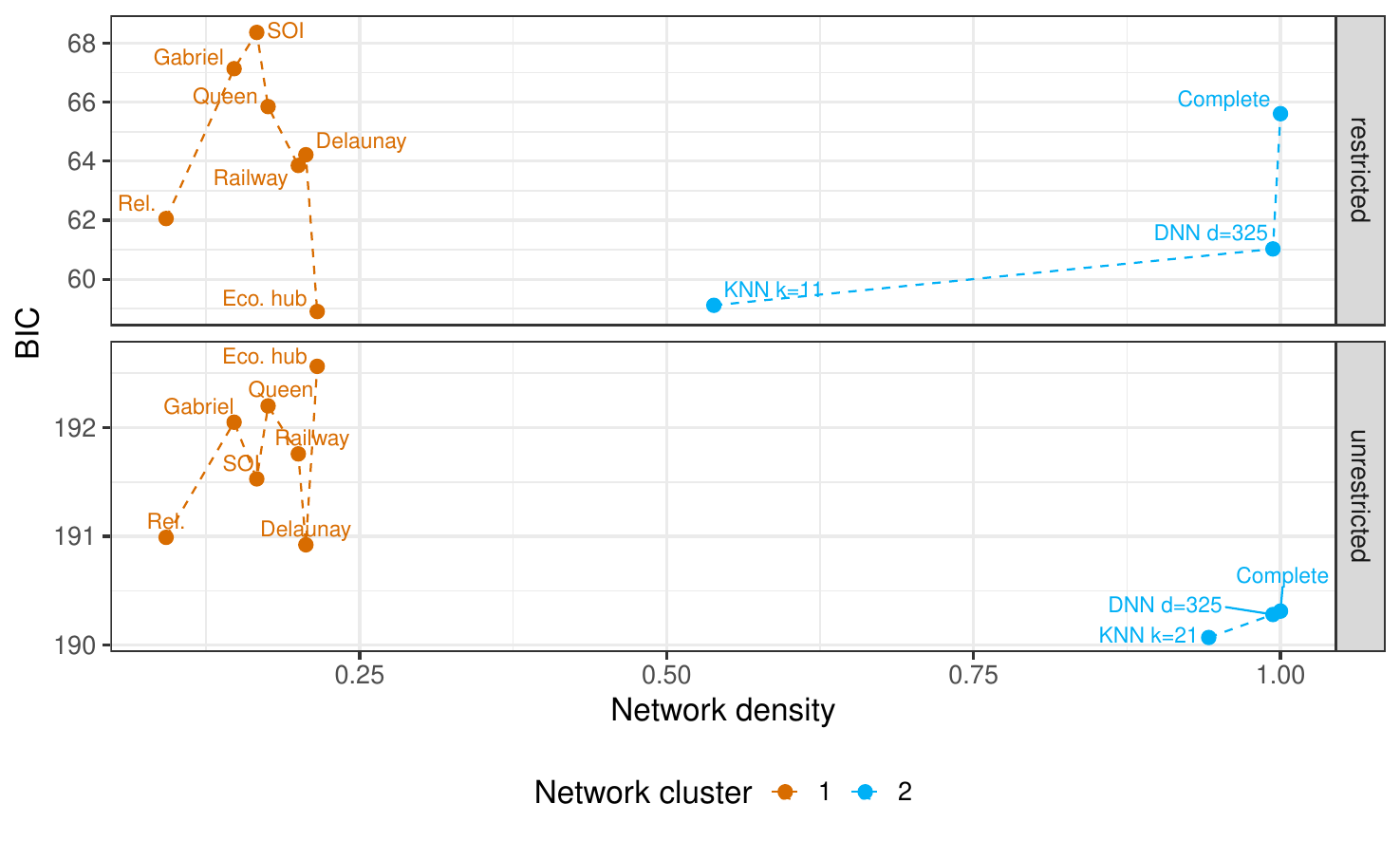}
    \caption{Relationship between network density and BIC value for the network-specific optimal GNAR model for the restricted and unrestricted data set; \as{clustering of networks according to kmeans based on network density and average local clustering coefficient}}
    \label{fig:bic_density}
\end{figure}

Figure \ref{fig:bic_density} implies that networks with high density obtain smaller BIC values than sparse networks for the unrestricted data set, with a minimum for the third densest network, the KNN network.
For the restricted data set, the optimal BIC values decrease as the network density increases, to reach its minimum for the \as{Economic hub} network, and increase for denser networks. 

\subsection{GNAR coefficient development for the optimal model: restricted versus unrestricted phase}
\label{app:gnar_coef_optimal}

The GNAR model $GNAR(\alpha = 5, \beta = (1, 1, 1, 1, 0))$ for the KNN network $k  = 21$, optimal on the entire data set, is fitted on the data for the restricted and unrestricted phases. 
The absolute value for $\alpha$ and $\beta$ coefficients increase for the unrestricted pandemic phase compared to the restricted phase. 

\begin{figure}[h!]
    \centering
    \includegraphics[width=0.8\textwidth]{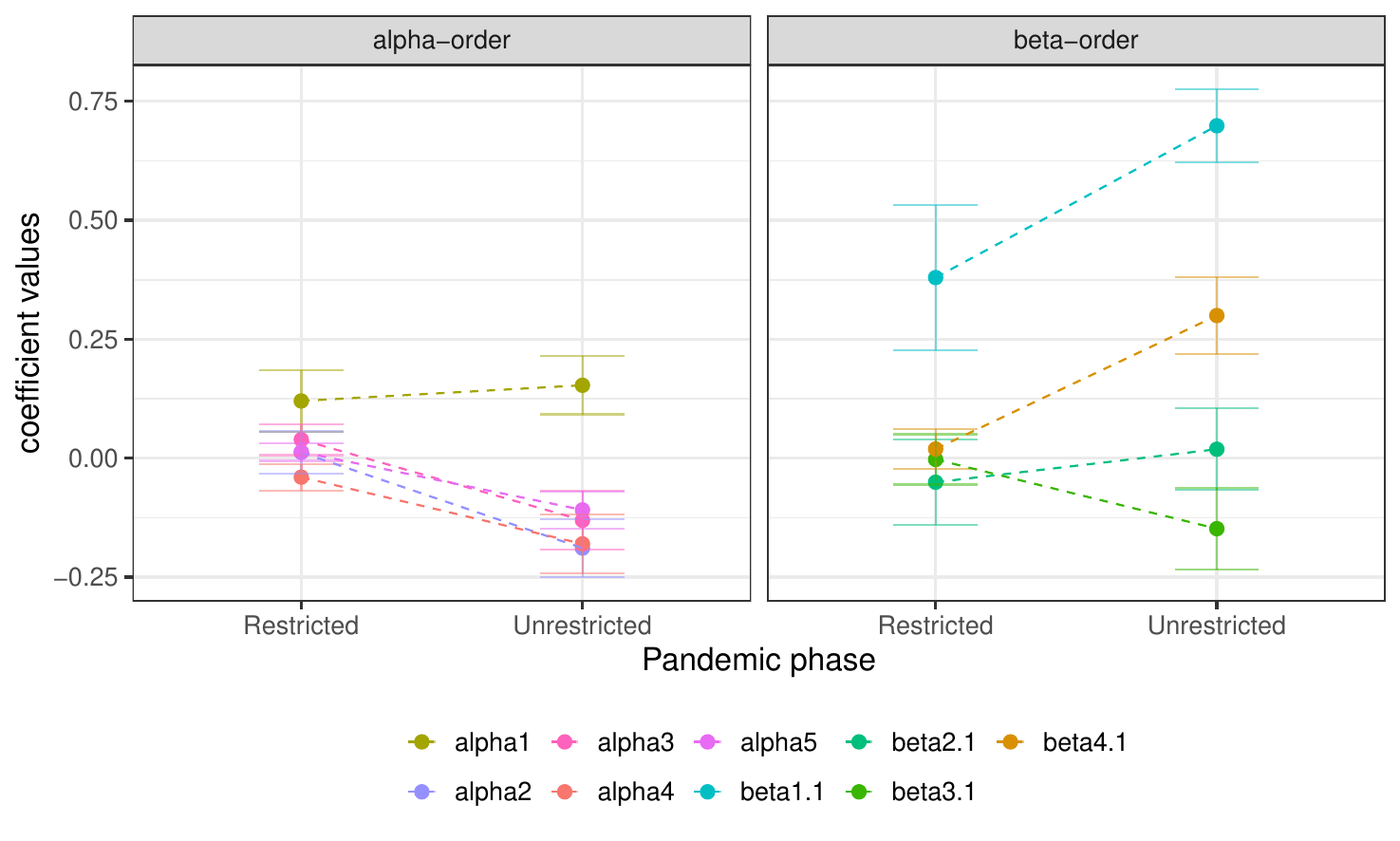}
\caption{Development of GNAR model coefficients for the restricted and unrestricted pandemic phase for the optimal GNAR model $GNAR(\alpha = 5, \beta = (1, 1, 1, 1, 0))$ for the KNN network with $k = 21$}
\label{fig:coef_change_optimal_model}
\end{figure}

\subsection{Prediction with GNAR models}
\label{app:predictions}
The predictive accuracy of the best performing model, fitted on the entire COVID-19 data set for each COVID-19 network \as{with SPL weighting}, is measured by predicting the lag-1 COVID-19 ID for the time period of \as{5} weeks, and computing the corresponding weekly mean absolute squared error (MASE) \cite{fahrmeir2016statistik, hyndman2006another, leeming2019new}. 

\begin{figure}[h!]
\centering
\subcaptionbox{{Delaunay triangulation}, {Gabriel}, {Relative neighbourhood}, {SOI} and {Railway-based} network}{\includegraphics[width=0.9\textwidth]{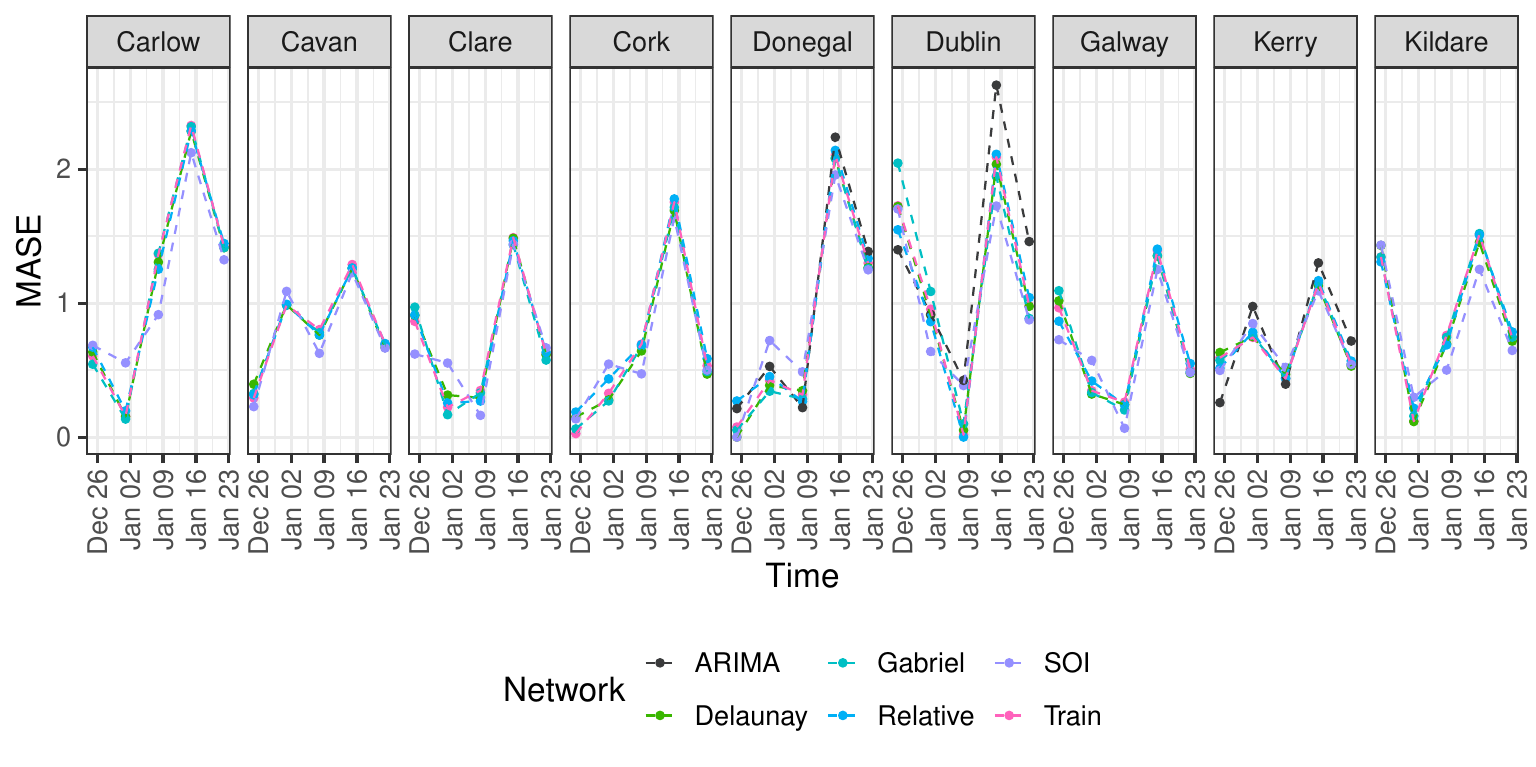}}%
\\
\subcaptionbox{{KNN}, {DNN}, {Complete}, {Queen's contiguity} and {Economic hub} network}{\includegraphics[width=0.9\textwidth]{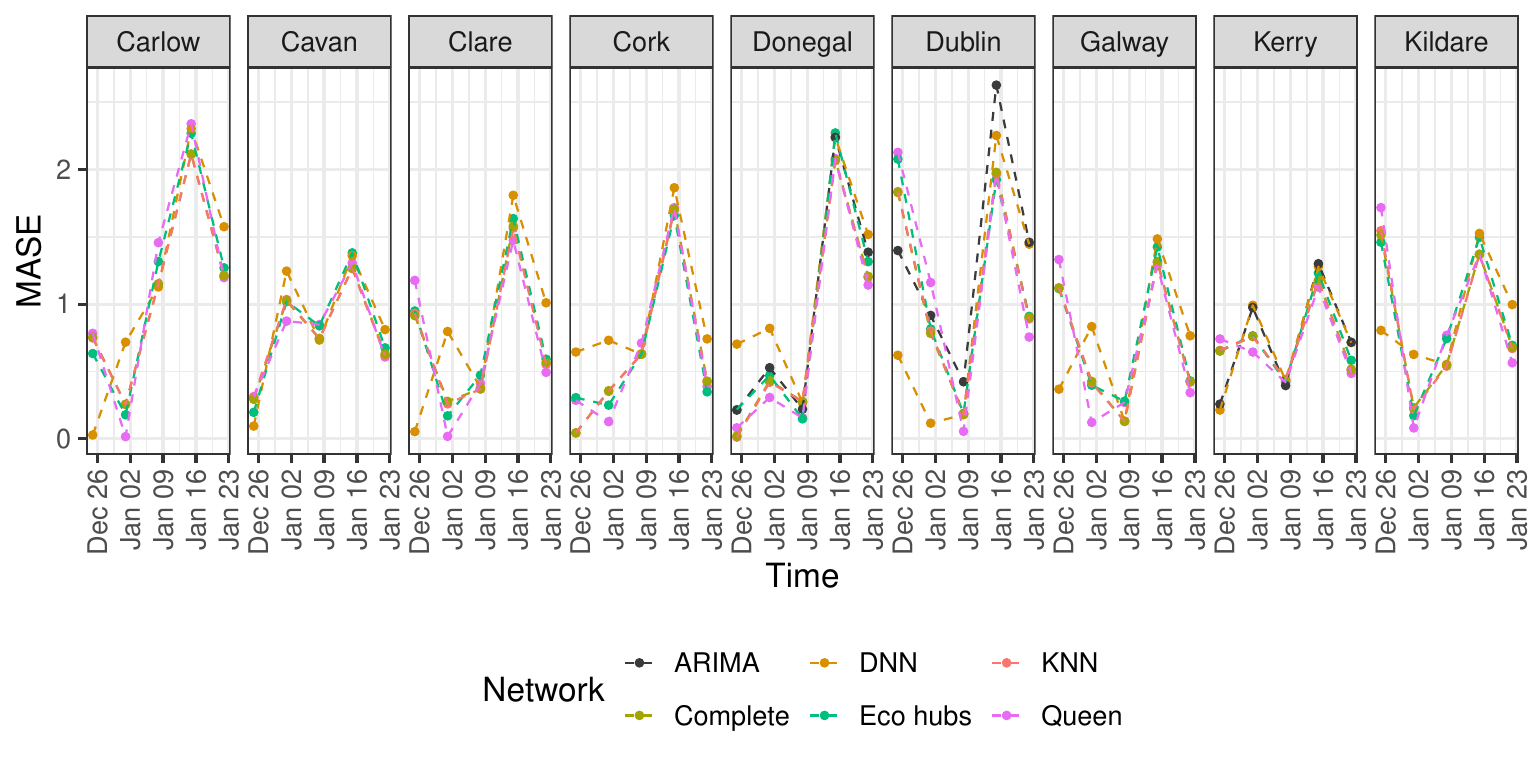}}%
\caption{MASE values for COVID-19 network, as well as for the county-specific {ARIMA} benchmark models}
\label{fig:mase}
\end{figure}

Figures \ref{fig:mase} include the MASE for the county-specific ARIMA models as a comparative benchmark.
The ARIMA models are comparable to any GNAR model in predictive accuracy. 
No network performs visibly better than any other over space and time, and are very similar. 

\begin{figure}[h!]
\centering
\subcaptionbox{{Delaunay triangulation}, {Gabriel}, {Relative neighbourhood}, {SOI} and {Railway-based} network}{\includegraphics[width=0.9\textwidth]{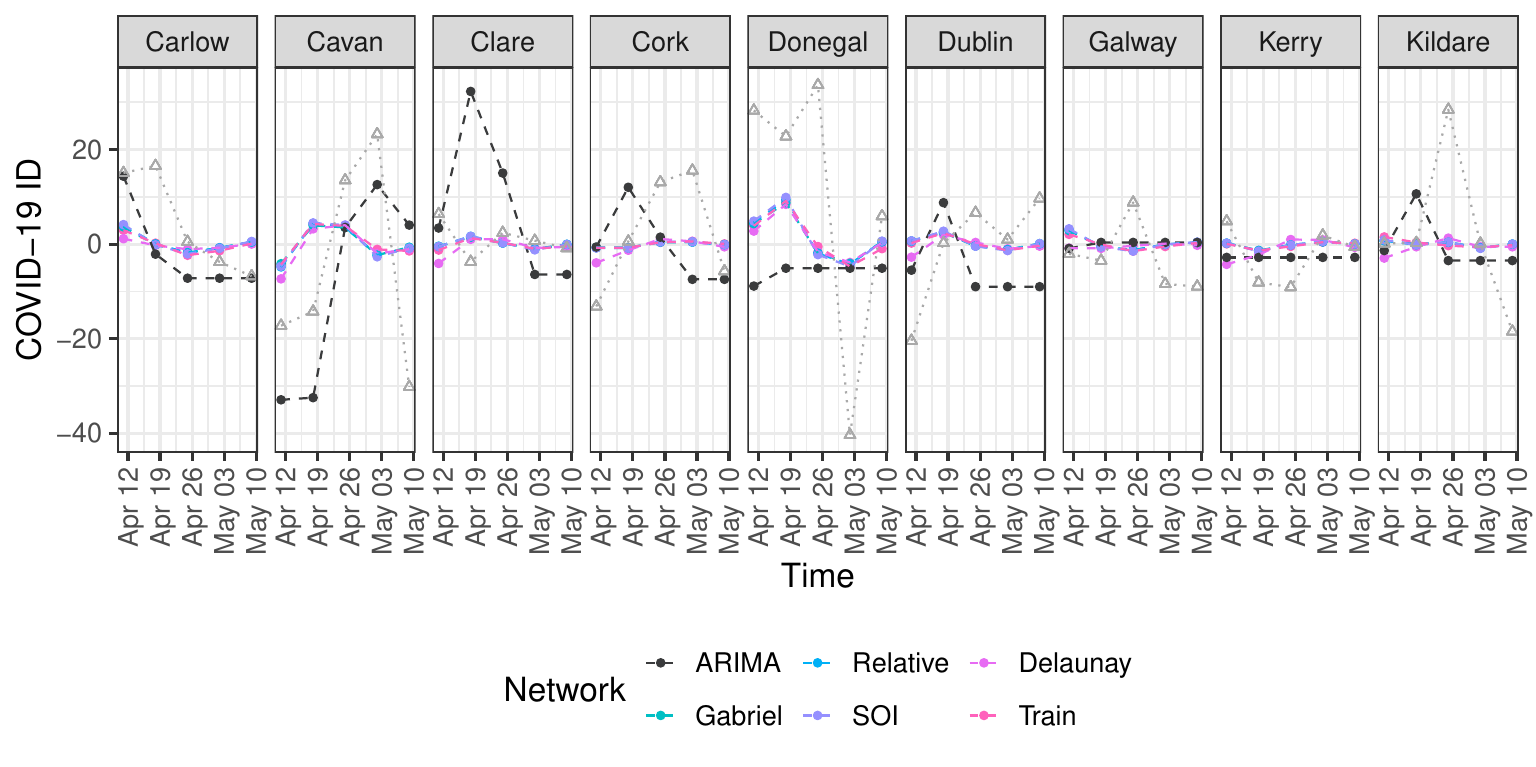}}%
\\
\subcaptionbox{{KNN}, {DNN}, {Complete}, {Queen's contiguity} and {Economic hub} network}{\includegraphics[width=0.9\textwidth]{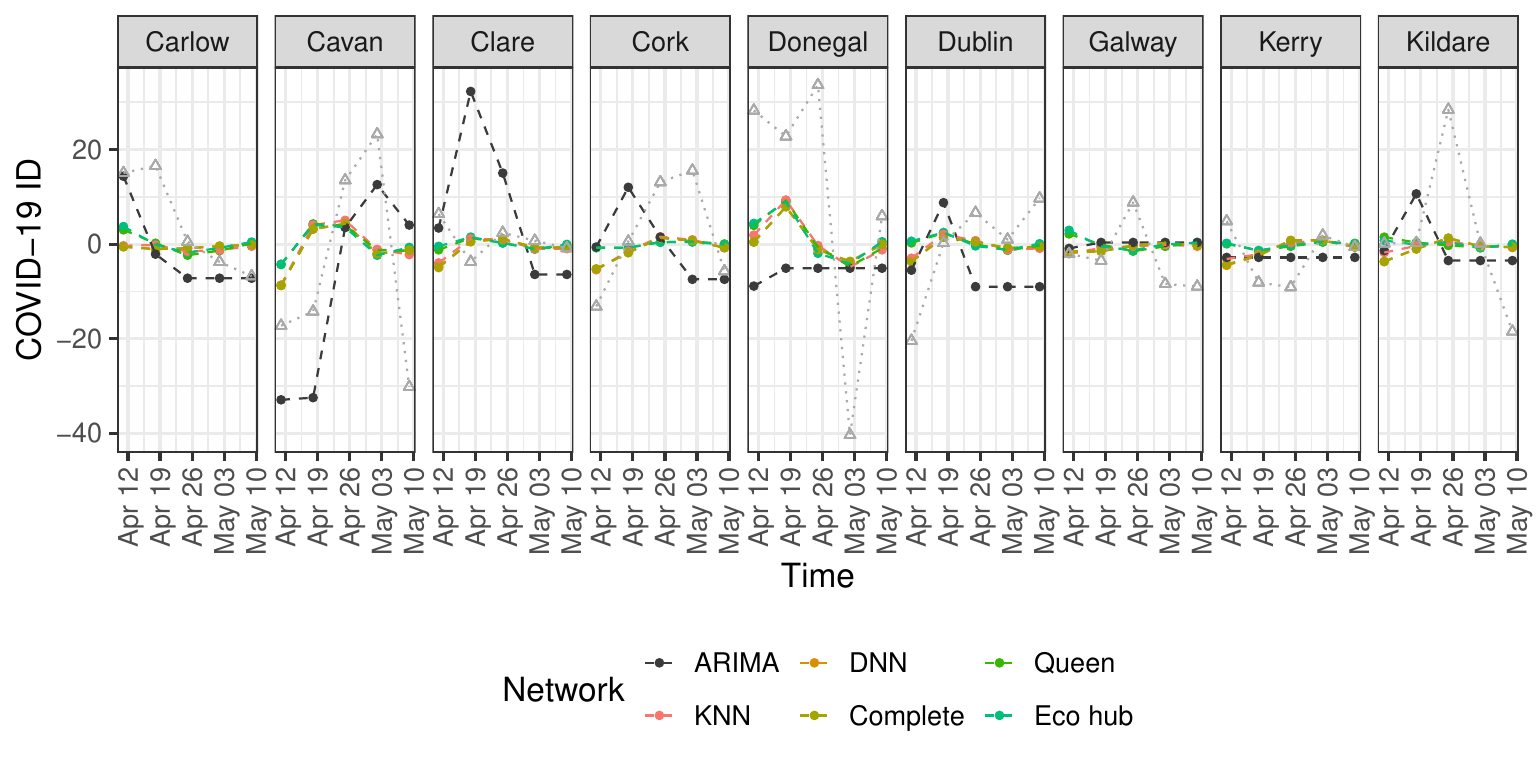}}%
\caption{Predicted COVID-19 ID for the restricted phase; true values in dotted grey with triangles.}
\label{fig:predicted_restricted_I}
\end{figure}

\begin{figure}[h!]
\centering
\subcaptionbox{{Delaunay triangulation}, {Gabriel}, {Relative neighbourhood}, {SOI} and {Railway-based} network}{\includegraphics[width=0.9\textwidth]{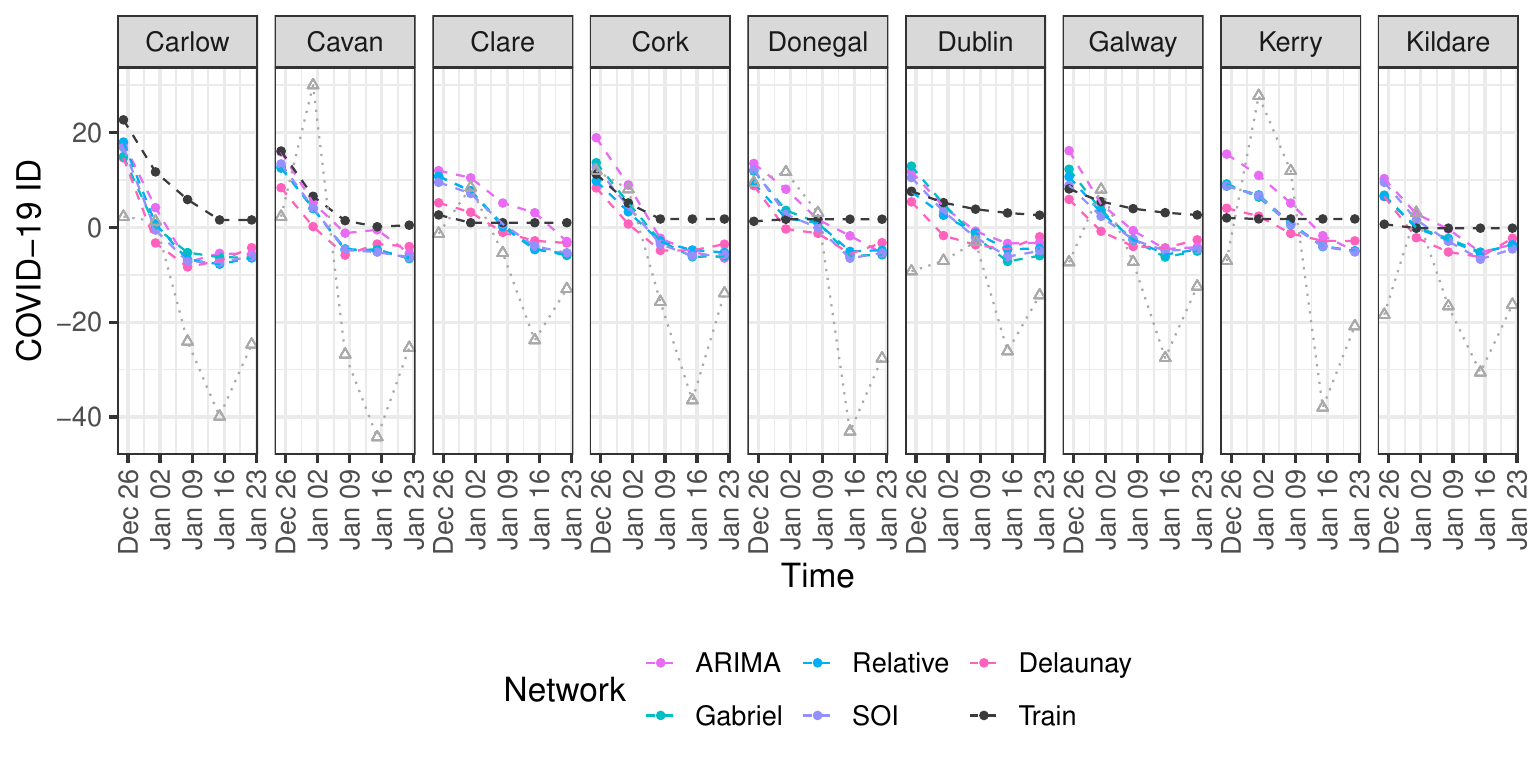}}%
\\
\subcaptionbox{{KNN}, {DNN}, {Complete}, {Queen's contiguity} and {Economic hub} network}{\includegraphics[width=0.9\textwidth]{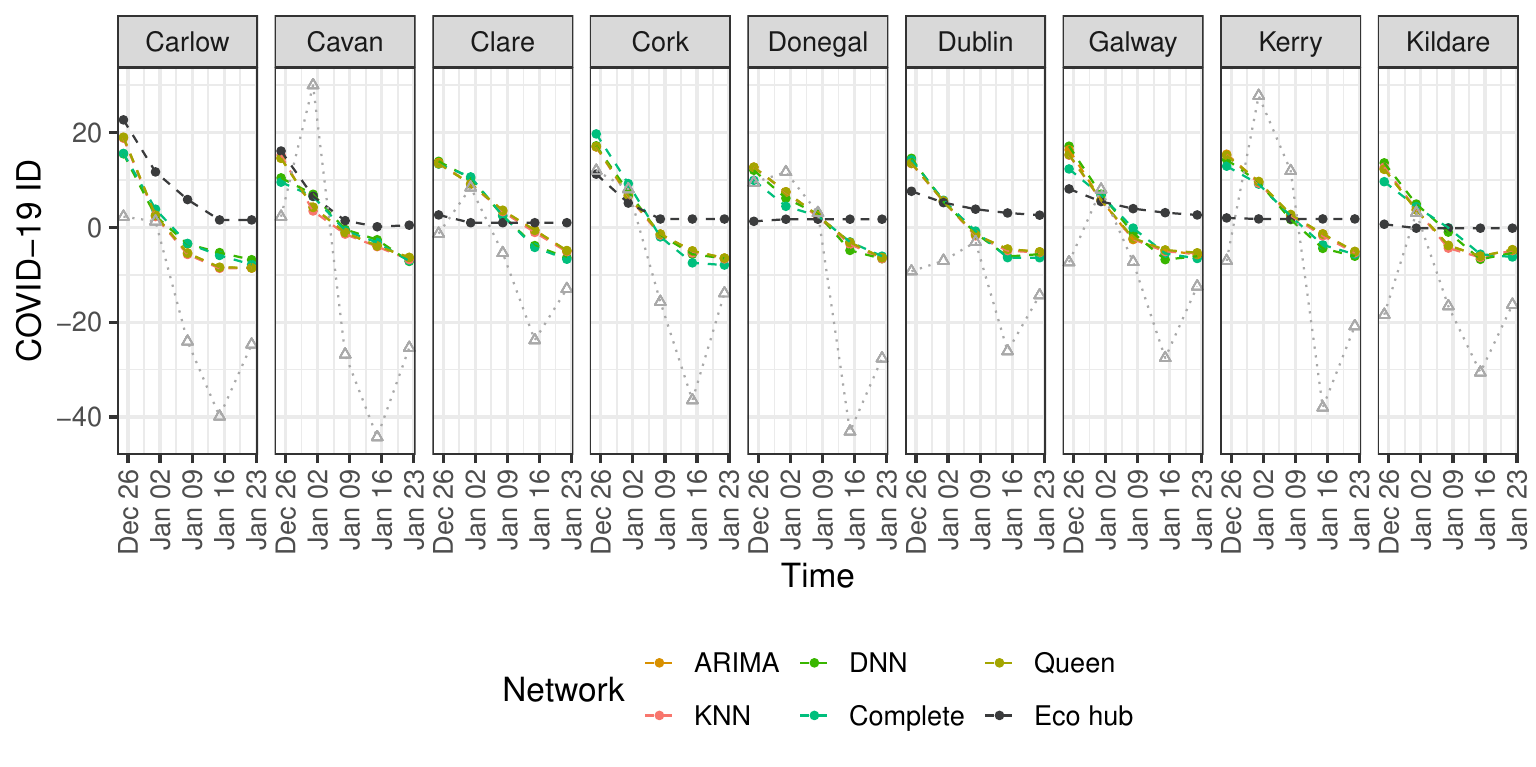}}%
\caption{Predicted COVID-19 ID for the unrestricted phase; true values in dotted grey with triangles.}
\label{fig:predicted_free_I}
\end{figure}

While the general fit may not be unreasonable, we find that GNAR models across networks, fit specifically to the restricted and unrestricted pandemic phase, do not pick up on the many peaks and dips in the COVID-19 data, as evident from plotting the predicted COVID-19 ID against the true COVID-19 ID in Figures \ref{fig:predicted_restricted_I} and \ref{fig:predicted_free_I}.
The Figures also include predictions based on the ARIMA model, fitted to each county separately. 
The predictive accuracy for the ARIMA models does not systematically surpass the GNAR models. 

\subsection{Gaussianity of residuals for GNAR models}
\label{app:qq_plots}
The Gaussianity of the residuals may be statistically tested in a Kolmogorov-Smirnov test. 
Table \ref{tab:mase_res_p_counties} summarises the fit of the GNAR model for the restricted and unrestricted data set for each county, including the p-value for the Kolmogorov-Smirnov test.
\begin{table}[ht]
\centering
\begin{tabular}{l|rrr | rrr}
  \toprule 
   & \multicolumn{3}{c}{\textbf{Restricted}} & \multicolumn{3}{c}{\textbf{Unrestricted}} \\
  County & \multicolumn{1}{c}{$\Bar{\varepsilon}$} & 
  \multicolumn{1}{c}{av. MASE}  & \multicolumn{1}{c}{p} & \multicolumn{1}{c}{$\Bar{\varepsilon}$} & 
  \multicolumn{1}{c}{av. MASE}  & \multicolumn{1}{c}{p} \\
 \midrule 
Carlow & 4.03 (9.9) & 1.31 (0.97) & 0.04 & -16.67 (10.73) & 1.16 (0.75) & 0.00 \\ 
  Cavan & -5.06 (22.37) & 0.82 (0.35) & 0.03 & -14.12 (24.91) & 0.82 (0.34) & 0.00 \\ 
  Clare & 0.96 (4.41) & 0.69 (0.52) & 0.05 & -10.97 (8.25) & 0.83 (0.63) & 0.00 \\ 
  Cork & 2.21 (11.78) & 0.75 (0.46) & 0.10 & -11.18 (12.45) & 0.66 (0.66) & 0.00 \\ 
  Donegal & 8.51 (27.44) & 0.67 (0.39) & 0.00 & -11.74 (18.34) & 0.76 (0.91) & 0.03 \\ 
  Dublin & -0.9 (12.14) & 0.81 (0.75) & 0.04 & -13.35 (8.87) & 1.3 (0.86) & 0.00 \\ 
  Galway & -3.1 (7.81) & 0.91 (0.39) & 0.00 & -10.91 (11.15) & 0.72 (0.6) & 0.00 \\ 
  Kerry & -2.06 (5.65) & 0.65 (0.52) & 0.31 & -9.24 (22.64) & 0.69 (0.34) & 0.03 \\ 
  Kildare & 2.05 (16.65) & 0.52 (0.68) & 0.84 & -15.95 (11.89) & 0.92 (0.68) & 0.00 \\ 
  Kilkenny & 3.26 (2.51) & 2.23 (1.39) & 0.00 & -17.01 (24.74) & 0.76 (0.26) & 0.00 \\ 
  Laois & -16.89 (12.15) & 2.09 (1.5) & 0.00 & -16.36 (19.04) & 0.68 (0.63) & 0.03 \\ 
  Leitrim & -0.95 (26.89) & 0.95 (0.39) & 0.03 & -13.87 (14.78) & 0.58 (0.62) & 0.00 \\ 
  Limerick & 5.35 (11.17) & 0.77 (0.49) & 0.01 & -15.08 (12.86) & 0.86 (0.65) & 0.00 \\ 
  Longford & -9.56 (26.08) & 0.76 (0.41) & 0.03 & -9.57 (12.54) & 0.67 (0.8) & 0.01 \\ 
  Louth & -3.99 (17.69) & 0.53 (0.2) & 0.03 & -16.2 (4.56) & 2.92 (0.82) & 0.00 \\ 
  Mayo & -4.97 (9.38) & 1.05 (0.71) & 0.03 & -9.83 (13.78) & 0.7 (0.33) & 0.00 \\ 
  Meath & -2.45 (4.43) & 0.54 (0.36) & 0.03 & -11.24 (14.17) & 0.59 (0.51) & 0.00 \\ 
  Monaghan & 2.31 (16.52) & 0.56 (0.31) & 0.03 & -11.6 (4.79) & 1.76 (0.73) & 0.00 \\ 
  Offaly & -19.45 (12.02) & 1.21 (0.75) & 0.00 & -11.6 (16.56) & 0.89 (0.42) & 0.00 \\ 
  Roscommon & 12.76 (24.28) & 0.9 (0.94) & 0.04 & -10.41 (23.4) & 0.52 (0.42) & 0.03 \\ 
  Sligo & -2.33 (17.32) & 0.67 (0.42) & 0.00 & -15.46 (30.67) & 0.56 (0.58) & 0.03 \\ 
  Tipperary & -4.07 (30.18) & 0.9 (0.17) & 0.03 & -15.21 (17.21) & 0.8 (0.33) & 0.00 \\ 
  Waterford & 0.16 (6.42) & 0.66 (0.32) & 0.09 & -21.24 (13.07) & 1.13 (0.7) & 0.00 \\ 
  Westmeath & -2.64 (21.5) & 0.51 (0.44) & 0.03 & -14.29 (13.36) & 0.84 (0.75) & 0.00 \\ 
  Wexford & 0.02 (20.03) & 1.23 (0.74) & 0.03 & -14.67 (11.33) & 1.05 (0.61) & 0.00 \\ 
  Wicklow & -4.81 (11.05) & 0.74 (0.36) & 0.00 & -12.3 (6.16) & 1.31 (0.66) & 0.00 \\ 
   \bottomrule 
\end{tabular}
\caption{Average residual $\Bar{\varepsilon}$ and average (av.) MASE value
with standard deviation in brackets (sd), as well as Kolmogorov-Smirnov p-value (p) for each county for restricted and unrestricted 
pandemic phase} 
\label{tab:mase_res_p_counties}
\end{table}

\begin{figure}[h!]
\centering
\subcaptionbox{Delaunay}{\includegraphics[width=0.25\textwidth]{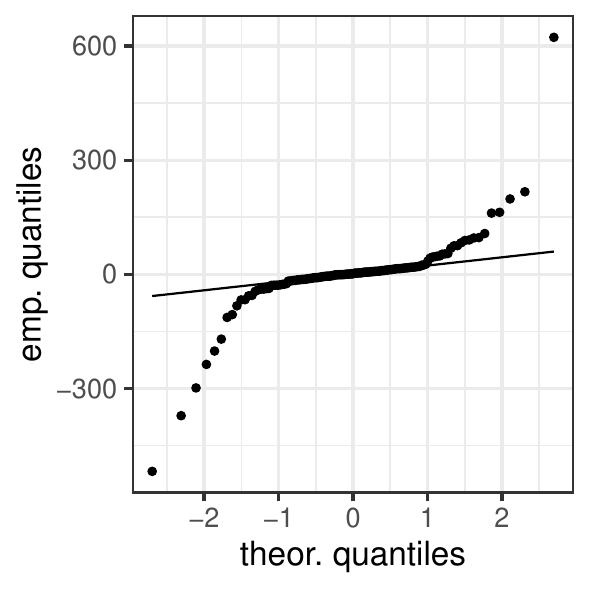}}%
\hspace{0.7cm}
\subcaptionbox{Gabriel}{\includegraphics[width=0.25\textwidth]{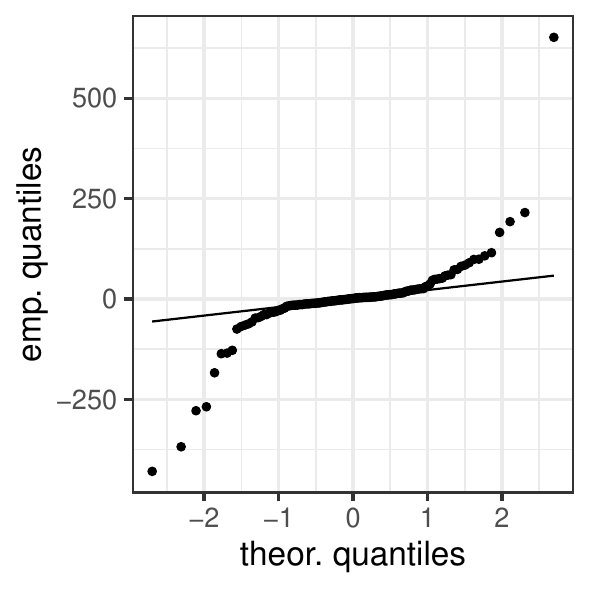}}%
\hspace{0.7cm}
\subcaptionbox{SOI}{\includegraphics[width=0.25\textwidth]{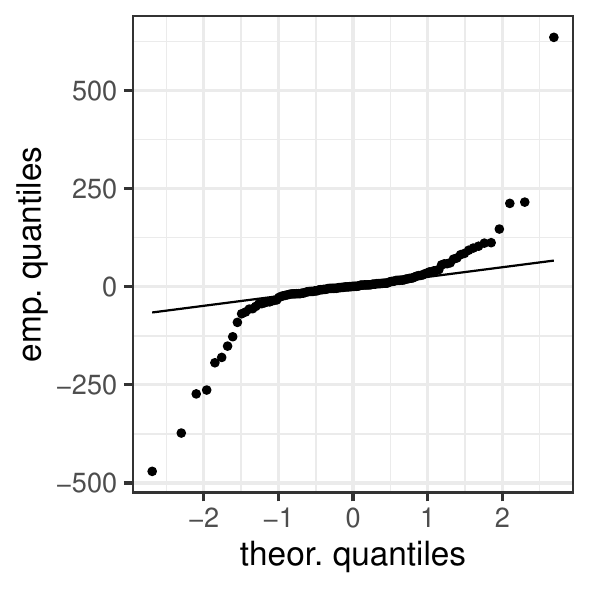}}%
\\
\subcaptionbox{Relative neighbourhood}{\includegraphics[width=0.25\textwidth]{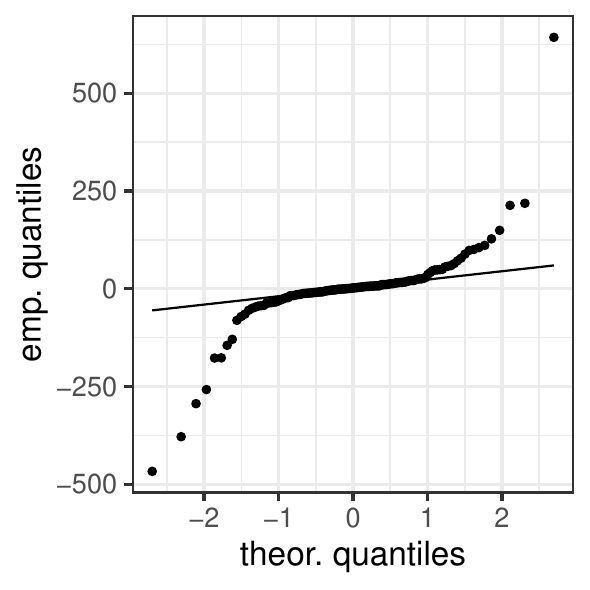}}%
\hspace{0.7cm}
\subcaptionbox{Queen}{\includegraphics[width=0.25\textwidth]{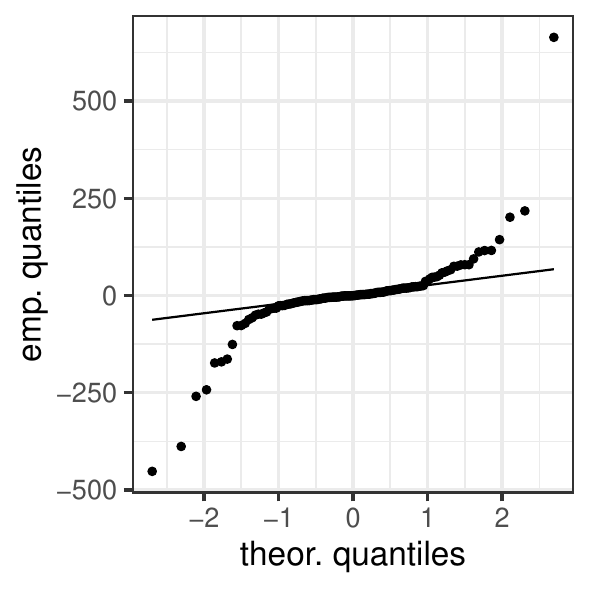}}%
\hspace{0.7cm}
\subcaptionbox{Economic hub}{\includegraphics[width=0.25\textwidth]{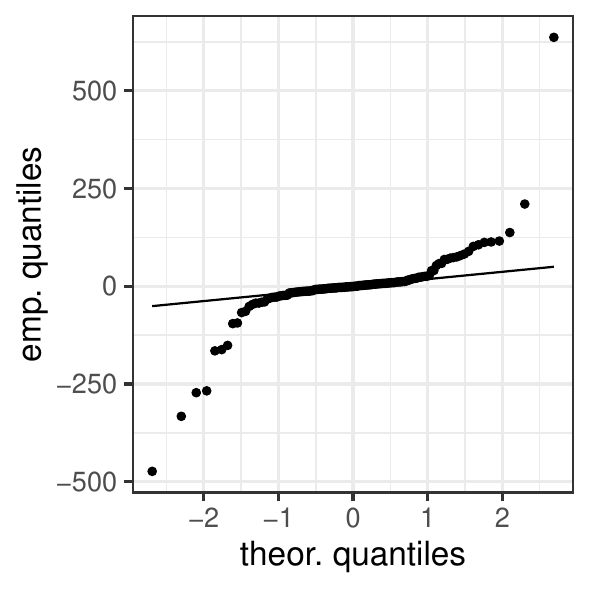}}%
\\
\subcaptionbox{Railway-based}{\includegraphics[width=0.25\textwidth]{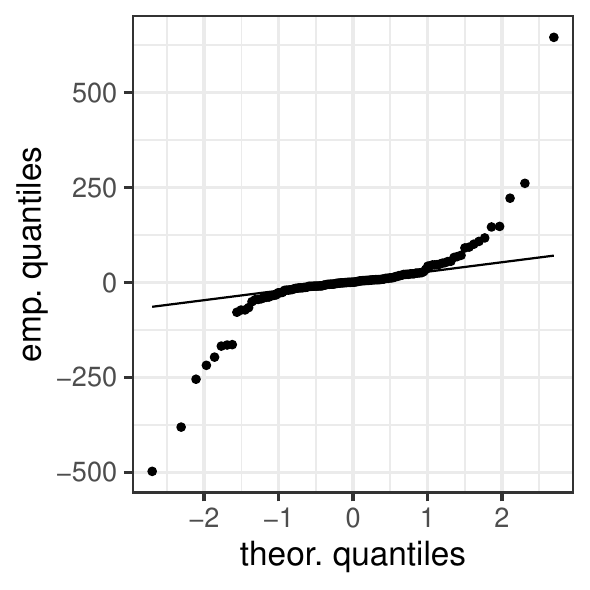}}%
\hspace{0.7cm}
\subcaptionbox{KNN $(k = 21)$}{\includegraphics[width=0.25\textwidth]{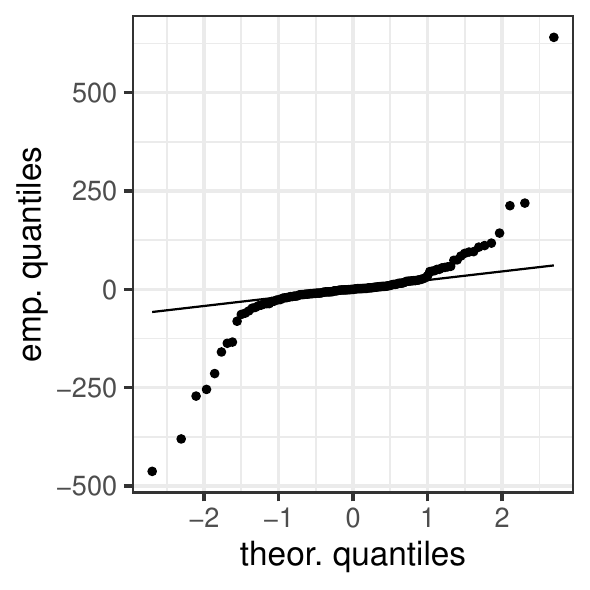}}%
\hspace{0.7cm}
\subcaptionbox{DNN $(d = 325)$}{\includegraphics[width=0.25\textwidth]{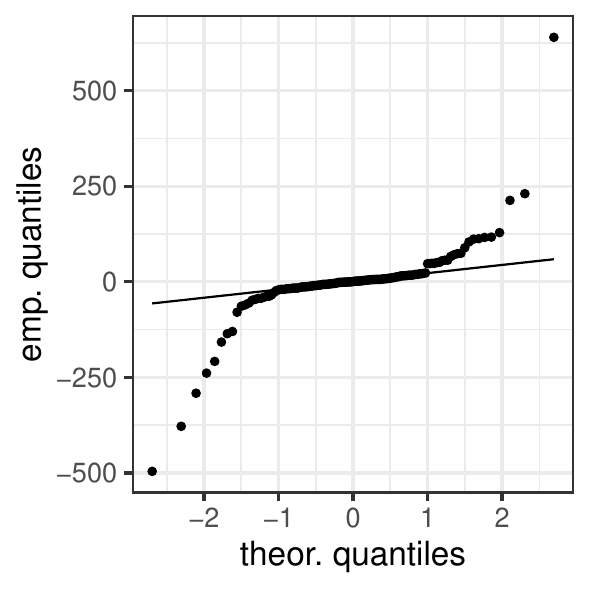}}%
\\
\subcaptionbox{Complete}{\includegraphics[width=0.25\textwidth]{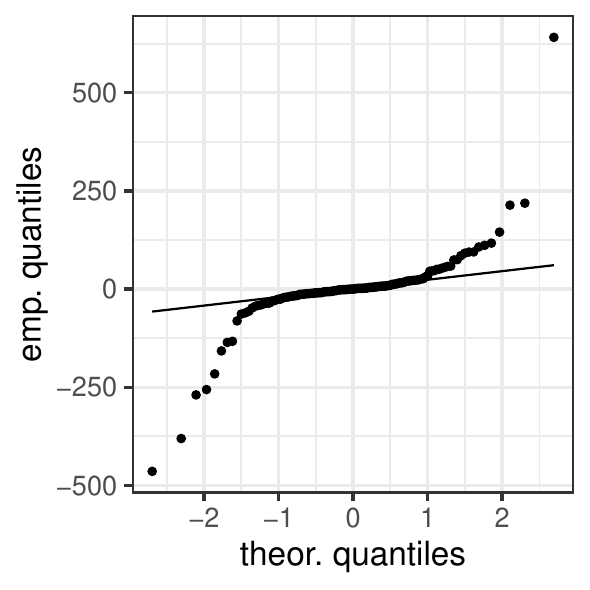}}%
\caption{QQ-plot for the residuals from the best performing GNAR models for county {Dublin} for each network on the entire COVID-19 data set}
\label{fig:qq_dublin_entire_dataset}
\end{figure}

The conclusions based on the Kolmogorov-Smirnov tests are verified by the QQ-plots.
The residual QQ-plots, which we show for the county Dublin as an illustration, indicate non-Gaussian residuals for the GNAR models across all networks, see Figure \ref{fig:qq_dublin_entire_dataset}. 
This conclusion is corroborated when applying the Kolmogorov-Smirnov test across counties. 
The residual variance is low for time periods with low 1-lag COVID-19 ID and high for time periods with high 1-lag COVID-19 ID. 
Thus, there may be some heteroscedastic noise in the data. 

\begin{figure}[h!]
\centering
\subcaptionbox{Restricted}{\includegraphics[width=0.2\textwidth]{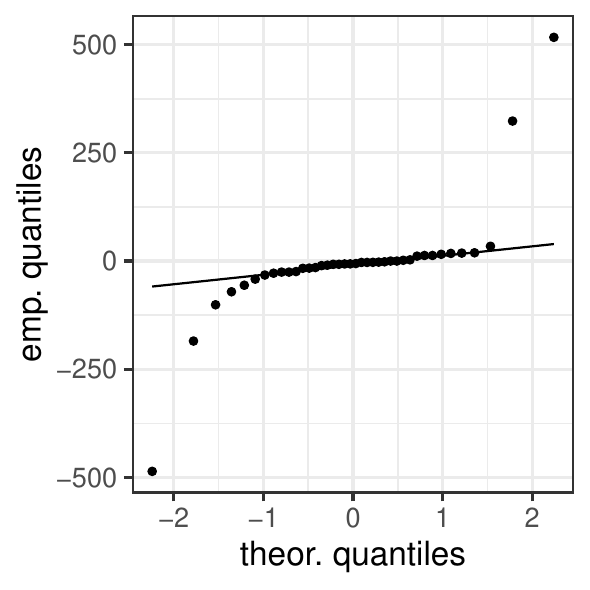}}%
\hspace{0.5cm}
\subcaptionbox{Unrestricted}{\includegraphics[width=0.2\textwidth]{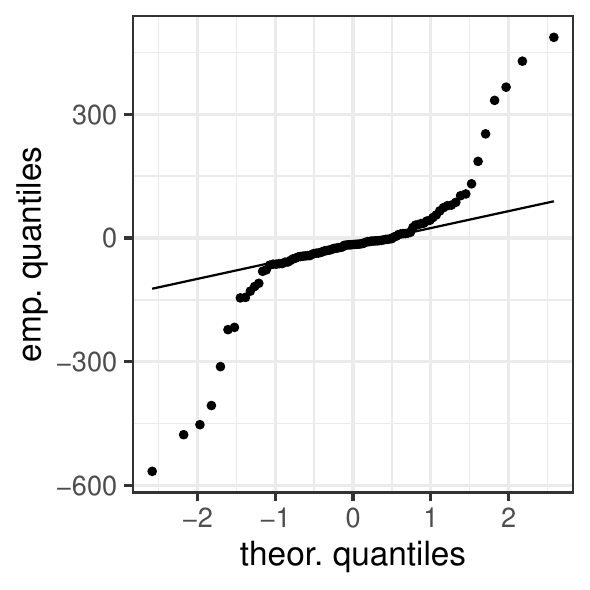}}%
\hspace{0.5cm}
\subcaptionbox{Restricted}{\includegraphics[width=0.2\textwidth]{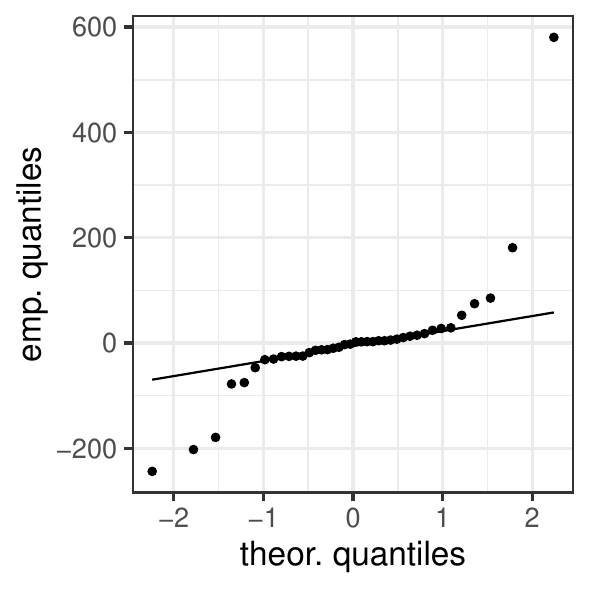}}%
\hspace{0.5cm}
\subcaptionbox{Unrestricted}{\includegraphics[width=0.2\textwidth]{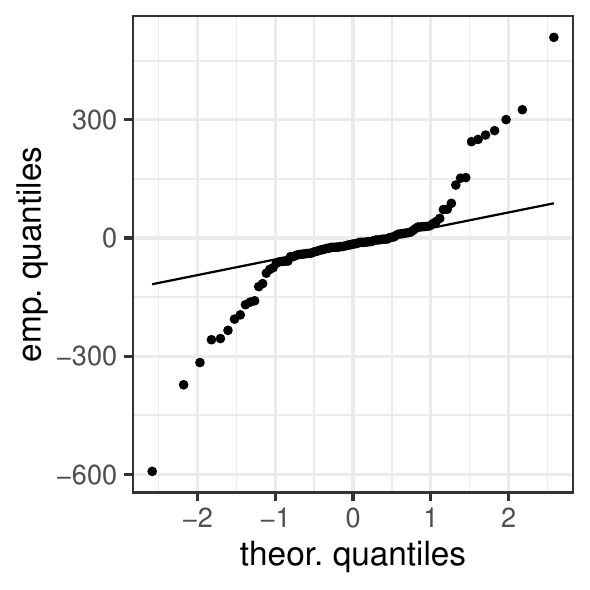}}%
\caption{QQ-plot for the residuals from the best performing GNAR model and network for restricted and unrestricted pandemic phase for county {Carlow} (left) and county {Cavan} (right)}
\label{fig:qq_carlow}
\end{figure}

In contrast to the combined data set, when separating the data set into restricted and unrestricted phases of the pandemic, the QQ-plots for each GNAR model indicate Gaussian residuals for the restricted pandemic phases and residuals which deviate from the normal distribution for the unrestricted pandemic phases. 
The QQ-plots are shown in Figure \ref{fig:qq_carlow} for the counties Dublin, Carlow and Cavan; the remaining counties show similar residual plots.


\end{document}